\documentclass[aps,prd,final,superscriptaddress,showpacs,showkeys,floatfix,preprintnumbers,nofootinbib,letterpaper]{revtex4-2}
\usepackage{graphicx}
\usepackage{float}
\usepackage{dcolumn}

\usepackage[latin1]{inputenc}                    
\usepackage{latexsym}                            
\usepackage{amsfonts}                            
\usepackage{amssymb}                             
\usepackage{amsmath}                             
\usepackage[mathscr]{eucal}                      
\usepackage{dcolumn}                             
\usepackage{subcaption}
\usepackage[english]{babel}
\usepackage{theorem}                            
\usepackage{xcolor}
\usepackage{hyperref}
\usepackage{cancel}
\usepackage{slashed}

\newcommand{\MSbar}{\ensuremath{\overline{\text{MS}}}}
\newcommand{\chip}{{\chi^\prime}}
\newcommand{\MOMqbarq}{\text{MOM}}
\newcommand{\SMOMqbarq}{\text{SMOM}}
\newcommand{\SMOMg}{\text{SMOM}_{\gamma_\mu}}
\newcommand{\MOMpd}{\text{MOM3q}}
\newcommand{\SYMpd}{\text{SYM3q}}
\newcommand{\GeV}{\mathrm{GeV}}
\newcommand{\mcA}{{\mathcal A}}
\newcommand{\mcB}{{\mathcal B}}

\newcommand{\mcO}{{\mathcal O}}
\newcommand{\mcP}{{\mathcal P}}
\newcommand{\mcS}{{\mathcal S}}
\newcommand{\tsep}{{t_2}}
\newcommand{\tins}{{t_1}}
\newcommand{\tskip}{t_\text{skip}}
\newcommand{\tmin}{{t_\text{min}}}
\newcommand{\tmax}{{t_\text{max}}}

\newcommand{\Tr}{\mathrm{Tr}}
\newcommand{\lp}{\left}
\newcommand{\rp}{\right}
\newcommand{\lla}{\langle\langle}
\newcommand{\rra}{\rangle\rangle}

\begin{document}

\title{Proton decay matrix elements on the lattice at physical pion mass}%

\author{Jun-Sik Yoo}
\affiliation{Department of Physics and Astronomy, Stony Brook University, 
  Stony Brook, New York 11794, USA}
\affiliation{Theory Center, Institute of Particle and Nuclear Studies,
  High Energy Accelerator Research Organization (KEK), Tsukuba 305-0801, Japan}
\author{Yasumichi Aoki}
\affiliation{RIKEN Center for Computational Science, Chuo-ku, Kobe, Japan}
\author{Peter Boyle}
\affiliation{Physics Department, Brookhaven National Laboratory, Upton, NY 11973, USA}
\affiliation{University of Edinburgh, Edinburgh EH9 3JZ, UK}
\author{Taku Izubuchi}
\affiliation{Physics Department, Brookhaven National Laboratory, Upton, NY 11973, USA}
\affiliation{RIKEN-BNL Research Center, Brookhaven National Lab, Upton, NY, 11973, USA}
\author{Amarjit Soni}
\affiliation{Physics Department, Brookhaven National Laboratory, Upton, NY 11973, USA}
\author{Sergey Syritsyn}
\affiliation{Department of Physics and Astronomy, Stony Brook University, 
  Stony Brook, New York 11794, USA}
\affiliation{RIKEN-BNL Research Center, Brookhaven National Lab, Upton, NY, 11973, USA}
\date{\today}%
\begin{abstract}
Proton decay is a major prediction of Grand-Unified Theories (GUT) and its observation would indicate
baryon number violation that is required for baryogenesis.
Many decades of searching for proton decay have constrained its rate and ruled out some of the
simplest GUT models.
Apart from the baryon number-violating interactions, this rate also depends on transition amplitudes
between the proton and mesons or leptons produced in the decay, which are matrix elements of
three-quark operators.
We report nonperturbative calculation of these matrix elements for the most studied two-body decay
channels into a meson and antilepton done on a lattice with physical light and strange quark masses
and lattice spacings $a\approx0.14$ and $0.20$ fm.
We perform nonperturbative renormalization and excited state analysis to control associated
systematic effects.
Our results largely agree with previous lattice calculations done with heavier quark masses and thus
remove ambiguity in ruling out some simple GUT theories due to quark mass dependence of hadron
structure.
\end{abstract}
\preprint{RBRC-1333}
\preprint{KEK-CP-0385}

\maketitle
\tableofcontents


\section{Introduction
  \label{sec:intro}}
Proton decay is a $|\Delta B|=1$ baryon number-violating process that has been predicted by Grand
Unified Theories (GUT)\cite{Pati:1973uk,Georgi:1974sy,Fritzsch:1974nn} but has not been observed so far.
The Standard Model Lagrangian does not contain baryon number-violating interactions, and although
sphaleron processes can convert baryons into antileptons, such transitions are highly suppressed at
temperatures below the electroweak phase transition~\cite{Espinosa:1989qn,Ambjorn:1990pu}.
Discovery of proton decay may potentially fulfil one of the three prerequisites to explain the
Baryon asymmetry in the Universe~\cite{Sakharov:1967dj}\footnote{
  There are viable alternatives such as leptogenesis~\cite{Fukugita:1986hr}.}, 
and also demand extension of the Standard
Model to accomodate baryon number violation~\cite{Lee:1994vp}, potentially involving
supersymmetry~\cite{Hisano:1992jj,Murayama:2001ur}.

There have been several experiments aimed at observing proton decay:
KOLAR~\cite{Krishnaswamy:1986id}, NUSEX~\cite{Bellotti:1982sw}, Frejus~\cite{Deuzet:1985di}, 
SOUDAN~\cite{Litchfield:1990cq}, Kamiokande~\cite{Hirata:1989kn}, IMB~\cite{McGrew:1999nd}, 
and Super Kamiokande~\cite{Miura:2016krn}.
The most recent experiment, Super Kamiokande, has been operating for more than two decades and has
set proton partial lifetime limits $\tau/Br(p\rightarrow e^+ \pi^0) \ge 1.6 \times 10^{34}$ 
years~\cite{Miura:2016krn} and $\tau/Br(p\rightarrow \nu K^+) \ge 5.9 \times 10^{33}$
years~\cite{Super-Kamiokande:2014otb}. 
The next generation of experiments to look for decay of the proton, DUNE~\cite{Acciarri:2016ooe} and
Hyper~Kamiokande~\cite{Abe:2018uyc}, are expected to start observation in 2024 and 2027,
respectively, and to improve these limits by roughly an order of magnitude.
Hyper-Kamiokande is a water Cherenkov detector and is best suited to constrain the pion decay mode;
it will improve the bound on $\tau/Br(p\rightarrow e^+\pi^0)$ to $\gtrsim 10^{35} $ in 8 years of
operation (4 Mt*year exposure)~\cite{Abe:2011ts}.
DUNE, a LArTPC detector, is expected to produce the best limit on the kaon decay mode 
$\tau/Br(p\rightarrow \bar\nu K) \gtrsim 6 \times 10^{34}$ years~\cite{Acciarri:2015uup} in two
decades of running.
JUNO is another neutrino experiment to be installed in China~\cite{JUNO:2015zny}.
It is a 20-kt liquid scintillator (LS) detector buried 700 m under the granite mountain, which
can detect proton decays in the $p\to K^+\bar\nu$ channel.
First it detects $K^+$ decay with kinetic energy of 105 MeV from proton decay, and then
it traces the subsequent $\mu$ decay. 
With the efficiency of 64\% and the background of 0.5 event per 20kt·year, it can reach sensitivity
of $1.9\cdot10^{34}$ years in 10 years of operation.

Grand Unified Theories~\cite{Pati:1973uk,Georgi:1974sy,Fritzsch:1974nn} and Supersymmetric Grand
Unified Theories (SUSY-GUTs)~\cite{Dimopoulos:1981zb,Witten:1981nf,Lucas:1996bc} hypothesize
existence of larger gauge groups that unify all the interactions at some energy scale 
$\Lambda_\text{GUT} \approx 10^{16}\,\mathrm{GeV}$ that may lead to effective quark-lepton
interactions causing the proton to decay.
At the hadronic scale, these effective interactions are the lowest, dimension-6 operators comprised
of four fermion fields~\cite{Weinberg:1979sa,Wilczek:1979hc},
\begin{equation}
\label{eqn:pdecay_int}
\begin{gathered}
\mathcal{L}_{\text{eff}} = \sum_{I}C_I \mcO_I\,+\, \mathrm{h.c.}\,,
\\
\mcO_I = \epsilon^{abc}   (\bar{q}^{a\text{C}} P_{\chi_I} q^{b}) 
  (\bar{\ell}^{\text{C}}P_{\chip_I} q^c)\,,
\end{gathered}
\end{equation}
where $\{\bar{q},\bar\ell\}^C=\{q,\ell\}^T C$ are charge-conjugated fields
~\footnote{
  Throughout the paper, we use Euclidean conventions for $\gamma$-matrices (see, e.g.,
  Ref.~\cite{Rinaldi:2019thf}), so that $C=\gamma_2\gamma_4$.
},
and the chirality projectors $P_{\chi^{(\prime)}=R,L}=\frac{1\pm\gamma_5}2$.
The interacting quark fields $q=u,d,s$ and the Wilson coefficients $C_I$ depend on the character and
the scale of an underlying unified theory.
In the simplest case, such interaction describes proton decay into a lepton and one or more mesons.
Neglecting for now the lepton mass $m_{\bar{\ell}}\ll m_N$, the partial decay width of the channel
$p \longrightarrow \Pi \bar{\ell}$ is equal to
\begin{equation}
\label{eqn:decayrate}
\Gamma (p \rightarrow \Pi \bar{\ell}  ) = 
  \frac{m_N}{32\pi} \Big[ 1 - \Big( \frac{m_\Pi}{m_N}\Big)^2 \Big]^2 
   \,\Big| \sum_I C_I W^I_{\bar\ell}\Big|^2,
\end{equation}
where the meson states are $ \Pi  = \pi,\,K $, 
the final leptons are  $\bar\ell = e^+,\,\bar\nu,\,\mu^+$, 
and the $ W_{\bar\ell} $ are the $p\to\Pi$ transition matrix elements of the quark component of the
operators $\mcO_I$~(\ref{eqn:pdecay_int}), which are classified below.
The decay rates are determined by effective interactions induced by particular GUT hypotheses at
scale $\Lambda_{(\text{GUT})}$ and encoded in the Wilson coefficients 
$C_I = \frac{\tilde c_I}{\Lambda^2}$, where $\tilde c_I$ is a dimensionless $O(1)$ coupling
renormalized to the nuclear scale. 
However, the hadronic matrix elements $\langle \Pi\ell |\mcO_I|p\rangle$  are determined by
nonberturbative quark dynamics and have to be evaluated either in a model or, preferably, in an ab
initio QCD calculation.
From dimensional analysis, $W^I_\ell\propto\Lambda_\text{QCD}^2$ and the proton decay rate is
suppressed as $\Gamma \propto m_N |c_I|^2 (\Lambda_\text{QCD}/\Lambda_{(\text{GUT})})^4
\propto \Lambda_\text{QCD}^5 / m_X^4$ where $m_X$ is the mass of a unified-theory boson.
Using $\Lambda_{QCD}=0.2\,\text{GeV}$ yields a reasonable estimate for the form factors value
$|W_{\bar\ell}|\approx 0.04\,\text{GeV}^2$ and an estimate for the partial lifetime 
\begin{equation}
\tau/Br(p\to\pi\bar\ell) \approx 1.4 \cdot 10^{33}\,\text{years} \cdot
  \Big(\frac{ \Lambda_\text{GUT} }{ 10^{15}\,\text{GeV} }\Big)^4 
  \cdot \frac1{|\tilde c_I|^2} \,.
\end{equation}

Prior to lattice QCD, matrix elements of these effective operators were estimated using the
non-relativistic quark model of the nucleon~\cite{Gavela:1981cf}, the chiral
lagrangian~\cite{Kaymakcalan:1983uc}, and the MIT bag model~\cite{Okazaki:1982eh,Martin:2011nd}.
Eliminating model uncertainty requires ab initio QCD calculations on a lattice, which have been
pursued with improving methodology since nucleon structure calculations became possible.
Amplitudes of transitions from a nucleon to a meson state can be approximated using
proton-to-vacuum (annihilation) decay constants determined on a lattice and Chiral perturbation
theory (ChPT)~\cite{Claudson:1981gh} (so-called ``indirect method'').
Alternatively, these amplitudes can be computed on a lattice directly, which enables better control of
systematic effects.
The former method was used quenched-QCD calculations with Wilson valence quark action~\cite{Hara:1986hk,
Bowler:1987us,Tsutsui:2004qc} and Domain Wall fermion (DWF) action~\cite{Aoki:2006ib}, 
as well as in unitary QCD with dynamical DWF action~\cite{Aoki:2008ku}.
Direct calculation of proton-to-meson transition matrix elements was performed in quenched QCD with
Wilson valence quarks~\cite{Aoki:1999tw} and DWF quarks~\cite{Aoki:2006ib}, 
as well as in unitary QCD with $N_f=2+1$ dynamical domain wall fermions~\cite{Aoki:2013yxa,Aoki:2017puj}.
In Refs.~\cite{Aoki:1999tw,Aoki:2006ib,Aoki:2017puj}, results from the indirect method were also
reported.

Although significant progress has been made in improving calculations of proton decay amplitudes, 
some important systematic uncertainties are still remaining.
The most recent direct calculation~\cite{Aoki:2017puj} reports uncertainty $20-40\%$ in the proton
decay amplitudes, and also reports disagreement between the direct and the indirect methods.
Since the indirect method relies on chiral perturbation theory (ChPT), it is plausible that the pion
masses $m_\pi\gtrsim340\,\mathrm{MeV}$ used in that calculation were too heavy for the ChPT to work.
However, direct-method transition amplitudes computed with unphysical heavy pion masses also
require chiral extrapolation, which may also result in systematic uncertainty.
In particular, in the framework of the chiral-bag proton model, it has been suggested that
the proton decay matrix elements may depend dramatically on the quark mass~\cite{Martin:2011nd}.
If this is the case, some GUT models (e.g., SUSY- and regular $SU(5)$) may evade constraints even
with the presently available experimental data.
It is also important to note that the effective proton decay operators~(\ref{eqn:pdecay_int})
contain chiral quark fields, and preserving chiral symmetry is particularly challenging in lattice
calculations.
Some of the (valence) quark actions used in earlier
calculations~\cite{Hara:1986hk,Bowler:1987us,Aoki:1999tw,Tsutsui:2004qc} break chiral symmetry
explicitly.

In this work, we study proton decay matrix elements using chirally symmetric dynamical $N_f=2+1$ and
valence Domain Wall fermions with physical quark masses.
We compute these matrix elements with both the direct and indirect methods and compare their
results.
A formidable progress has been made towards lattice calculations with chiral fermions at the
physical point~\cite{Blum:2014tka,Boyle:2015exm,Blum:2019ugy}.
We use two ensembles with lattice spacings $a=0.20$ and $a=0.14\,\mathrm{fm}$ and explore different
kinematics in order to obtain reliable interpolation to the physical decay kinematic points.
Together with nonperturbative renormalization and analysis of nucleon and meson excited states, 
our calculation is aimed to eliminate common lattice QCD systematic effects.

Another potential proton decay channel is into three leptons or a lepton and one or more
photons~\cite{Silverman:1980ha,Hambye:2017qix,Girmohanta:2019xya}.
Such processes can occur either through effective dimension-9 operators~\cite{Hambye:2017qix}, which
may only be relevant if the BSM physics scale $\Lambda_{BSM}\ll\Lambda_{GUT}$, 
or through emission of a photon from quark or charged lepton involved into effective dimension-6
interaction~(\ref{eqn:pdecay_int}).
Decays into three leptons have also been constrained with data from Super Kamiokande; 
for example, $\tau/Br(p\rightarrow \ell\bar{\ell}^{\prime}\bar{\ell}^{\prime\prime}) 
\geq 0.9\ldots(3.4)\cdot10^{34}$ years for $\ell=e\,\mu$~\cite{Tanaka:2020emn}.
Rates of such processes depend on the same proton decay constants $\langle\text{vac}|\mcO_I|p\rangle$
as the ones in the indirect calculation of proton decay amplitudes $p\to\Pi\ell$ mentioned above.
We report results of nonperturbative lattice calculations of both proton-meson and proton-vacuum
amplitudes that are important to proton decay phenomenology.

Another baryon-number violating process that could be responsible for baryogenesis is the six-quark
interaction leading to nonconservation of the $(B-L)$ number and $|\Delta B|=2$ transitions.
Such events are potentially observable as neutron-antineutron oscillations~\cite{Phillips:2014fgb},
and their matrix elements have been recently computed on a lattice with chirally symmetric action at
the physical point~\cite{Rinaldi:2018osy,Rinaldi:2019thf}.

The paper is organized as follows.
In Section~\ref{sec:method}, we introduce our conventions and notations, describe our methodology and
lattice QCD setup for computing nucleon-meson matrix elements.
A detailed discussion of nonperturbative renormalization methodology and results is presented in
Sec.~\ref{sec:renorm}.
In Section~\ref{sec:result}, we show details of our analysis and present our results 
for proton and meson spectra, proton decay amplitudes, and proton decay constants obtained on the 
two lattice QCD ensembles as well as in the continuum limit.
Finally, in Sec.~\ref{sec:concl} we compare our results to previous calculations, discuss systematic
errors in our calculation, discuss the impact of our results, and suggest further directions to
improve systematic uncertainties.

\section{Methodology
  \label{sec:method}}
\subsection{Operator definitions}
The minimal complete set of the lowest-dimension effective proton decay operators symmetric under
$SU(3)_c\times SU(2)_{EW} \times U(1)$ has been constructed in
Refs.~\cite{Weinberg:1979sa,Wilczek:1979hc,Abbott:1980zj}.
Using notation of Ref.~\cite{Abbott:1980zj,Aoki:1999tw}, these operators are
\begin{align}
\mcO^{(1)}_{abcd} &= (\bar{D}^C_{ia}\,U_{jb})_R\,(\bar{Q}^C_{\alpha k c} L_{\beta d})_L
  \, \epsilon_{ijk} \epsilon_{\alpha\beta}
\,,\\
\mcO^{(2)}_{abcd} &= (\bar{Q}^C_{\alpha i a}\,Q_{\beta jb})_L\,(\bar{U}^C_{k c}l_{d})_R
  \, \epsilon_{ijk} \epsilon_{\alpha\beta}
\,,\\
\bar\mcO^{(4)}_{abcd} &= 
  (\bar{Q}^C_{\alpha i a}\,Q_{\beta jb})_L\,(\bar{Q}^C_{\gamma k c} L_{\delta d})_L
  \, \epsilon_{ijk} \epsilon_{\alpha\delta} \epsilon_{\beta\gamma}
\,,\\
\mcO^{(5)}_{abcd} &= (\bar{D}^C_{i a}\,U_{jb})_R\,(\bar{U}^C_{k c}l_{d})_R
  \, \epsilon_{ijk}
\,,
\end{align}
where $a,b,c,d$ are generation indices, $i,j,k$ are $SU(3)_c$ indices,
$\alpha,\beta,\gamma,\delta$ are the indices of the left-handed $SU(2)_{EW}$ fermion doublets, 
and 2-spinors of the indicated chirality ($R,L$) are contracted inside the parentheses.
From now on, we will omit the color indices and imply contraction with the antisymmetric tensor
$\epsilon_{ijk}$.
These operators conserve the $(B-L)$ number, and the outgoing antilepton ($e^+$, $\mu^+$, $\bar\nu$)
may have only electroweak interaction with quark fields present in the initial and final states,
which may be neglected at the hadronic scale.
The amplitude $p\to\Pi\ell$ may then be factorized into
\begin{equation}
\label{eqn:p2m_amplitude}
{\mathcal M}(p\to\Pi\ell) 
  = \bar{v}_{\ell\alpha}^C(\vec q) \, \langle\Pi(\vec p^\prime) |  
      (\bar{q_{1}}^C q_{2})_\chi \, q_{3\chip\alpha} | N(\vec k)\rangle
\end{equation}
where $\chi^{(\prime)}=R,L$ denotes chirality and $\bar{v}_{\ell}^C$ is the spinor of the antilepton
in the final state with momentum $\vec q=\vec k - \vec p$.
To avoid redundancy due to parity symmetry, we will only consider combinations $(\chi,\chip)=(L,L)$
and $(R,L)$ below.
The following quark combinations $q_{1,2,3}$ are possible for the initial proton $N=p$ and the 
energy-allowed final meson $\Pi=\pi^0$, $\pi^+$, $K^0$, $K^+$, and $\eta$ states:
\begin{equation}
\label{eqn:3qop_me_def}
\begin{aligned}
\langle \pi^0 | (\bar{u}^C d)_\chi \, u_L)|p\rangle & \,,\\
\langle \pi^+ | (\bar{u}^C d)_\chi \, d_L)|p\rangle &= U_{1\chi} \,,\\
\langle   K^0 | (\bar{u}^C s)_\chi \, u_L)|p\rangle &= S_{1\chi} \,,\\
\langle   K^+ | (\bar{u}^C s)_\chi \, d_L)|p\rangle &= S_{2\chi} \,,\\
\langle   K^+ | (\bar{u}^C d)_\chi \, s_L)|p\rangle &= S_{3\chi} \,,\\
\langle   K^+ | (\bar{d}^C s)_\chi \, u_L)|p\rangle &= S_{4\chi} \,,\\
\langle \eta  | (\bar{u}^C d)_\chi \, u_L)|p\rangle &= S_{5\chi} \,.
\end{aligned}
\end{equation}
Similar quark combinations can be enumerated for neutron decays, and their relation to the
proton matrix elements under isospin symmetry can be found in Ref.~\cite{Aoki:2013yxa}.
In addition, isospin symmetry requires that 
\begin{equation}
\langle \pi^0 | (\bar{u}^C d)_\chi \, u_L | p \rangle
  = \frac1{\sqrt2} \langle \pi^+ | (\bar{u}^C d)_\chi \, d_L | p \rangle 
  = \frac1{\sqrt2} U_{1\chi}\,,
\end{equation}
which is precisely fulfilled in our $SU(2)_f$-symmetric calculation at the contraction level.
Computing amplitudes of $\eta$-channel decays ($S_{5\chi}$) require disconnected quark contractions.
Such calculation is challenging with DW fermions at the physical point, and we omit these amplitudes
in the present work.

Matrix elements between the proton and the meson+lepton pair in Eq.~(\ref{eqn:p2m_amplitude}) 
may be decomposed into linear combinations of two form factors
$W_{0,1}$~\cite{Aoki:1999tw}\footnote{
  The conventions for on-shell nucleon $u_N$ and antilepton $\bar{v}_\ell^C$ states can be found
  in App.~\ref{sec:app_convent}.}:
\begin{equation}
\label{eqn:pdecay_ff_def}
\begin{aligned}
&\bar{v}^C_{\ell\alpha}(\vec q) \langle \Pi(\vec{p}) | \mcO^{\chi\chip}_\alpha(q) | N (\vec{k}) \rangle 
\\
&= \big(\bar{v}^C_\ell(\vec q) \, P_\chip \, \big[ W^{\mcO}_0(Q^2)  
    - \frac{i \slashed{q}}{m_N} W^\mcO_1(Q^2) \big]  u_N (\vec{k}) \big)
\\
&\approx \big[\bar{v}^C_\ell(\vec q) P_{\chip} u_N(\vec{k})\big] \, W_0(-m_\ell^2) + O(m_{\ell}/m_N) \,,
\end{aligned}
\end{equation}
where $Q^2 = -q^2 = -(E_N-E_\Pi)^2 + (\vec k - \vec p)^2$ and 
$\bar{v}^C_\ell(\vec q)$ is the spinor of the antilepton in the final state.
In the last line, the contribution of the form factor $W_1$ can be neglected in decays into
positrons and antineutrinos but not for antimuons with $m_\ell/m_N\approx 0.1$.
Also, unless the outgoing antilepton is ultrarelativistic, there is interference between 
left- and right-handed amplitudes, and the decay rate~(\ref{eqn:decayrate}) takes the form
\begin{equation}
\label{eqn:pdecay_rate_full}
\Gamma (p \rightarrow \Pi \bar{\ell}  )
  = \frac1{8\pi}\frac{E_\ell|\vec q_\ell|}{m_N} \Big(\mcA - \frac{m_\ell}{E_\ell} \mcB\Big)\,,
\end{equation}
where $E_\ell$ and $\vec q_\ell$ are the outgoing antilepton energy and momentum, and
\begin{align}
\mcA &= \big|W^L_{\bar\ell}\big|^2 + \big|W^R_{\bar\ell}\big|^2\,,\\
\mcB &= 2\,\mathrm{Re} \big[ W^L_{\bar\ell} \,\big(W^R_{\bar\ell}\big)^* \big]\,,
\end{align}
and the on-shell antilepton helicity matrix elements are~\footnote{
  The $O(W_1)$ correction to the two body decay amplitude given in Ref.\cite{Aoki:2017puj} is 
  oversimplified and cofusing. The correct formulae are given here.
}
\begin{equation}
\label{eqn:pdecay_lepton}
\begin{aligned}
W_{\bar\ell}^L = \sum_{i,\chi} \big[C^{i,\chi L} W_0^{i,\chi L} 
    - \frac{m_\ell}{m_N} C^{i,\chi R} W_1^{i,\chi R}\big]_{Q^2=-m_\ell^2}\,,\\
W_{\bar\ell}^R = \sum_{i,\chi} \big[C^{i,\chi R} W_0^{i,\chi R} 
    - \frac{m_\ell}{m_N} C^{i,\chi L} W_1^{i,\chi L}\big]_{Q^2=-m_\ell^2}\,
\end{aligned}
\end{equation}
(the summation over $i$ does not include the $\chi,\chip$ helicities~(\ref{eqn:p2m_amplitude}), 
unlike $I$ in Eqs.~(\ref{eqn:pdecay_int},\ref{eqn:decayrate})).
In the $m_\ell\to0$ limit, the interference contribution $\mcB$ disappears, and the decay 
rate~(\ref{eqn:pdecay_rate_full}) is simplified to Eq.~(\ref{eqn:decayrate}).

For indirect evaluation of the $p\to\Pi\bar{\ell}$ amplitudes using chiral perturbation theory, 
as well as computing $3\ell$- or $\ell\gamma$-channel decay amplitudes,
one needs the nucleon decay constants from the following matrix elements 
\begin{equation}
\label{eqn:pdecay_lec}
\begin{aligned}
\big\langle\mathrm{vac}\big|\, (\bar{u}^C d)_R \, u_L \, \big|N\big\rangle 
&= \alpha P_L u_N\,,
&& 
\big\langle\mathrm{vac}\big|\, (\bar{u}^C d)_L \, u_R \, \big|N\big\rangle 
&= -\alpha P_R u_N\,,
\\
\big\langle\mathrm{vac}\big|\, (\bar{u}^C d)_L \, u_L \, \big|N\big\rangle 
  &= \beta P_L u_N\,,
&&
\big\langle\mathrm{vac}\big|\, (\bar{u}^C d)_R \, u_R \, \big|N\big\rangle 
  &= -\beta P_L u_N\,.
\end{aligned}
\end{equation}
Combinations of these constants 
\begin{align}
\big\langle\mathrm{vac}\big|\, (\bar{u}^C \gamma_5 d) \, u \, \big|N\big\rangle
&= (\alpha -\beta) u_N\,,
\\
\big\langle\mathrm{vac}\big|\, (\bar{u}^C d) \, \gamma_5 u \, \big|N\big\rangle
&= -(\alpha +\beta) u_N\,,
\end{align}
yield the overlap of the positive-parity nucleon ground state with nonrelativistic
(scalar diquark and upper $u$-quark) and relativistic (pseudoscalar diquark and
lower $u$-quark) nucleon interpolating fields, respectively.
In the nonrelativistic limit corresponding to calculations with unphysical heavy $u,d$ quark
masses, it is expected that $|\alpha+\beta|\ll|\alpha-\beta|$.
In the ``indirect'' method, proton decay amplitudes are combinations of the low-energy constants $\alpha$,
$\beta$, quark contributions to the baryon spin, and the meson decay constants
$f_{\pi,K}$~\cite{Aoki:1999tw,Aoki:2017puj}.
These formulas are collected in Appendix~\ref{sec:app_indirect} for completeness.

\subsection{Lattice Setup}
For our calculation, we use physical-point ensembles of gauge fields on $24^3 \times
64$~\cite{Blum:2014tka} and $32^3 \times 64$~\cite{Arthur:2012yc} lattices ensembles with spatial
volumes $\approx (4.8\text{ fm} )^3$ and $(4.6 \text{ fm})^3$, respectively.
These ensembles have been generated by the RBC/UKQCD collaboration using I-DSDR gauge action and
$N_f=2+1$ flavors of dynamical quarks with M\"obius Domain Wall Fermion (MDWF) action.
These MDWF fermions possess chiral symmetry due to the additional, fifth dimension of $L_5 = 24$ and
$L_5 =12$, respectively, which are sufficient to suppress chiral symmetry breaking effects otherwise
present in lattice fermion actions.
To soften explicit chiral symmetry breaking effects due to the relatively large lattice spacing,
these ensembles also employ the dislocation-suppressing-determinant-ratio
(DSDR)~\cite{Renfrew:2008zfx}.
Lattice spacings, bare quark masses, pseudoscalar meson masses, and other parameters are summarized
in Tab.~\ref{tab:latt_param}.
The masses of mesons and of the proton are reported below in Sec.\ref{sec:result_spectrum}.
As our lattices are nearly precisely at the physical point, our results below will not require
chiral extrapolation. 
Slight deviations of the pion and kaon masses from their physical values can be, in principle,
rectified by ChPT-inspired corrections to our results, but the precision we aim for in this work
does not warrant such a step.

\begin{table*}[t]
\caption{\raggedright
  Lattice parameters for the 24ID and 32ID ensembles. 
  Both ensembles have I-DSDR gauge and (zMobius) Domain Wall fermion actions.
  The pion and kaon masses are determined in our analysis (see Sec.~\ref{sec:result_spectrum}).
  Lattice spacings and residual masses are computed elsewhere~\cite{Blum:2019ugy,MurthyD:PhD:2017,Tu:2020vpn}.
  In the last group of columns, we show the number of light-quark deflation eigenvectors, the
  numbers of CG iterations to compute light- and strange-quark propagators for approximate samples,
  and the number of gauge configurations analyzed.  
  \label{tab:latt_param}}
\begin{tabular}{lll|cllll|llc|ccc|c}
\hline \hline
$L_x^3\times L_t$   & $a^{-1}\text{ [GeV]}$   & $\beta$   
  & $L_{5f}(L_{5s})$  & $M_5$ & $a m_\text{res}$   & $a m_l$   & $a m_s$   
  & $a m_\pi$ & $a m_K $ & $m_\pi L$     
  & $N_\text{EV}(N_\text{basis})$   & $N_\text{CG}^{u/d}$ & $N_\text{CG}^s$  & $N_\text{cfg}$ \\
\hline
$24^3 \times 64$    & 1.023(2)  & 1.633   
  & 32/12   & 1.8   & 0.00228(1) & 0.00107   & 0.0850 
  & 0.1378(7) & 0.5004(25) & 3.31   & 2000(1000)    
  & 300 & 200  & 140\\
\hline        
$32^3 \times 64$    & 1.378(5)  & 1.75    
  & 12/12   & 1.8   & 0.00189(1) & 0.0001    & 0.0450 
  & 0.1008(5) & 0.3543(6)   & 3.25   
  & 2000(250)    & 200  & 200 & 112 \\
\hline \hline 
\end{tabular}
\end{table*}

In order to make the numerical calculation affordable, we perform \emph{``all-mode-averaging''
(AMA)} sampling~\cite{Blum:2012uh}, in which we approximate the light and strange quark propagators
with truncated solutions to the MDWF operator~\cite{Brower:2009sb}.
On the 24ID ensemble, the MDWF operator itself is approximated with ``z-M\"obius'' operator, 
in which complex coefficients $b_5,c_5$ are varied along the fifth dimension so that it can be
reduced to $L_{5s}=12$ while keeping the residual mass $m_{res}$ the same. 
For a better approximation of the low-eigenmode space of the light-quark Dirac operator, we augment
the truncated Conjugate-Gradient solver with deflation using a combination of exact and
coarse-blocked eigenvectors, which are computed with multigrid Lanczos
algorithm~\cite{Clark:2017wom}.
On each gauge configuration, we compute 32 approximate (``sloppy'') samples with such truncated
quark propagators.
In order to correct for any potential bias, we recompute one sample on each configuration using
exact quark propagators.
We find that with our parameters the AMA approximation is very efficient, i.e., the statistical
variance of the difference between the approximate and exact samples is negligible, and the
statistical precision is always dominated by fluctuations in the approximate samples.

\subsection{Nucleon-meson correlators on a lattice}
In order to compute the matrix elements in Eqs.(\ref{eqn:pdecay_ff_def},\ref{eqn:pdecay_lec}) on a
lattice, we evaluate three-point correlation functions of proton creation $\bar{N}$, 
proton decay  $\mcO^{\chi}_\alpha = (\bar{q}_1^C q_2)_\chi q_{3L\alpha}$, and meson annihilation
$J_\Pi$ operators:
\begin{equation}
\label{eqn:pdecay_threept}
\begin{aligned}
&C_{\alpha\beta}^{\Pi\mcO N}(\vec p,\vec q;\tsep,\tins) 
  = \sum_{\vec y,\vec z} \, e^{-i\vec p\vec y - i\vec q\vec z + i\vec k\vec x} 
\langle J_\Pi(\vec y, x_4+\tsep) \, \mcO^{\chi\chip}_\alpha(\vec z, x_4+\tins) \,
  \bar{N}_\beta (x) \rangle\,.
\end{aligned}
\end{equation}
The spin indices are contracted with polarization matrices $\mcP$ 
\begin{equation}
\label{eqn:pdecay_threept_proj}
C_{\mcP}^{\Pi\mcO N} = \mcP_{\beta\alpha} C_{\alpha\beta}^{\Pi\mcO N}\,
\end{equation}
that yield nontrivial combinations of proton decay form factors $W_{0,1}$
The nucleon and meson interpolating operators are
\begin{align}
\label{eqn:interpolator_N}
&N = \epsilon^{ijk} (\tilde{u}^{iT} C \gamma_5 \tilde{d}^j) \tilde{u}^k \,, \\
\label{eqn:interpolator_piplus}
&J_{\pi^+} = \bar{\tilde d} \gamma_5 \tilde{u} \,, \\
\label{eqn:interpolator_pizero}
&J_{\pi^0} = \frac1{\sqrt2}\big(\bar\tilde{u} \gamma_5 \tilde{u} 
        - \bar{\tilde d} \gamma_5 \tilde{d}\big) \,, \\
\label{eqn:interpolator_Kplus}
&J_{K^+} = \bar{\tilde s} \gamma_5 \tilde{u} \,, \\
\label{eqn:interpolator_Kzero}
&J_{K^0} = \bar{\tilde s} \gamma_5 \tilde{d} \,, \\
\label{eqn:interpolator_eta}
&J_{\eta} = \frac1{\sqrt6}\big(\bar{\tilde u} \gamma_5 \tilde{u} 
        + \bar{\tilde d} \gamma_5\tilde{d} - 2 \bar{\tilde s} \gamma_5 \tilde{s} \big)\,, 
\end{align}
where the component quark fields $\tilde{u},\tilde{d},\tilde{s}$ are smeared with gauge-invariant
Wuppertal smearing~\cite{Gusken:1989qx} using APE-smeared gauge links.
The smearing parameters are collected in Tab.~\ref{tab:smear_param}.

\begin{table}
\caption{\raggedright
  Parameters for covariant Gaussian smearing of quark sources and APE smearing of the gauge fields
  used in their construction.\label{tab:smear_param}}
\begin{tabular}{c|cc|cc}
\hline\hline
 & $A_\text{APE}$  & $N_\text{APE}$  & $\alpha_\text{Wup}$  & $N_\text{Wup}$ \\
\hline
24ID   & 2.85    & 25    & 2.5   & 10 \\
32ID   & 2.85    & 25    & 2.5   & 40 \\
\hline\hline
\end{tabular}
\end{table}

In the case of $\pi$ and $K$ mesons in the final states, contractions of quark fields in the
operators~(\ref{eqn:pdecay_threept}) generate only connected diagrams, while in the case of the
$\eta$ meson, there are combinations of both connected and disconnected diagrams.
At the physical point, contributions from disconnected diagrams in the $\eta$ correlators 
can be large; since we do not evaluate disconnected contractions in this work, decays in the
$\eta$-channel are not studied here.
We use the standard sequential propagator technique to compute connected contributions to the
three-point function~(\ref{eqn:pdecay_threept}).
First, we compute a forward quark propagator from a smeared source located at the origin of a
particular sample $x$ on the time slice $t_0=x_4$.
Then, we compute a backward propagator from a sequential source that is constructed with one of
the momentum-projected meson interpolation operators 
$e^{-i\vec p\vec y}\,(\bar{\tilde q}_2\Gamma\tilde{q}_1)_{\vec y}$ restricted to the ``sink'' time slice
$t_0+\tsep$.
Finally, the backward propagator is contracted with the two forward propagators at the operator
insertion point, and the result is projected on momentum $\vec q$.

\begin{table}[ht]
\caption{\raggedright
  Initial and final momenta in the three-point functions selected 
  for close-to-physical kinematics $|q^2|\lesssim m_\ell^2$ using 
  masses determined from fits on a lattice (see Sec.~\ref{sec:result_spectrum}).
  \label{tab:kinematics}}
\centering
\begin{tabular}{cc|c|rr}
\hline \hline
$\Pi$   & $\vec{n}_\Pi$   & $\vec{n}_N$   
  & \multicolumn{2}{c}{$Q^2(\text{GeV}^2)$} \\
& & & (24ID)   & (32ID) \\
\hline
$\pi$   & [1 1 1]   & [0 0 0]   & $-0.011$    & $ 0.020$ \\
        & [1 1 1]   & [0 1 0]   & $-0.117$    & $-0.089$ \\
        & [0 0 2]   & [0 0 0]   & $ 0.120$    & $ 0.150$ \\
\hline
$K$     & [0 1 1]   & [0 0 0]   & $ 0.038$    & $ 0.047$ \\
        & [0 1 1]   & [0 1 0]   & $-0.057$    & $-0.052$ \\
        & [0 0 1]   & [0 0 0]   & $-0.074$    & $-0.070$ \\
\hline \hline
\end{tabular}
\end{table}

With two momentum projections in Eq.~(\ref{eqn:pdecay_threept}), the initial nucleon momentum $\vec
k=\vec p + \vec q$ is determined by the momentum conservation after averaging over a gauge ensemble.
We select kinematic points $(\vec p, \vec q)$ so that 
(1) the lepton is close to being on-shell, $|q^2|=|(k-p)^2|\lesssim m_\ell^2$, 
(2) the nucleon spatial momentum is small to minimize statistical fluctuations, and 
(3) the decay kinematic point $q^2\approx0$ is bracketed enabling a reliable \emph{interpolation}.
The selected initial and final state momenta combinations are shown in Tab.~\ref{tab:kinematics}.
Since the physical volumes on both lattice ensembles are very close, so are the quanta $2\pi/(aL)$
of spatial momentum, which result in identical selections of lattice momenta and only slightly
different $q^2$ values for the two ensembles.
In order to further reduce the cost of our computation, we use the ``coherent trick'',
in which backward propagators for two maximally separated samples are computed
simultaneously from the sum of their respective sequential sources.

Meson and nucleon two-point functions
\begin{align}
\label{eqn:twopt_Pi}
&C^{\Pi\Pi}(\vec k, t) 
  = \sum_{\vec x} e^{-i\vec p\vec x} \, \langle J_\Pi(x) \, J_\Pi^\dag(0) \rangle \,,\\
\label{eqn:twopt_N}
&\begin{aligned}
C_+^{N\bar N} &= \Tr\Big[\frac{1+\gamma_4}2 C^{N\bar N}\Big] \,,\\
C_{\alpha\beta}^{N\bar N}(\vec k, t) 
  &= \sum_{\vec x} e^{-i\vec k\vec x} \, \langle N_\alpha(x) \, \bar N_\beta(0)\rangle \,,
\end{aligned}
\end{align}
are also evaluated to compute their energies as well as overlaps of their interpolating
operators~(\ref{eqn:interpolator_N}-\ref{eqn:interpolator_eta}) with their respective ground states.
Similarly to the three-point functions, only the correlators with $\eta$ meson require disconnected
diagrams, which are not studied in the present work.
For the nucleon, we use the positive-parity projected spinor for all momenta $\vec k$.
Although with $\vec k\ne0$ the nucleon does not have definite parity, our momenta are small enough
for it to be a good approximation for the ground-state nucleon.

\subsection{Proton decay matrix elements}

In the large-time limit $\{\tins,(\tsep-\tins)\}\to\infty$, the three-point correlation 
functions~(\ref{eqn:pdecay_threept_proj}) are dominated by the ground-state proton-meson amplitude. 
However, in our lattice calculation the time separations may be not large enough to
neglect contributions from their excited states.
The spectral decomposition of a three-point correlation function yields:
\begin{equation}
\label{eqn:threept_exp}
\begin{aligned}
& C_{\alpha\beta}^{\Pi\mcO \bar N} (\vec p,\vec q; \tsep,\tins)
  = \sum_{m,n,s} \langle\Omega| J_\Pi| \Pi_m(\vec p) \rangle \,
    \frac{e^{-E_{\Pi,m}(\tsep-\tins)}}{2 E_{\Pi,m}} \cdot 
\\ &\quad    
\cdot \langle \Pi_m(\vec p) | \mcO_\alpha | N^{(s)}_n(\vec k) \rangle \,
\frac{e^{-E_{N,n}\tins}}{2 E_{N,n}} \,
\langle N^{(s)}_n(\vec k) | \bar{N}_\beta |\Omega\rangle\,,
\end{aligned}
\end{equation}
where indices $m,n$ denote the ground ($m,n=0$) and excited meson ($m>0$) and nucleon ($n>0$) states.
The ground state matrix elements
$M^{00}_{\alpha,s}(q) = \langle \Pi_0(\vec p) | \mcO_\alpha | N_0^{(s)}(\vec k)\rangle$ 
dominate this sum for $\{\tins,(\tsep-\tins)\}\to\infty$.
Lattice interpolating fields for the meson $J_\Pi$ and the nucleon $N$ may have arbitrary
normalizations due to quark smearing, which are reflected in their \emph{overlap} factors $Z_\Pi$
and $Z_N$,
\begin{align}
\label{eqn:Zmeson}
\langle\Omega| J_\Pi| \Pi_0(\vec p) \rangle &= \sqrt{Z_{\Pi}(\vec p)} 
\,,\\
\label{eqn:Znucleon}
\langle N^{(s)}_0(\vec k) | \bar{N}_\alpha |\Omega\rangle
  &= \bar{u}^{(s)}_{\alpha} \,\sqrt{Z_N(\vec k)}
\,,
\end{align}
These momentum-dependent factors may obtained from
ground-state terms in their respective two-point correlation functions,
\begin{align}
\label{eqn:twopt_Pi_exp}
C^{\Pi\Pi}(\vec p, t)\Big|_\text{g.s.}  &= \frac{Z_{\Pi}(\vec p)}{2E_\Pi} e^{-E_\Pi t}\,,\\
\label{eqn:twopt_N_exp}
C^{N\bar{N}}_+(\vec k, t)\Big|_\text{g.s.}
  &= \frac{Z_{N}(\vec k)}{2 E_N}\,\Tr[\mcP_+ (-i\slashed{k} + m_N) ] \, e^{-E_N t}\,,
\\\nonumber &= Z_N(\vec k)\frac{E_N+m_N}{E_N} e^{-E_N t}\,.
\end{align}
We perform two-state fits to $C^{\Pi\Pi}(t)$ and $C_{+}=\Tr[\mcP_+ C^{N\bar N|}(t)$ 
\begin{align}
\label{eqn:c2fit_Pi}
C^{\Pi\Pi}(\vec p; t) = C_{\Pi,0} e^{-E_{\Pi,0} t} + C_{\Pi,1} e^{-E_{\Pi,1} t}\,,\\
\label{eqn:c2fit_N}
C^{N\bar{N}}_+(\vec k; t) = C_{N,0} e^{-E_{N,0} t} + C_{N,1} e^{-E_{N,1} t}\,,
\end{align}
independently for each momentum $\vec p$, $\vec k$
and extract the ground-state overlap factors for the meson and the nucleon
\begin{equation}
\label{eqn:Zdef}
Z_\Pi(\vec p) = 2 E_{\Pi,0} C_{\Pi,0}\,, \quad 
Z_N(\vec k)   = \frac{E_{N,0}}{E_{N,0} + m_N} C_{N,0}\,,
\end{equation}

In order to compute the form factors~(\ref{eqn:pdecay_ff_def}), we project the three-point function
$\Tr[\mcP C^{\Pi\mcO\bar{N}}] =  C^{\Pi\mcO\bar{N}}_\mcP$ with a set of suitable projectors 
\begin{equation}
\{\mcP\}= \{ P_+, P_+\gamma_j \}\,,
\text{ where } j=1,2,3, \; P_+ = \frac{1+\gamma_4}{2}\,.
\end{equation}
Similarly to the meson and nucleon two-point functions~(\ref{eqn:twopt_N_exp},\ref{eqn:twopt_N_exp}), 
the ground-state contribution to a spin-projected three-point function~(\ref{eqn:pdecay_threept_proj})
can be written as
\begin{equation}
\label{eqn:threept_proj}
C_\mcP^{\Pi\mcO \bar N} (\vec p, \vec q ; \tsep, \tins)\Big|_\text{g.s.}
  = \frac{\sqrt{Z_\Pi Z_N}}{4 E_\Pi E_N} 
    e^{-E_\Pi(\tsep-\tins) - E_N\tins}
\, \Tr\Big[\mcP \mcP_\chi \big(W^\mcO_0 -\frac{i\slashed{q}}{m_N} W^\mcO_1\big) 
            \big(-i\slashed{k} + m_N\big)\Big].
\end{equation}
where the matrix element $M^{00}_{\alpha,s}(q)$ is decomposed into decay form factors $W_{0,1}$. 
However, before these form factors can be extracted, the ground-state matrix element
$M^{00}_{\alpha,s}(q)$ must be isolated from excited-state contamination.
For this purpose, we study the time dependence of the projected three-point
function~(\ref{eqn:threept_proj}) with two methods described below.

\textit{Plateau method} is based on a ratio of correlation functions
\begin{equation}
\label{eqn:ratio}
R^\mcO_\mcP(\vec p, \vec q ; \tsep, \tins)
  = \frac{ \sqrt{ Z_\Pi(\vec p) Z_N(\vec k)}  
            C^{\Pi\mcO\bar{N}}_\mcP(\vec p, \vec q ; \tsep, \tins)}
         { C^{\Pi\Pi}(\vec p, \tsep-\tins) \, C^{N\bar{N}}_+ (\vec k, \tins)}
\end{equation}
where the $Z_{\Pi,N}$ overlap factors~(\ref{eqn:Zdef}) are extracted from the fits to the two-point 
functions~(\ref{eqn:c2fit_Pi},\ref{eqn:c2fit_N}).
The values of this ratio near the center of the plateau, $\tins\approx\tsep/2$, 
must converge to the ground-state matrix element for large time separation $\tsep$.
Deviations from the ground state are suppressed as 
$\mcO(e^{-\Delta E_\Pi (\tsep - \tins)},e^{-\Delta E_N \tins})$, 
and the plateaus are expected to converge to ground-state values faster for $\tins>\tsep/2$,
i.e., closer to the meson sink rather than the proton source because of the larger energy gap of the
former.
However, the noise is also expected to be larger in this region due to the much larger nucleon mass.
For each value of $\tsep$, we calculate the average of 2 or 3 central plateau points $\overline{R}(\tsep)$ 
and estimate statistical errors using Bootstrap.
Convergence of the $\overline{R}(\tsep)$ values with increasing $\tsep$ indicates suppression of
excited states and allows us to estimate related systematic effects.

\textit{Two-state fit method} is intended to take into account the excited states in a systematic
fashion and reduce the bias associated with them.
Similarly to the two-point function fits~(\ref{eqn:c2fit_Pi},\ref{eqn:c2fit_N}), we include single
excited states for both the meson and the proton, and perform correlated least-$\chi^2$ fits
\begin{equation}
\label{eqn:threept_exp2}
C^{\Pi\mcO\bar{N}}_\mcP(\vec p, \vec q ; \tsep, \tins) 
  = \sum_{m,n=0,1} C^{\Pi\mcO\bar{N}}_{\mcP,mn} e^{-E_{\Pi,m}(\tsep-\tins) -E_{N,n}\tins}\,.
\end{equation}
Discarding the excited state contributions, we define the equivalent of the ratio~(\ref{eqn:ratio})
that contains only the ground-state contributions
\begin{equation}
\label{eqn:ratio_fit}
R^{\Pi\mcO\bar{N}}_{\mcP,00}(\vec p,\vec q) 
  = \frac{\sqrt{ Z_\Pi(\vec p) Z_N(\vec k)}  C^{\Pi\mcO\bar{N}}_{\mcP,00}}
         {C_{\Pi,0} C_{N,0}}
\end{equation}
and must be equal to the converged value of Eq.~(\ref{eqn:ratio}) at
$\{\tins,(\tsep-\tins)\}\to\infty$.
Systematic uncertainties in this method are estimated by comparing fit results performed in
ranges $\tskip^N \le \tins < (\tsep-\tskip^\Pi)$ with varying $\tskip^{N,\Pi}$.
$\tskip=2$ and $\tskip=3$ , all of which yielded reasonable $\chi^2$ values.

These methods are applied independently to each combination of initial and final momenta in
Tab.~\ref{tab:kinematics} and for all nontrivial spin projections $\mcP$ of the tree-point
correlation functions.

\subsection{Proton decay form factors}
While only the $W_0$ form factor is necessary for computing width of decays into $e^+$ and $\bar\nu$,
the $W_1$ form factor is also necessary for decays into $\mu^+$.
In order to disentangle form factors $W_{0,1}$, one needs at least two independent matrix elements 
$M^{00}_{\alpha,s}(q)$ in Eq.~(\ref{eqn:threept_exp}) or, equivalently, two independent nontrivial 
projections of the three-point functions~(\ref{eqn:threept_proj}).
Evaluating the spin traces in Eq.~(\ref{eqn:threept_proj}) leads to the following (ground-state)
contributions to the ratios~(\ref{eqn:ratio},\ref{eqn:ratio_fit})
\begin{align}
\label{eqn:ff_coeff}
R_{P_+}^{\Pi\mcO\bar{N}} 
  &= \frac12 W_0 + \Big(\frac{\Delta E}{2m_N} - \frac{\vec k\cdot\vec q}{2m_N(E_N+m_N)}\Big) W_1\,, \\
\label{eqn:ff_coeff2}
R_{P_+\gamma_i}^{\Pi\mcO\bar{N}} 
  &= \frac{-i k_i}{2(E_N+m_N)} W_0 
  + \frac{-i(E_N+m_N)q_i + i\Delta E k_i \pm (\vec q\times\vec k)_i}{2 m_N (E_N+m_N)} W_1\,,
\end{align}
where $\Delta E=E_N-E_\pi$
and the $(\pm)$ sign corresponds to the decay operator helicity $\chip=R,L$, respectively.
These equations take into account nonzero nucleon momentum $\vec k$, which is useful for a better
approximation of the physical kinematic point $q^2\approx0$.
All previous proton decay calculations were done with zero nucleon momentum 
$\vec k=0$~\cite{Aoki:2013yxa}.

We take a projection of Eq.(\ref{eqn:ff_coeff2}) on the spatial vector $\vec q$ in order to simplify
computing the form factors and obtain
\begin{equation}
\label{eqn:ff_coeff3}
R_{P_+(i\vec q\cdot\vec\gamma)} = \sum_i i\vec q_i R_{P_+\gamma_i}
  = \frac{\vec k\cdot\vec q}{2(E_N+m_N)}W_0 
  + \Big(\frac{\vec q^2}{2m_N} 
      - \frac{\Delta E(\vec k\cdot\vec q)}{2m_N(E_N+m_N)}\Big)W_1\,.
\end{equation}
While it is possible to perform ``overdetermined'' fits by considering $P_+\gamma_i$ polarization
projections separately, doing so would accomplish only a check of the rotational symmetry.

\section{Renormalization
  \label{sec:renorm}}
\subsection{Nonperturbative renormalization scheme}
The bare hadronic matrix elements computed on a lattice have to be converted to a continuum
renormalization scheme such as $\MSbar$ that is used in proton decay phenomenology.
Operators are defined on a lattice at a relatively low scale of $a^{-1}\approx1-2\,\mathrm{GeV}$,
where the strong coupling $\alpha_S$ is large.
Nonperturbative renormalization avoids major systematic effects due to truncation of
perturbative series on a lattice and is required to achieve reliable and precise results.
In a typical approach called Rome-Southampton method~\cite{Martinelli:1994ty}, one computes correlators
of a bare operator with bare \emph{external} quark and gluon fields carrying large virtual
momenta in a fixed gauge and compares their behavior to a perturbative prediction, resulting in a
finite conversion factor from lattice to a perturbative renormalization scheme, e.g., to $\MSbar$.
Landau gauge is typically employed as straightforward to implement consistently between lattice and
continuum.

Such an intermediate scheme requires additional perturbative conversion to the $\MSbar$ scheme; in
addition, lattice field correlators may have nonperturbative infrared contributions.
Systematic effects from both of these sources depend on the configuration of external field momenta.
In the case of some quark-bilinear operators, selecting a \emph{non-exceptional} momentum
configuration (``$\SMOMqbarq$'' scheme) is crucial for avoiding large nonperturbative effects that
may appear if the operator carries zero momentum (``$\MOMqbarq$'' scheme)~\cite{Sturm:2009kb}.
In the case of the three-quark operators, momenta can be arranged in even more ways. 
The two choices discussed in the literature are either with all three quarks carrying the same
momentum $p$~\cite{Aoki:2006ib} or carrying momenta of the same magnitude $p^2=k^2=r^2$ that add to
zero vertex momentum $p+k+r=0$~\cite{Gracey:2012gx};
below we will refer to these momentum arrangements as ``$\MOMpd$'' and ``$\SYMpd$'', respectively,
to discriminate from the schemes used for quark-bilinear operators.

In order to avoid both discretization and nonperturbative effects, the momenta of the fields must
satisfy the ``scale window'' condition 
\begin{equation}
\label{eqn:scale_window}
\Lambda_\text{QCD} \ll p \ll (\pi/a)\,.
\end{equation}
The three-quark operator in the correlator following the $\MOMpd$ scheme will carry momentum
$(3p)^2$ and may require a wider scale window, which is challenging on coarse lattice ensembles 
that are used in the present work.
Additionally, the large vertex momentum may result in large perturbative conversion factors to the
$\MSbar$ scheme and, consequently, larger systematic uncertainties.
Indeed, the amputated Green's function of the 3-quark operator at the $O(\alpha_S)$ order is larger
\footnote{
  It is worth noting that the complete conversion factors may also include perturbative corrections
  due to the quark fields depending on their renormalization scheme.
  In particular, the $\SMOMg$ scheme that we use below requires $O(\alpha_S)$ correction comparable
  in magnitude to the one in Eq.~(\ref{eqn:pertC_3q_MOM}), while $\SMOMqbarq$ and $\MOMqbarq$ do not.}
in the $\MOMpd$ scheme~\cite{Aoki:2006ib,Aoki:2006ib-ERRATUM} 
compared to the $\SYMpd$ scheme\cite{Gracey:2012gx,Pivovarov:1991nk}
(see also Appendix~\ref{sec:app_renorm})
\begin{align}
\label{eqn:pertC_3q_MOM}
\Big[\Lambda_{3q}^{\overline{MS}}\Big]_{\MOMpd}
  &\approx 1 + (-4.060)\frac{\alpha_s}{4\pi} + O(\alpha_S^2)\,,
\\
\label{eqn:pertC_3q_symm}
\Big[\Lambda_{3q}^{\overline{MS}}\Big]_{\SYMpd}
  &\approx  1 + (0.989) \frac{\alpha_s}{4\pi} + O(\alpha_S^2)\,.
\end{align}
The $\SYMpd$ Green's function~(\ref{eqn:pertC_3q_symm}) is available up to $O(\alpha_S^2)$
order~\cite{Gracey:2012gx}, while the $\MOMpd$ Green's function~(\ref{eqn:pertC_3q_MOM}) is
available only up to $O(\alpha_S)$~\cite{Aoki:2006ib,Aoki:2006ib-ERRATUM}.
The large difference at the $O(\alpha_S)$ order indicates that the unknown $O(\alpha_S^2)$ correction
to the former may also be larger compared to the latter, which has been computed and can be used for
more accurate perturbative matching.
On the other hand, in the $\SYMpd$ scheme, the vertex carries zero momentum $q=p+k+r=0$, which might
result in a nonperturbative contribution from the nucleon pole $\sim(q^2 + m_N^2)^{-1}$.
However, the overlap of a point-localized three-quark operator with the nucleon state is 
suppressed due to the nonzero nucleon size.
Since most of the nucleon mass comes from the glue (as shown by the momentum sum-rule in deep
inelastic scattering experiments~\cite{Barger:1993my}) and the nucleon remains massive in the chiral 
limit, such a pole contribution should be negligible.
Therefore, we select the $\SYMpd$ scheme because it enables better control of these systematic
uncertainties.

\subsection{Renormalization of decay operators}

In order to determine nonperturbative renormalization factors, we compute Green's functions of
operators~(\ref{eqn:3qop_me_def}) with three external quark fields carrying definite Euclidean 4-momenta.
There are two flavor structures
\begin{align}
\label{eqn:3qop_uds}
\lp[ \mcO_{\Gamma \Gamma'}^{(ud)s} \rp]_\delta 
  &= \epsilon^{abc} (\bar{u}^{T\,a} \Gamma d^{b}) \Gamma' s^c_\delta\,,\\
\label{eqn:3qop_udd}
\lp[ \mcO_{\Gamma \Gamma'}^{(ud)d} \rp]_\delta
  &= \epsilon^{abc} (\bar{u}^{T\,a} \Gamma d^{b}) \Gamma' d^c_\delta\,,
\end{align}
with 10 linearly-independent Lorentz-invariant choices of 
$(\Gamma\otimes\Gamma')=\{SS,PP,AA,VV,TT,SP,PS,AV,VA,TQ\}$\footnote{
  Operators with permuted quark fields can be reduced to the 
  forms~(\ref{eqn:3qop_uds},\ref{eqn:3qop_udd}) using Fierz identities.
},
of which 5 are positive- and 5 are negative-parity.
The labels $S,P,V,A,T,Q$ stand for
$\Gamma^{(\prime)}=1,\gamma_5,\gamma_\mu,\gamma_\mu\gamma_5,\sigma_{\mu\nu},\sigma_{\mu\nu}\gamma_5$,
respectively, with Lorentz indices $\mu,\nu$ contracted in $(\Gamma\otimes\Gamma')$.

\begin{table}[ht]
\caption{\raggedright
  Classification of operators by parity and diquark symmetry~\cite{Aoki:2006ib}.
  Switching symmetry is determined by $\Gamma$: $\mcS(\Gamma=S,P,A)=-1$ and
  $\mcS(\Gamma=V,T)=+1$, while parity $\mcP=-\mcP(\Gamma)\mcP(\Gamma^\prime)$ is determined by both 
  $\mcP(\Gamma^{(\prime)}=S,V,T)=+1$ and $\mcP(\Gamma^{(\prime)}=P,A,Q)=-1$.
  \label{tab:3qop_classif}}
\begin{tabular}{l|c|c}
\hline\hline
          & $\mcS=-1$ & $\mcS=+1$ \\
\hline
$\mcP=-1$ & $SS$, $PP$, $AA$ & $VV$, $TT$ \\
$\mcP=+1$ & $SP$, $PS$, $AV$ & $VA$, $TQ$ \\
\hline\hline
\end{tabular}
\end{table}

We perform calculations with equal, $SU(3)_f$ symmetric quark masses and extrapolate to the chiral
limit $m_{u,d,s}\to0$.
The 10 $\mcO^{(ud)s}$ operators can then be further classified by the symmetry of the diquark
factor~\cite{Aoki:2006ib} (see Tab.~\ref{tab:3qop_classif}).
Since the relevant operators are
\begin{equation}
\label{eqn:O3q_pm_basis}
\begin{aligned}
\mcO^{3q}_{RR} &= \frac14\lp(\mcO^{3q}_{SS} + \mcO^{3q}_{SP} + \mcO^{3q}_{PS} + \mcO^{3q}_{PP}\rp)
  = \mcO^{3q}_+ \,,\\
\mcO^{3q}_{RL} &= \frac14\lp(\mcO^{3q}_{SS} - \mcO^{3q}_{SP} + \mcO^{3q}_{PS} - \mcO^{3q}_{PP}\rp)
  = \mcO^{3q}_- \,,\\
\end{aligned}
\end{equation} 
where $R,L$ correspond to to $\Gamma^{(\prime)}=\frac12(1\pm\gamma_5)$,
the only potential mixing is between operators $\{\mcO_{SS},\,\mcO_{PP},\,\mcO_{AA}\}$.
Equivalently, one can study mixing and renormalization of operators
$\{\mcO_{SP},\,\mcO_{PS},\,-\mcO_{AV}\}$, which is identical\footnote{
  We note that discussion of renormalization and mixing is more natural in the spin-structure basis
  $\mcO_{SS},\,\mcO_{PP},\,\mcO_{AA}$, in which the symmetry constraints are straightforward.
  The results, however, are reported in the phenomenological ``permutation'' basis 
  $\{(ud)s,(us)d,(sd)u\}_{RR,RL}$~(\ref{eqn:3qop_me_def}).
  The latter basis is not orthogonal, which would complicate the pattern of nonperturbative mixing,
  e.g., due to chiral symmetry breaking by lattice discretization.
}.
In the case of  $\mcO^{(ud)d}$ operators, Fierz identities reduce the number of independent 
operators to the following four,
\begin{align}
\mcO^{(ud)d}_{AA} &= \mcO^{(ud)d}_{PP} - \mcO^{(ud)d}_{SS} = \mcO^{(ud)d}_{VV} \,, \\
\mcO^{(ud)d}_{AV} &= \mcO^{(ud)d}_{SP} - \mcO^{(ud)d}_{PS} = \mcO^{(ud)d}_{VA} \,, \\
\mcO^{(ud)d}_{TT} &= \mcO^{(ud)d}_{SS} + \mcO^{(ud)d}_{PP} \,, \\
\mcO^{(ud)d}_{TQ} &= \mcO^{(ud)d}_{SP} + \mcO^{(ud)d}_{PS} \,, \\
\end{align}
and any potential mixing is respectively simplified.

\begin{figure}
\centering
\begin{subfigure}[b]{.49\textwidth}
  \includegraphics[width=\textwidth]{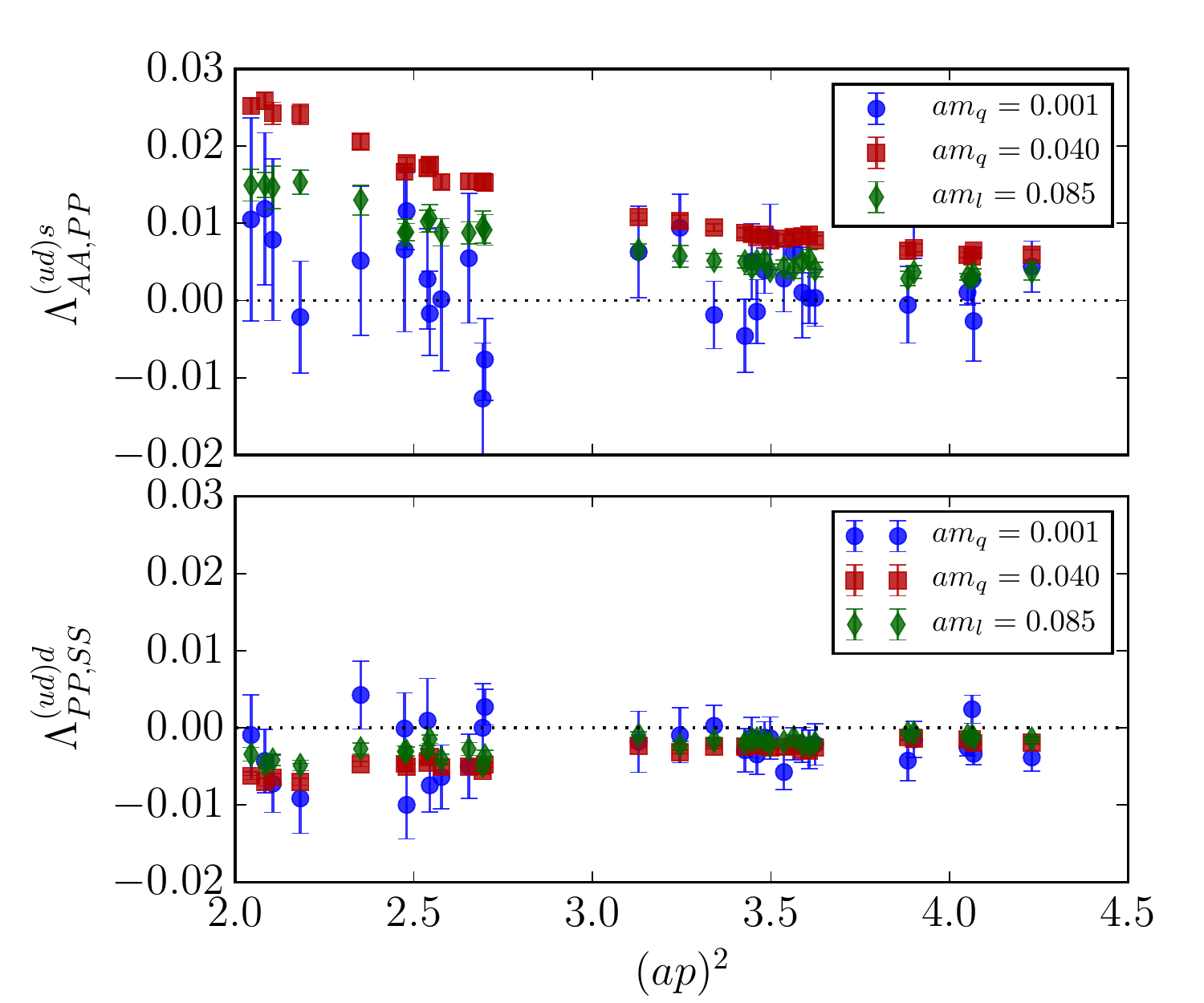}
  \caption{24ID}
\end{subfigure}
\begin{subfigure}[b]{.49\textwidth}
  \includegraphics[width=\textwidth]{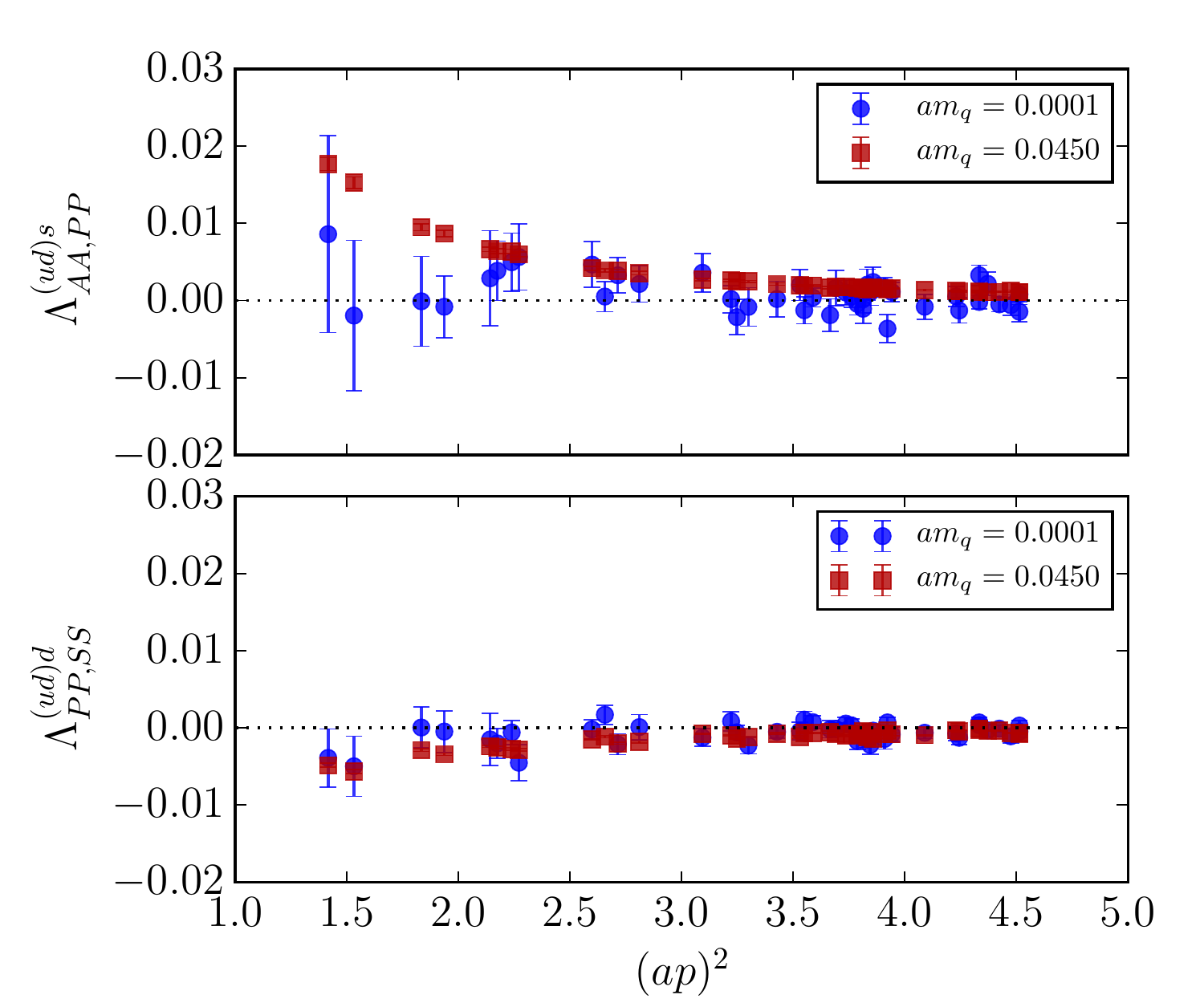}
  \caption{32ID}
\end{subfigure}
\caption{\raggedright
  Quark mass and scale dependence of the mixing Green's functions $\Lambda^{(ud)s}_{AA,PP}$ (top)
  and $\Lambda^{(ud)d}_{PP,SS}$ (bottom).
  Only two quark masses, the highest and the lowest, are shown for the 32ID ensemble.
  \label{fig:Z_mq_dep}}
\end{figure}

The nonperturbative Green's functions are computed using quark propagators with point-sources
and momentum-projected sinks in the Landau gauge,
\begin{equation}
S_q(x, p) = \sum_y e^{ip(x-y)}\langle q(x) \bar q(y)\rangle\,,
\end{equation}
which are contracted at the source $x$ (spin and color indices are omitted)
\begin{equation}
\label{eqn:npr_UDS}
\begin{aligned}
& G^{3q}_{\Gamma\Gamma'}(x; p,k,r) = \sum_{y_1 y_2 y_3} \, e^{iq x -ip y_1 -ik y_2 -ir y_3} \,\cdot 
\langle \mcO^{3q}_{\Gamma\Gamma'}(x) \bar{s}(y_3) \bar{d}(y_2) \bar{u}(y_1) \rangle\,,
\end{aligned}
\end{equation}
where $q=p+k+r$, and equivalently for the $\mcO^{(ud)d}$ operators.
These Green's functions are then amputated with the same point source propagators after averaging 
over an entire ensemble,
\begin{equation}
\label{eqn:lambda_UDS}
\Lambda^{(ud)s}_{\Gamma\Gamma'}(p,k,r) = 
  \lla G^{(ud)s}_{\Gamma\Gamma'}(p,k,r) \rra \, \cdot \lp[\lla S_{s}(r) \rra^{-1} \, 
  \lla S_{u/d}(k) \rangle^{-1} \, \lla S_{u/d}(p) \rra^{-1}\rp] \,,
\end{equation}
We evaluate 32 low-precision samples and 1 high-precision sample per configuration to correct for
potential bias and use jackknife resampling to estimate statistical uncertainty.
We perform these calculations with three values of equal quark masses $m_{u,d,s}$ varied between
$m_{u/d}$ and $m_s$ on each ensemble (see Tab.~\ref{tab:NPRmass}).
We observe very weak quark mass dependence in the diagonal renormalization constants.
The figures below refer to the lightest quark mass, and final renormalization results are obtained 
by linear extrapolation with $m_{u,d,s}\to0$.
The only statistically significant mixing is observed between $AA$ and $PP$ operators, which
vanishes towards the massless quark limit indicating that chiral symmetry is preserved (see
Fig.~\ref{fig:Z_mq_dep}).

\begin{table}[ht]
\centering
\caption{\raggedright
  Quark masses used for computing light and strange quark propagators for nonperturbative renormalization. 
  The numbers of configurations used are shown in the second column.
  \label{tab:NPRmass}
}
\begin{tabular}{c|cc|cc|cc}
\hline\hline
Ensemble  & $m_q^\text{NPR(1)}$   & $N_{\text{cfg}}$
          & $m_q^\text{NPR(2)}$   & $N_{\text{cfg}}$ 
          & $m_q^\text{NPR(3)}$   & $N_{\text{cfg}}$ \\
\hline
24ID & 0.00107 & 16   & 0.04 & 18  & 0.085 & 27 \\ 
\hline
32ID &  0.0001 & 22   & 0.02 & 0   & .045 & 21 \\ 
\hline\hline	
\end{tabular}
\end{table} 

The tree-level vertices of the three-quark operators $\mcO_{SS,PP,AA}$ have the following spin/color
structure 
\begin{equation}
\label{eqn:NPR_basis}
\begin{aligned}
\big[\Lambda^{3q}_{SS}\big]_{\alpha\beta\gamma\delta}^{abc} 
  &= \epsilon^{abc} \, (C)_{\alpha\beta} \, (\mathbf{1})_{\gamma\delta}\,,\\
\big[\Lambda^{3q}_{PP}\big]_{\alpha\beta\gamma\delta}^{abc} 
  &= \epsilon^{abc} \, (C \gamma_5)_{\alpha\beta} \, (\gamma_5)_{\gamma\delta}\,,\\
\big[\Lambda^{3q}_{AA}\big]_{\alpha\beta\gamma\delta}^{abc} 
  &= \epsilon^{abc} \, (C \gamma_\mu \gamma_5)_{\alpha\beta} \, (\gamma_\mu\gamma_5)_{\gamma\delta}\,.
\end{aligned}
\end{equation}
and the corresponding projectors for the amputated Green's functions
\begin{equation}\begin{aligned}
\label{eqn:proj_new}
\big[P^{3q}_{SS}\big]_{\alpha\beta\gamma\delta}^{abc}
  &= \frac{1}{96} \epsilon^{abc} \, ( C^{-1})_{\beta\alpha} \, (\mathbf{1})_{\delta\gamma}\,,\\
\big[P^{3q}_{PP}\big]_{\alpha\beta\gamma\delta}^{abc}
  &= \frac{1}{96} \epsilon^{abc} \, (\gamma_5 C^{-1})_{\beta\alpha} \, (\gamma_5)_{\delta\gamma}\,\\
\big[P^{3q}_{AA}\big]_{\alpha\beta\gamma\delta}^{abc}
  &= \frac{1}{384} \epsilon^{abc} \, (\gamma_5\gamma_\mu C^{-1})_{\beta\alpha} \, 
      (\gamma_5\gamma_\mu)_{\delta\gamma}
\end{aligned}\end{equation}
are used to form a $3\times3$ amputated Green's function matrix 
\begin{equation}
\label{eqn:renorm_mat}
\Lambda^{3q}_{XY}(p,k,r) = \big(\Lambda^{3q}_X(p,k,r)\big) \cdot  P^{3q}_Y\,,
\end{equation}
where $X,Y = \{SS,PP,AA\}$ and the dot indicates summation in all color and spin indices.
From this matrix, the nonperturbative renormalization/mixing matrix $Z^{3q}_{XY}$ is determined as
\begin{equation}
\label{eqn:RScheme}
Z_q^{-3/2} \, Z^{3q}_{XY} \, \Lambda^{3q}_{YZ} = \delta_{XZ}\,.
\end{equation}
The arrangement of momenta $(p,k,r)$ defines a particular subtraction scheme for the three-quark
operator; we will use notation $|p|$ to indicate the subtraction point defining the scale.
It is difficult to select lattice momenta satisfying the $\SMOMqbarq$ and $\SYMpd$ conditions
exactly.
We select momenta with the following criteria (1) $p^2$, $k^2$, $r^2$ values are within 10\% of each
other, and (2) $p$, $k$, $r$ momenta satisfy the ``democratic''
orientation~\cite{Alexandrou:2015sea}
\begin{equation}
\frac{\sum_\mu p_\mu^4}{\big(\sum_\mu p_\mu^2)^2} \le 0.4\,.
\end{equation}
This ratio estimates the ``diagonality'' of the momentum, and the constraint avoids directions close
to a single axis, which are expected to produce larger discretization effects.
For the $\MOMqbarq$ and $\MOMpd$ schemes, we explore momenta along axes as well as 2-, 3-, and 4-d
diagonals of the lattice.
Wherever possible, amputated correlators are averaged over reflections and rotations of the external momenta.

\begin{figure}
\centering
\begin{subfigure}[b]{.49\textwidth}
  \includegraphics[width=\textwidth]{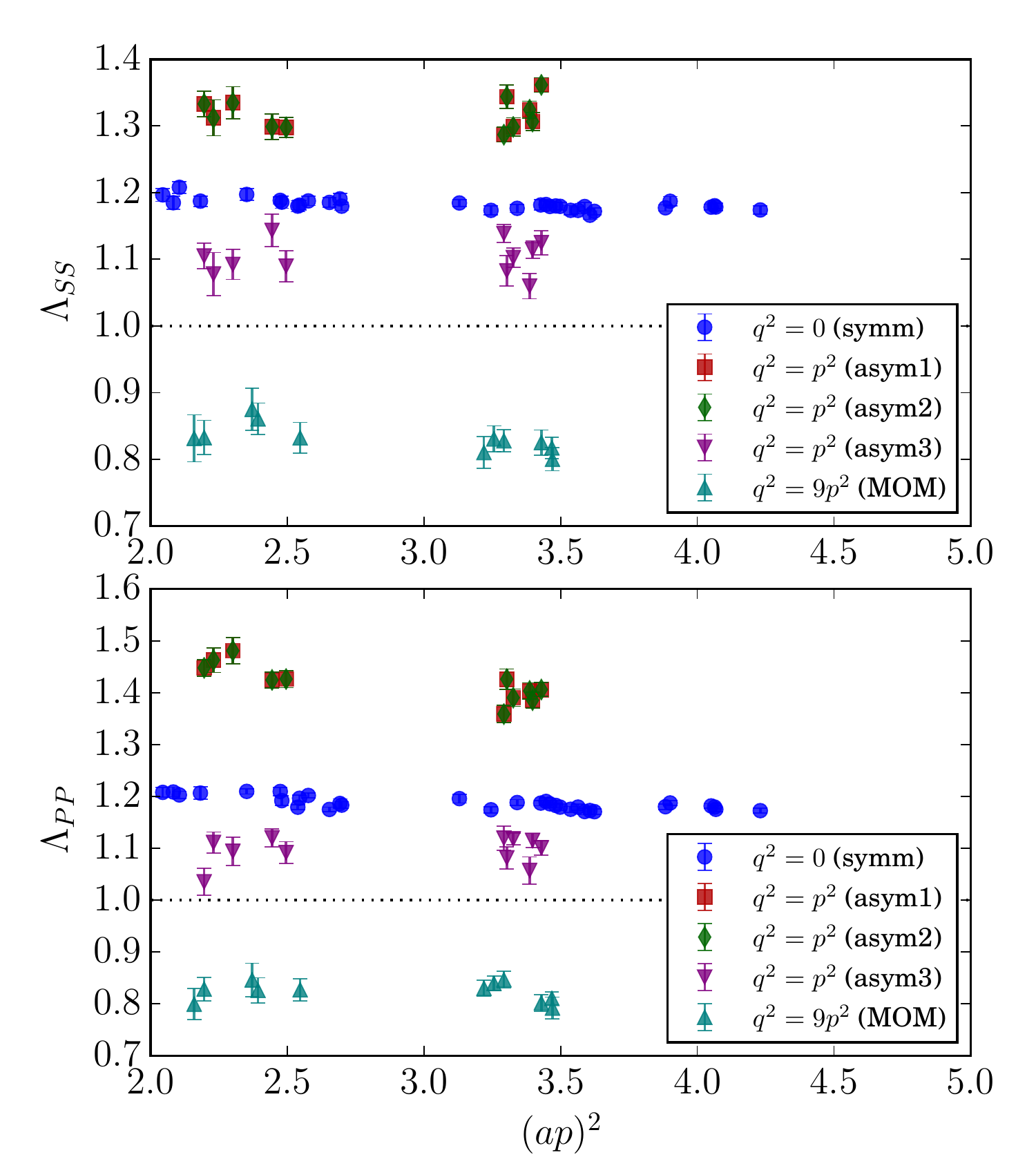}
  \caption{24ID}
\end{subfigure}
\begin{subfigure}[b]{.49\textwidth}
  \includegraphics[width=\textwidth]{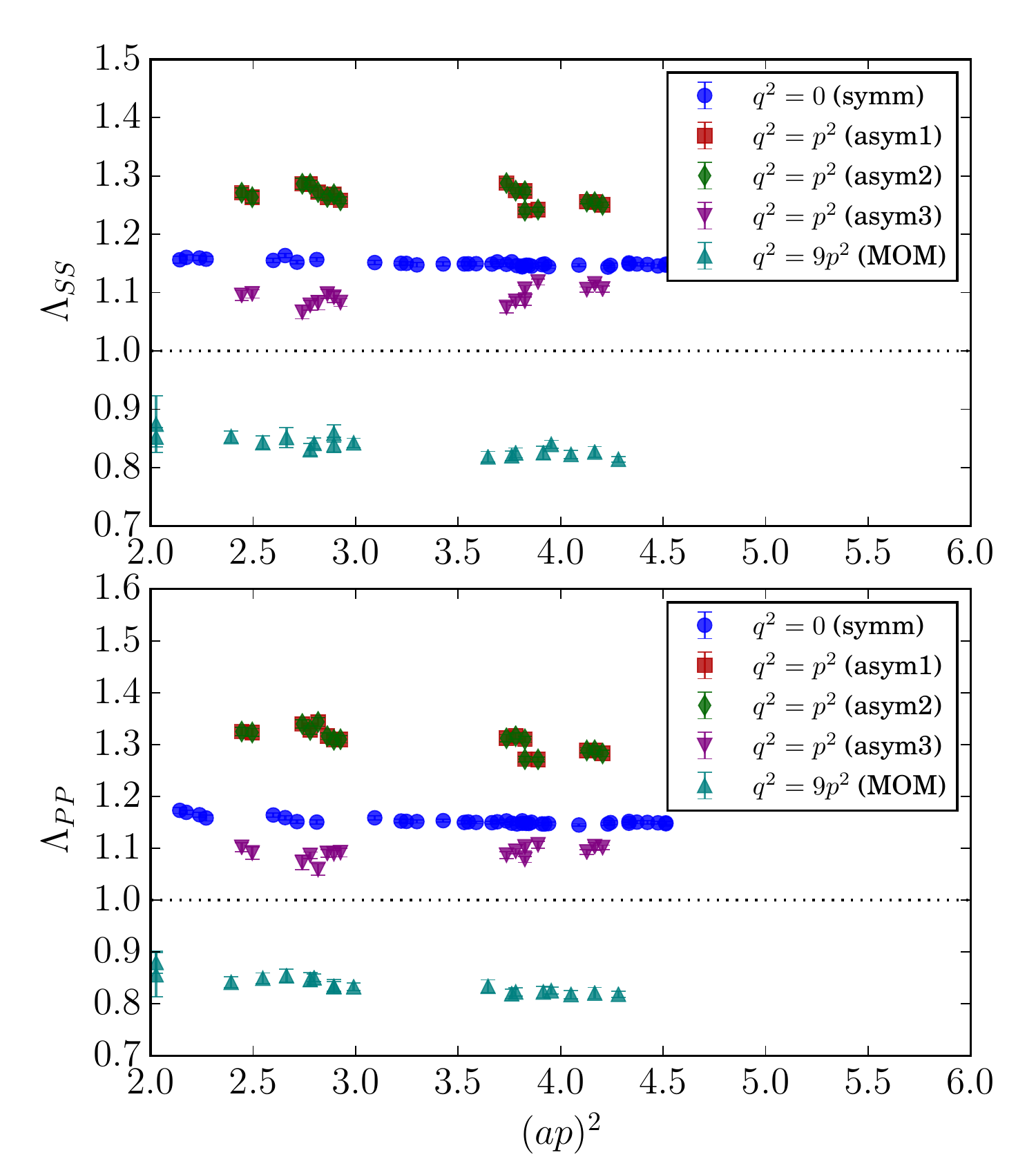}
  \caption{32ID}
\end{subfigure}
\caption{\raggedright
  Amputated Green's functions 3-quark operators $\mcO^{(ud)s}_{SS}$ and $\mcO^{(ud)s}_{PP}$ with
  varying virtual momentum of the operator.
  Only diagonal entries (i.e., projected on $SS$ and $PP$, respectively) are shown.
  \label{fig:nprV_3q_cmp}}
\end{figure}

In Figure~\ref{fig:nprV_3q_cmp}, we compare the amputated and projected Green's functions of the
three-quark operators in $\MOMpd$ and $\SYMpd$ momentum schemes and find that they are different by
$\approx 30-35\%$.
Although some difference is expected due to kinematics, it turns out to be substantially larger than 
expected from perturbative calculations~(\ref{eqn:pertC_3q_MOM},\ref{eqn:pertC_3q_symm}), 
which is $\approx12\%$ at $|p|=2\,\mathrm{GeV}$ (see Eqs.~(\ref{eqn:pertC_3q_MOM},\ref{eqn:pertC_3q_symm})).
Since we observe only weak dependence of Green's functions on the momentum scale as $|p|\to0$, this
discrepancy is unlikely to be caused by nonperturbative effects such as a nucleon pole $(p^2+m_N)^{-1}$, 
and may indicate large $O(\alpha_S^2)$ perturbative corrections in the $\MOMpd$ scheme.
We have also briefly explored Green's functions in alternative schemes with quark momenta 
$p=\pm k =\pm r$ and $|p+k+r|=|p|$ (``asym1,2,3'') shown in Fig.~\ref{fig:nprV_3q_cmp}, which
confirm the strong dependence of the vertex functions on the external quark momentum configuration. 
These observations validate our choice of the $\SYMpd$ scheme for renormalizing the
three-quark operators.

\begin{figure}[ht!]
\centering
\begin{subfigure}[b]{.49\textwidth}
  \includegraphics[width=\textwidth]{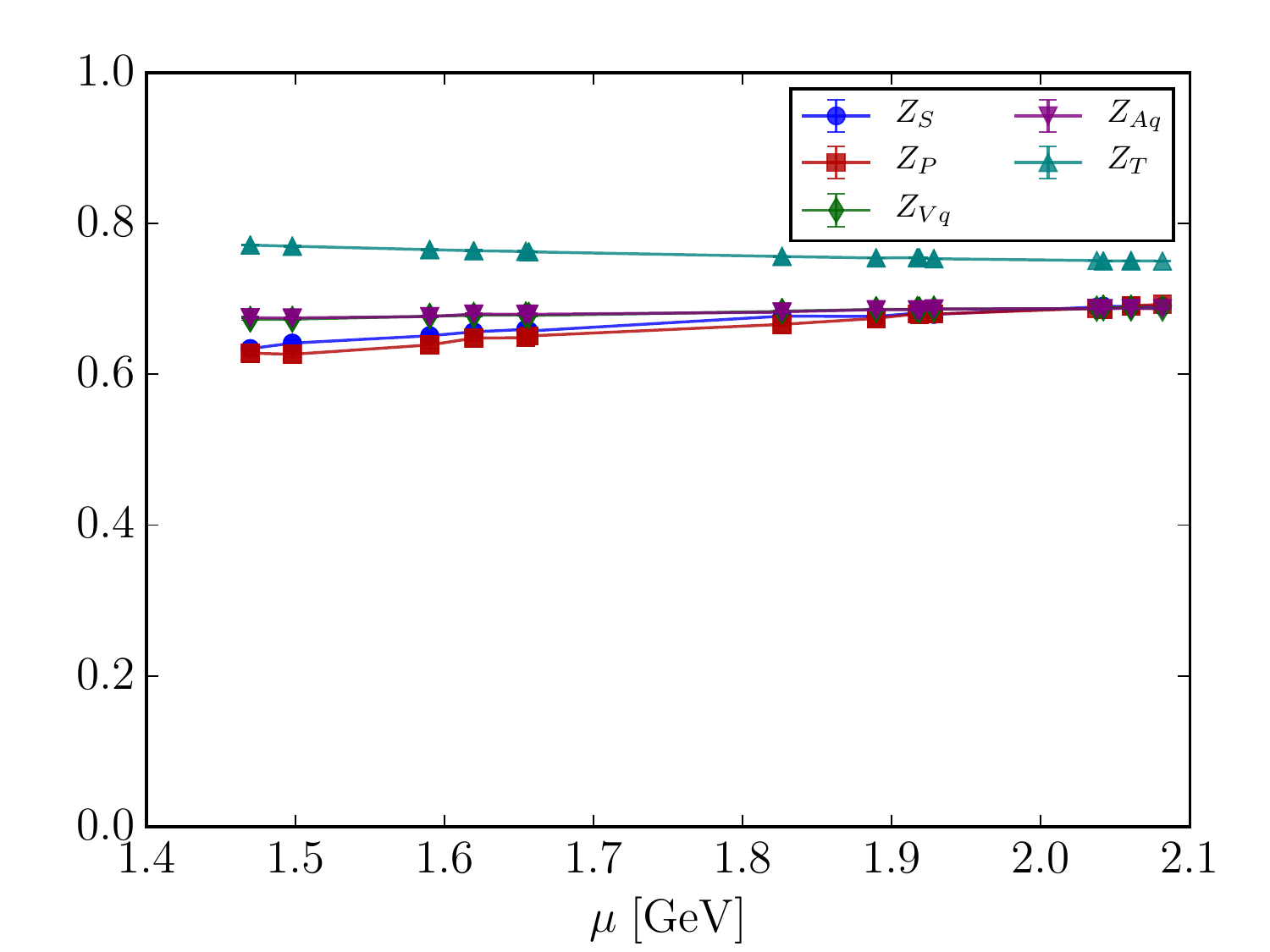}\\
  \caption{24ID}
\end{subfigure}
\begin{subfigure}[b]{.49\textwidth}
  \includegraphics[width=\textwidth]{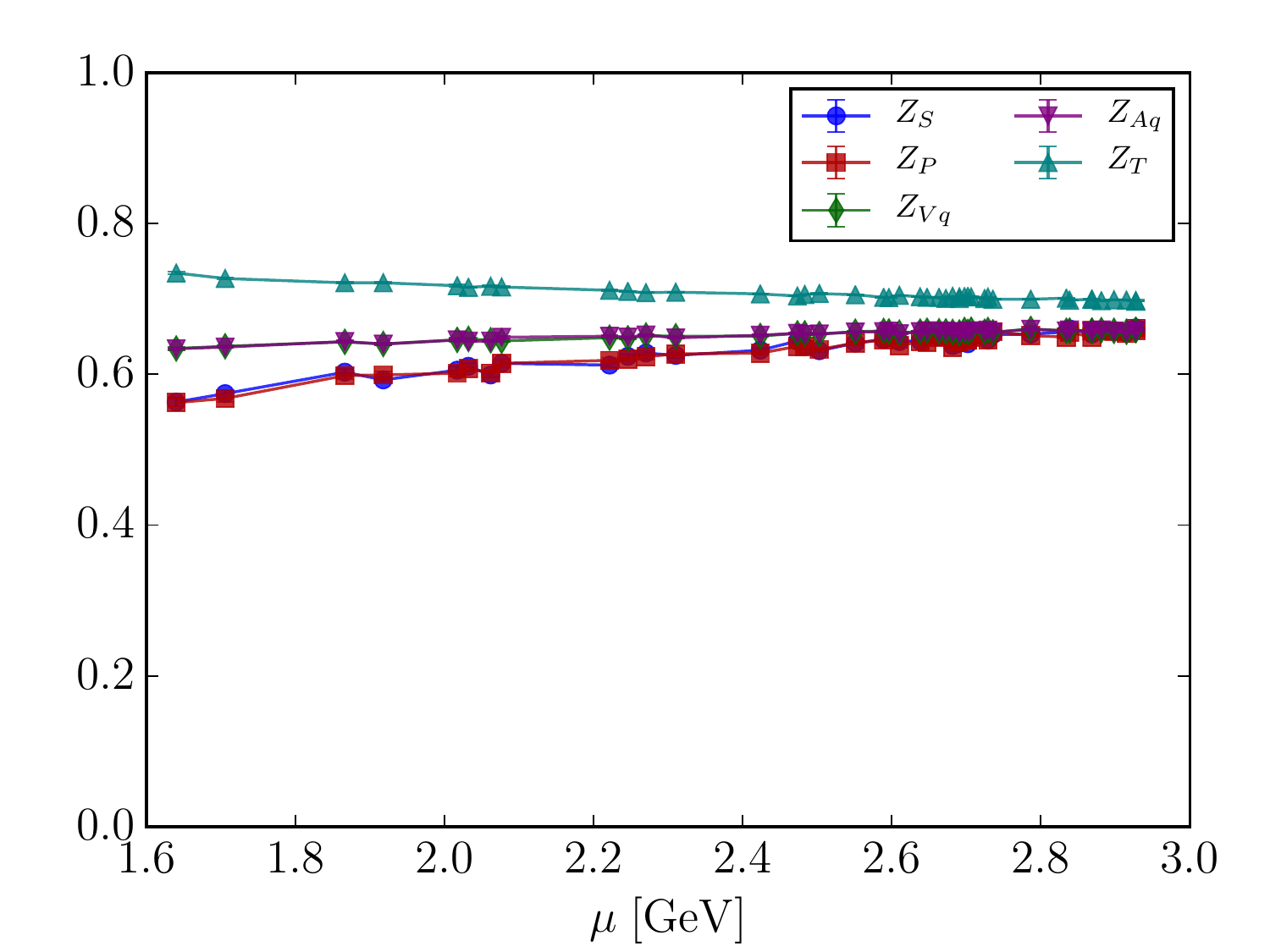}\\
  \caption{32ID}
\end{subfigure}
\caption{\raggedright
  Lattice renormalization factors of quark-bilinear operators~(\ref{eqn:NPR_qbarqZ})
  in the $\SMOMqbarq$ scheme.
  \label{fig:nprZ_SPVAT}}
\end{figure}


To eliminate the quark field renormalization factor $Z_q$, we use Green's functions of quark axial-vector
current 
\begin{align}
\label{eqn:NPR_bilinear}
\Lambda_{\Gamma}(p, p^\prime)
  &=\langle [\bar{q} \Gamma q] \, q(p) \, \bar q(p^\prime)\rangle_\text{amp.}
\end{align}
with non-exceptional momenta $p^2=p^{\prime2}=(p-p^\prime)^2$.
This Green's function is projected according to the ``$\SMOMg$'' scheme,
\begin{align}
\label{eqn:NPR_ZA2Zq}
Z^{\SMOMg}_q(|p|) &= [\Lambda_A]_{\SMOMg}  Z_A^{WI}
  = \frac1{48} \sum_\mu \mathrm{Tr}\big[\gamma_5\, \gamma_\mu\,
      \Lambda_{\gamma_\mu\gamma_5} \big]\,Z_A^{WI}\,.
\end{align}
While such scheme is incompatible with the Ward identity for the axial current~\cite{Sturm:2009kb},
it is more practical on a lattice because it does not depend on components of virtual external quark
momenta and accompanying discretization effects.
We use the values of the renormalization factors $Z_A^{WI}=0.73457(11)$ (24ID) and $0.68779(11)$ (32ID) 
determined in Refs.~\cite{MurthyD:PhD:2017,Tu:2020vpn} and perturbative calculations in the $\SMOMg$
scheme~\cite{Almeida:2010ns} (see App.~\ref{sec:app_renorm} for details).
In Figure \ref{fig:nprZ_SPVAT}, we show lattice renormalization constants of quark bilinears in
the $\SMOMg$ scheme
\begin{equation}
\label{eqn:NPR_qbarqZ}
Z_\Gamma = Z_A \frac{\Lambda^{\SMOMg}_A(|p|)}{\Lambda_\Gamma(|p|)}
\end{equation}
for $\Gamma=1(S),\gamma_5(P),\gamma_\mu(V),\gamma_\mu\gamma_5(A),\sigma_{\mu\nu}(T)$ in
Fig.~\ref{fig:nprZ_SPVAT}.

\begin{figure}
\centering
\begin{subfigure}[b]{.49\textwidth}
  \includegraphics[width=\textwidth]{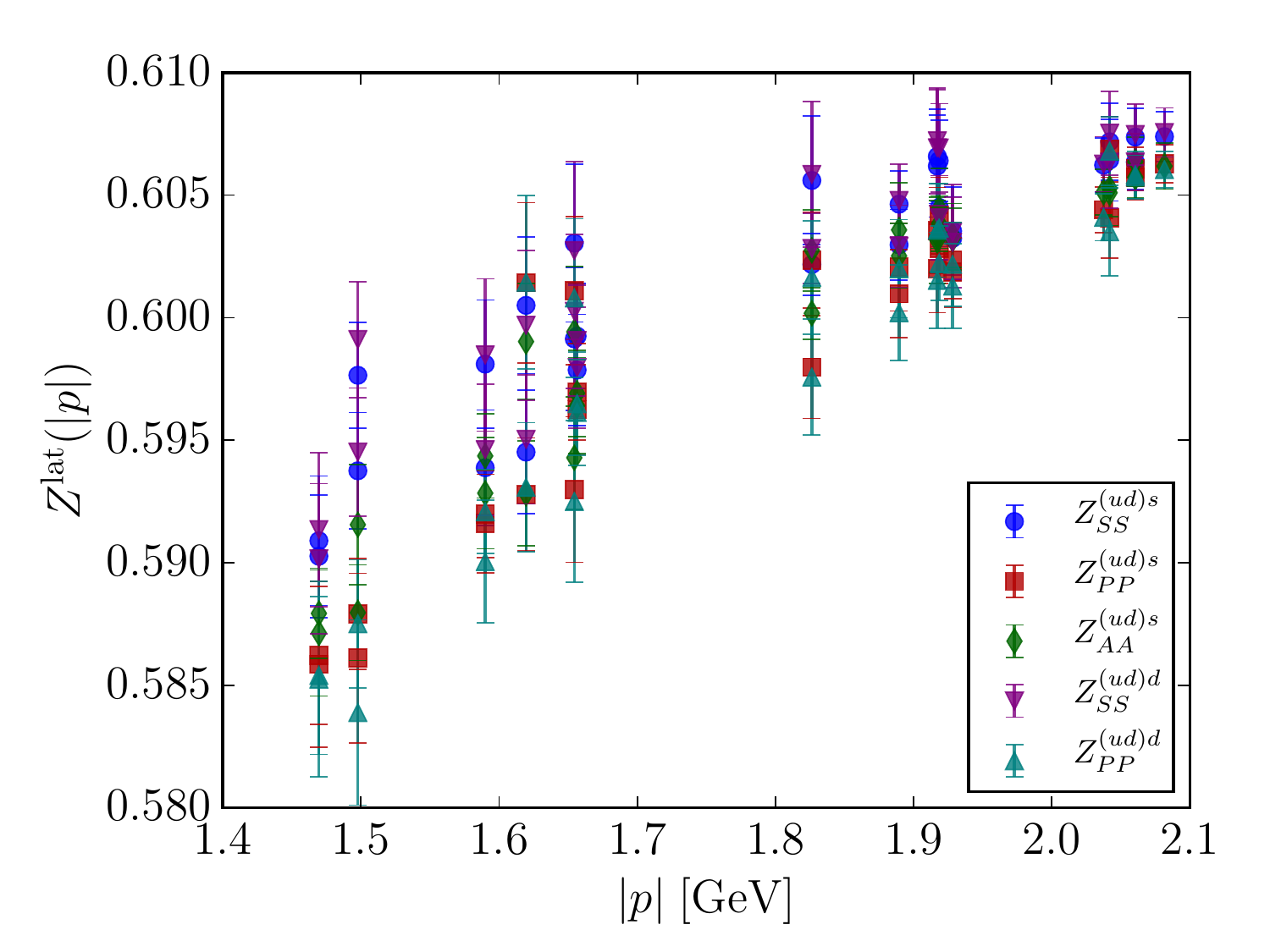}\\
  \includegraphics[width=\textwidth]{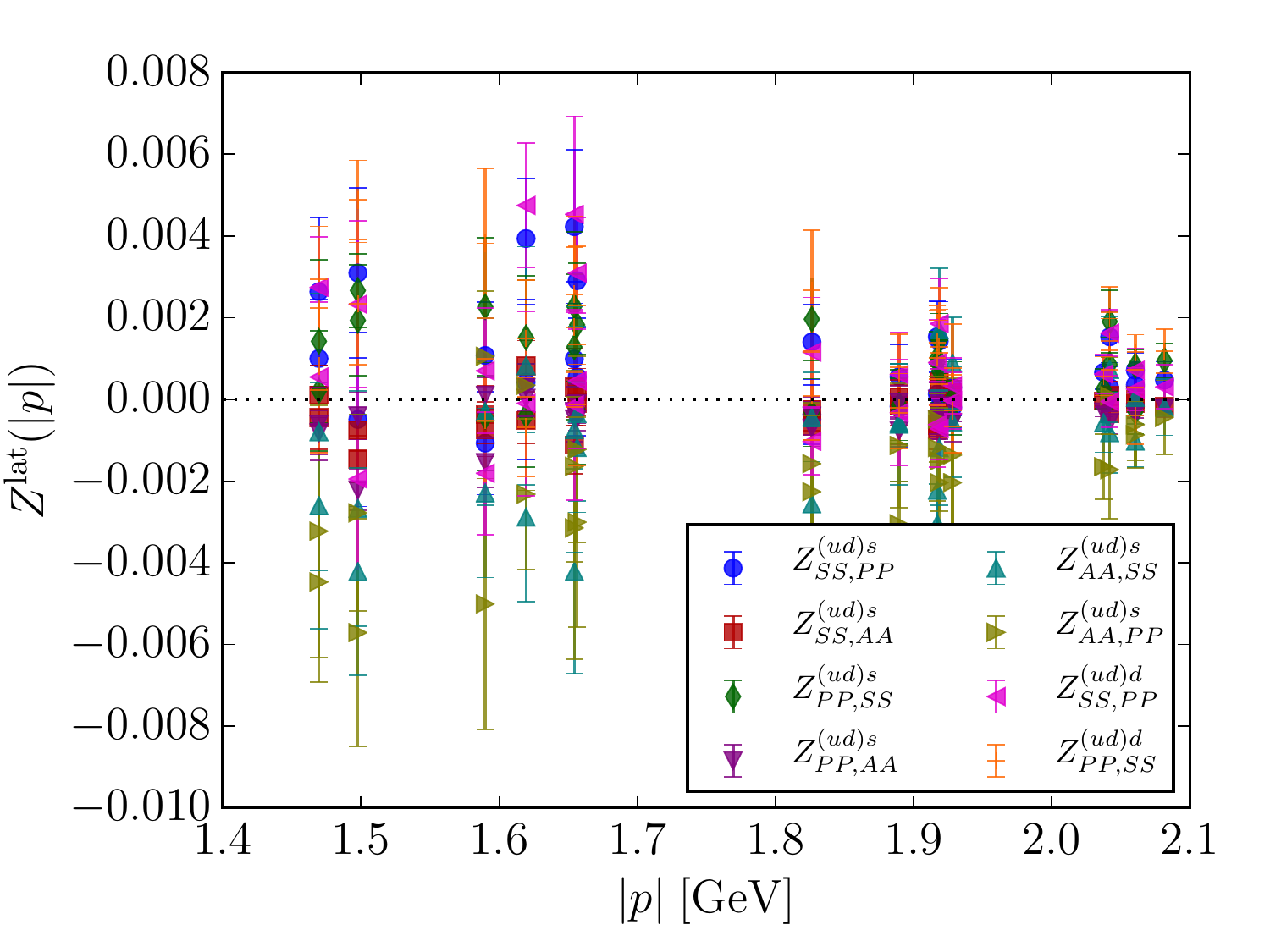}\\
  \caption{24ID}
\end{subfigure}
\begin{subfigure}[b]{.49\textwidth}
  \includegraphics[width=\textwidth]{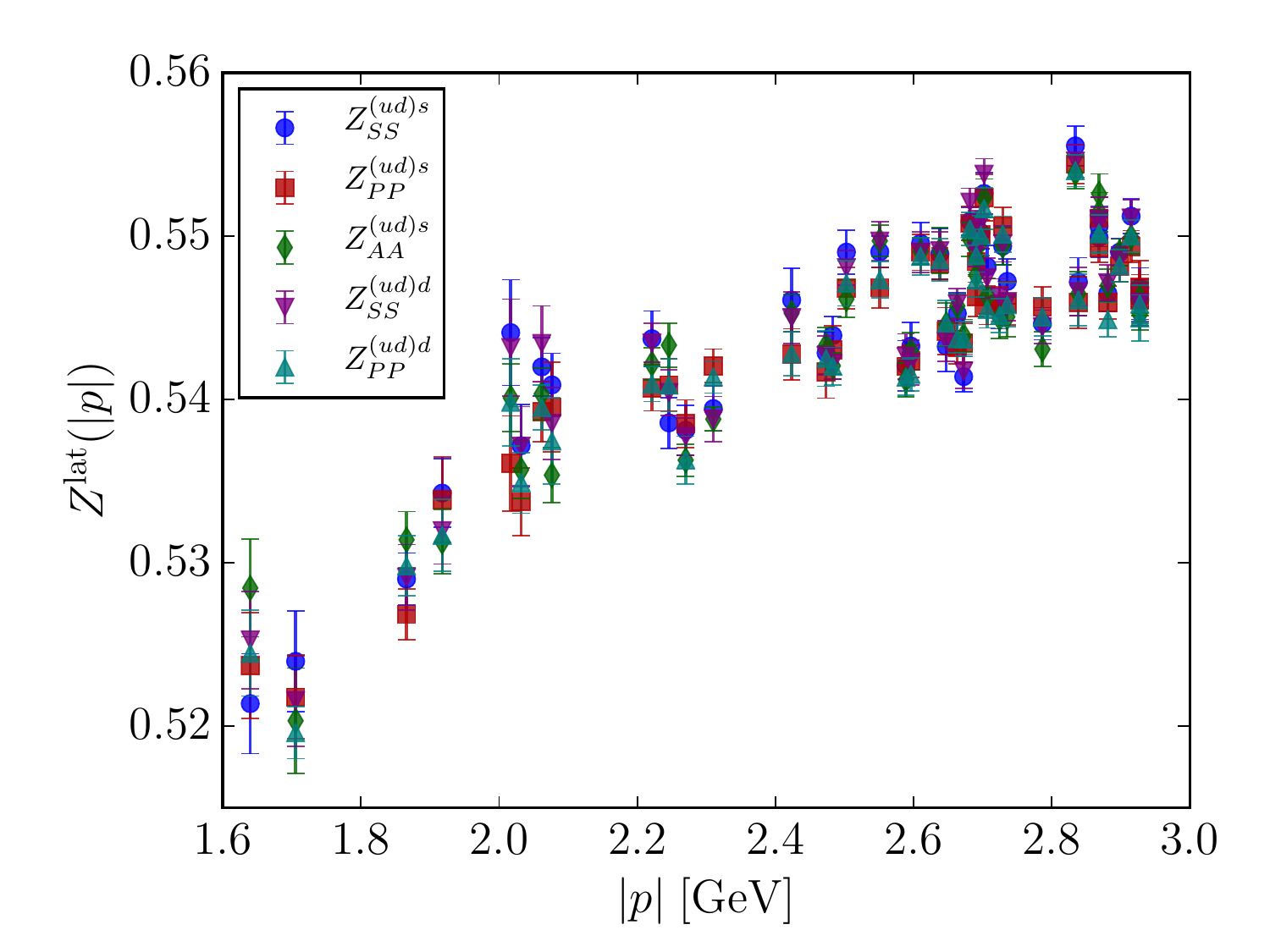}\\
  \includegraphics[width=\textwidth]{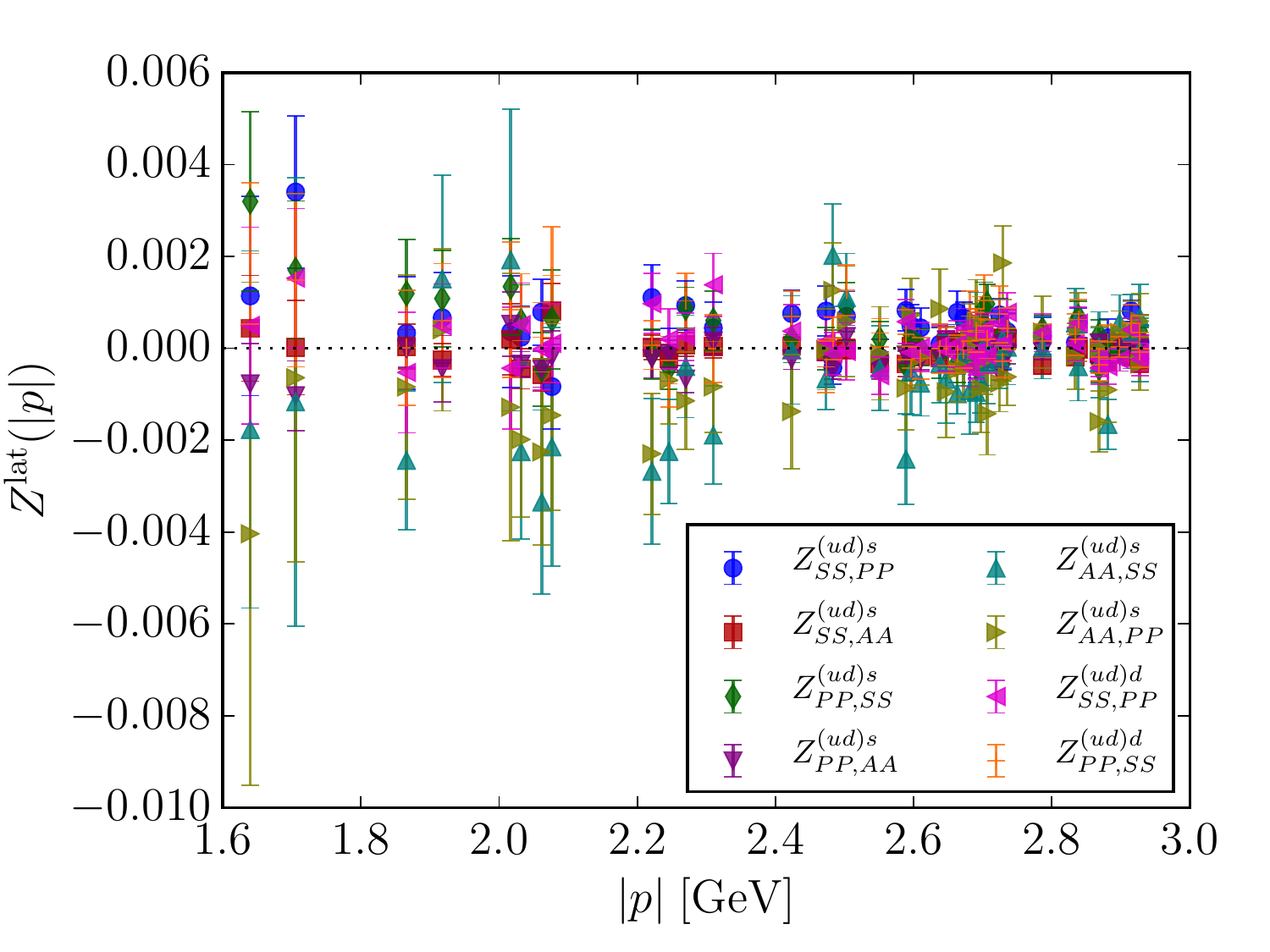}\\
  \caption{32ID}
\end{subfigure}
\caption{\raggedright
  Diagonal renormalization (top) and off-diagonal mixing (bottom) components of the $\SYMpd$
  renormalization matrix~(\ref{eqn:nprZ_3q}).
  \label{fig:nprZ_3q}}
\end{figure}
Combining Eq.~(\ref{eqn:RScheme}) with the axial-vector renormalization~(\ref{eqn:NPR_ZA2Zq}), 
we find the $\SYMpd$ renormalization matrix of three-quark operators
\begin{equation}
\label{eqn:nprZ_3q}
Z^{\text{lat}}_{X,Y}(|p|) 
  = \big[Z_A \Lambda_A(|p|)\big]^{3/2} \, \big[\Lambda^{3q}(|p|)\big]_{X,Y}^{-1}\,,
\end{equation}
and show the results for the diagonal and off-diagonal components in Fig.~\ref{fig:nprZ_3q}.
In previous calculations, $\MOMqbarq$ scheme with exceptional momenta was used to renormalize the
quark fields. 
To emphasize the difference, we will refer to our scheme as $\SYMpd/\SMOMg$, and to the previously
used scheme in Refs.~\cite{Aoki:2008ku,Aoki:2013yxa,Aoki:2017puj} as $\MOMpd/\MOMqbarq$.
All the off-diagonal components are negligible compared to the diagonal components.
The most important observation is that mixing with the $\mcO^{3q}_{AA}$ operator may be neglected
as its matrix elements have not been computed.
Below, we study only the diagonal factors $Z_{X,X}$ and refer to them simply as $Z_X$ for
$X=SS,PP,AA$.

\begin{figure}
\centering
  \includegraphics[width=\textwidth]{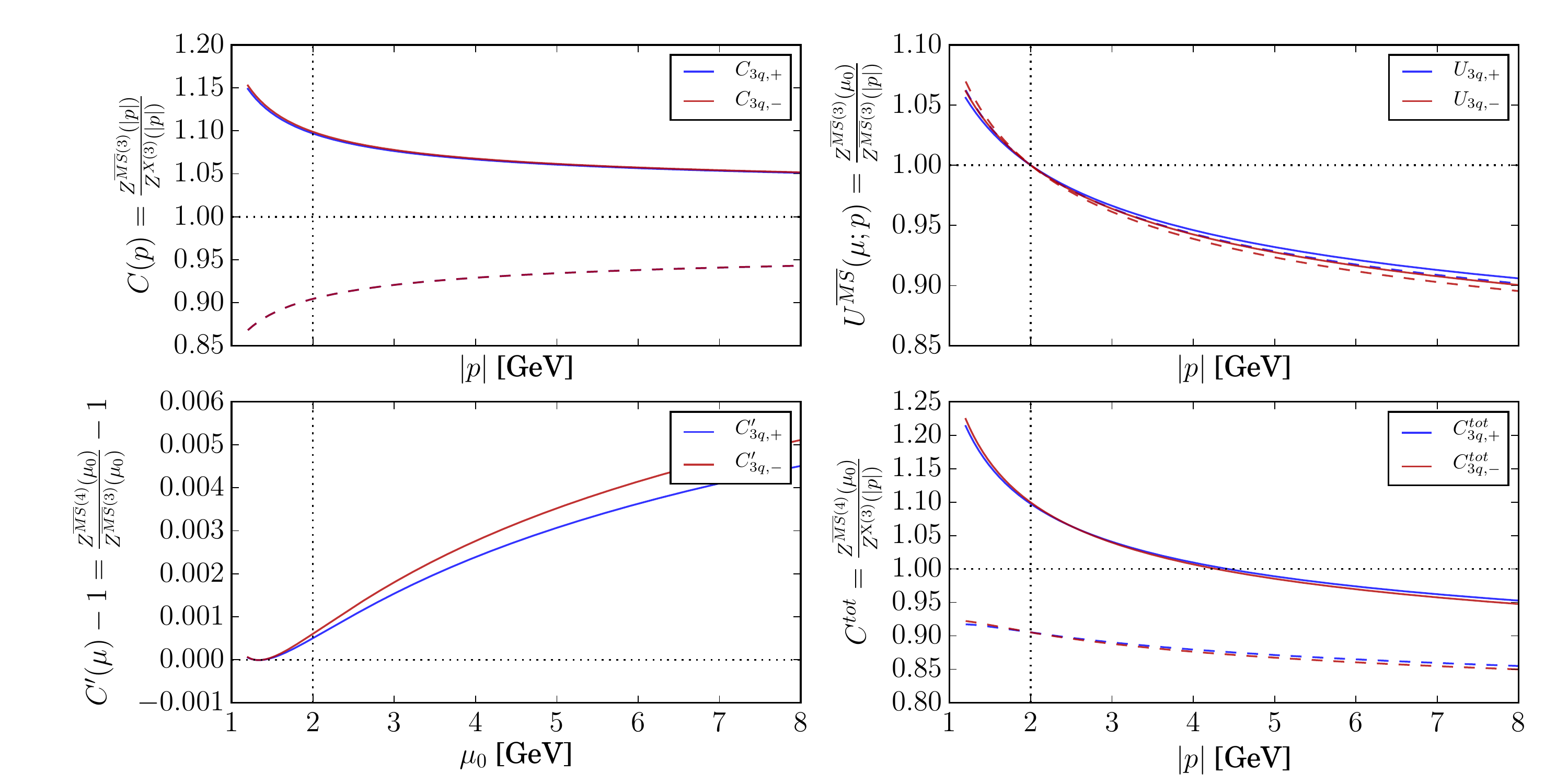}\\
\caption{\raggedright
  (Top left) perturbative running and conversion from $\SYMpd/\SMOMg$ (solid lines) and $\MOMpd/\MOMqbarq$
  (dashed lines) to $\MSbar$ scheme at scale $|p|$;
  (top right) $\MSbar$ running from scale $|p|$ to $\mu_0=2\,\GeV$ at $O(\alpha_S^3)$ used in this
  work (solid) and $O(\alpha_S^2)$ used previously (dashed);
  (bottom left) conversion from $N_f=3$ to $N_f=4$ flavors;
  (bottom right) all factors collected~(\ref{eqn:pert_running}).
  \label{fig:pert_running}}
\end{figure}

\subsection{Perturbative matching}

In order to extract lattice to $\MSbar$-conversion coefficients, we divide $\SYMpd/\SMOMg$
renormalization factors~(\ref{eqn:nprZ_3q}) by their perturbative evolution.
Specifically, we use continuum-QCD conversion factor from $\SYMpd/\SMOMg$ scheme with $N_f=3$
flavors at the momentum-subtraction point $|p|$ to our final $\MSbar$ scheme with $N_f=4$ flavors 
at scale $\mu_0=2\,\GeV$,
\begin{equation}
\label{eqn:pert_running}
\begin{aligned}
&C^{tot}(\mu_0;|p|) = \frac{ Z^{\MSbar(4)}(\mu_0) }{ Z^{\SYMpd/\SMOMg(3)}(|p|) }
\\&\quad  
  = \lp(\frac{ Z^{\MSbar(4)}(\mu_0) }{ Z^{\MSbar(3)}(\mu_0) }\rp) \cdot
    \lp(\frac{ Z^{\MSbar(3)}(\mu_0) }{ Z^{\MSbar(3)}(|p|) }\rp) \cdot
    \lp(\frac{ Z^{\MSbar}(|p|)   }{ Z^{\SYMpd/\SMOMg}(|p|) }\rp) \,,
\end{aligned}
\end{equation}
where the last factor is computed with $N_f=3$ flavors as summarized in
Appendix~\ref{sec:app_renorm}.

Since there is perturbative mixing between $SS$ and $PP$ operators starting at $O(\alpha_S^2)$, the
anomalous dimension matrix has to be diagonalized.
This results in two different anomalous dimensions $\gamma_\pm$ for
operators~(\ref{eqn:O3q_pm_basis})~\cite{Pivovarov:1991nk,Gracey:2012gx}\footnote{
  Using parity, we take the average the opposite chirality operators 
  $\mcO_+=\frac12(\mcO_{RR}+\mcO_{LL})$ and $\mcO_+=\frac12(\mcO_{LR}+\mcO_{RL})$.
}.
We integrate these $O(\alpha_S^3)$ anomalous dimensions~\cite{Gracey:2012gx} using the 4-loop
$\alpha_S^{\MSbar}(\mu)$ running derived from $\alpha_S^{\MSbar}(M_Z)=0.11823(81)$ and matched to
the $N_f=3$ QCD at $\bar{m}_b$ and $\bar{m}_c$ thresholds~\cite{Zyla:2020zbs}.
The complete conversion factor is shown in Fig.~\ref{fig:pert_running}, as well as all the factors
separate factors in the r.h.s. of Eq.~(\ref{eqn:pert_running}).
The correction from $N_f=3$ dynamical flavors used in lattice calculations to $N_f=4$ active flavors
at scale $\mu_0$ is smaller than $10^{-3}$ and thus may be neglected.
Also, the difference in evolution of operator normalization with $O(\alpha_S^2)$ and $O(\alpha_S^3)$ 
anomalous dimensions are very small.
Finally, we compare the complete conversion factors in the $\SYMpd/\SMOMg$ and $\MOMpd/\MOMqbarq$
schemes in Fig.~\ref{eqn:pert_running} (bottom right).
The $\alpha_S$ and $\alpha_S^2$ orders contribute respectively $\approx7.1\%$ and $\approx2.6\%$ 
to perturbative $\SYMpd\to\MSbar$ conversion factors~\cite{Gracey:2012gx}, and we estimate the
perturbative systematic uncertainty as half of the $O(\alpha_S^2)$ contribution at $\approx1.3\%$.
Such uncertainty is negligible compared to uncertainties from other sources, in particular,
stochastic and discretization effects.

\begin{figure}
\centering
\begin{subfigure}[b]{.49\textwidth}
  \includegraphics[width=\textwidth]{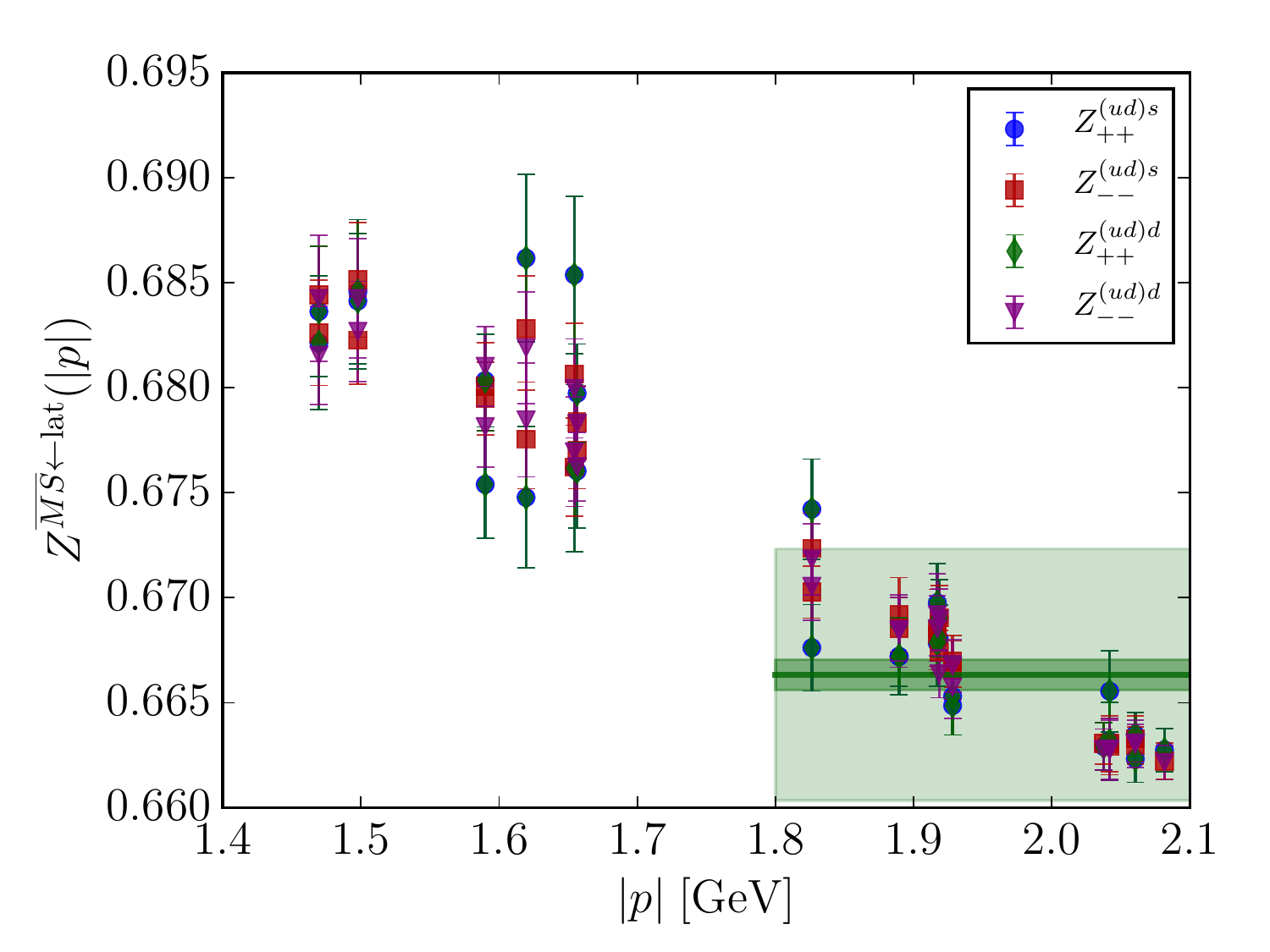}\\
  \caption{24ID}
\end{subfigure}
\begin{subfigure}[b]{.49\textwidth}
  \includegraphics[width=\textwidth]{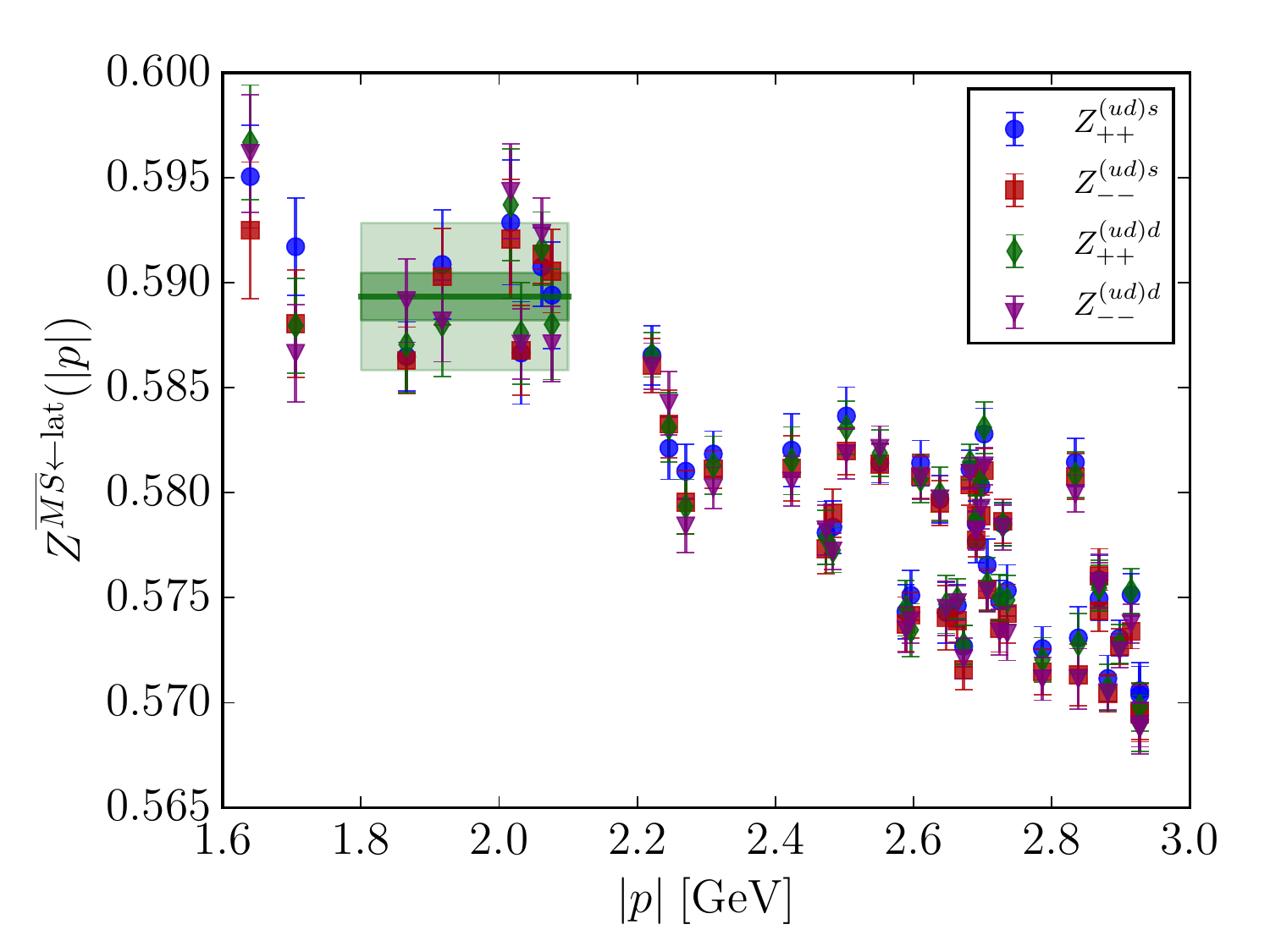}\\
  \caption{32ID}
\end{subfigure}
\caption{\raggedright
  Diagonal conversion factors from $\SYMpd/\SMOMg$ to $\MSbar(2\,\text{GeV})$
  scheme~(\ref{eqn:nprZSI_3q}), in which perturbative running with intermediate scale $|p|$ has been
  removed.
  In absence of discretization, nonperturbative, and higher-order perturbative effects, it should be
  independent of $|p|$.
  The green bands indicate averages over the same momentum range on both ensembles, which is
  necessary for consistent continuum extrapolation.
  \label{fig:nprZSI_3q}}
\end{figure}

Conversion factors from lattice to $\MSbar(N_f=4,\mu_0)$ are products of the perturbative
running~(\ref{eqn:pert_running}) and lattice renormalization factor~(\ref{eqn:nprZ_3q}) in the
$\mcO_\pm$ basis~(\ref{eqn:O3q_pm_basis}) 
\begin{equation}
\label{eqn:nprZSI_3q}
Z^{\MSbar\leftarrow\text{lat}}(\mu_0;|p|) 
  = C^{tot}(\mu_0;|p|) Z^{\text{lat}}(|p|) \,,
\end{equation}
which should be scale independent of the intermediate scale $|p|$ within the
window~(\ref{eqn:scale_window}).
Indeed, as shown in Fig.~\ref{fig:nprZSI_3q}, the variation of $Z^{\MSbar\leftarrow\text{lat}}(\mu_0;|p|)$ 
with the lattice scale $|p|$ is insignificant compared to our target precision.
Final renormalization numbers are determined as simple averages of central values in the range
$|p|=1.8\ldots2.1\,\GeV$ for both ensembles.
While data at larger scales are available for the 32ID ensemble, we use the same scale window in
physical units to ensure consistency of our continuum extrapolation below.
The systematic uncertainties are estimated as half of the maximal variation in the averaging range,
and are subdominant compared to the perturbative uncertainty discussed above; combined systematic
uncertainties are $\lesssim1.6\%$.
The statistical uncertainties are estimated with Jackknife resampling and are $\lesssim0.2\%$.
The final renormalization constants are collected in Tab.~\ref{tab:nprZ3q_final}.

\begin{table}[ht]
\centering
\caption{\raggedright
  Final renormalization factors from lattice to $\MSbar(N_f=4,\,\mu=2\,\GeV)$ with statistical(1), 
  and   systematic uncertainties from momentum scale(2) and perturbative matching(3).
  \label{tab:nprZ3q_final}}
\begin{tabular}{l | D..{16} D..{16} | D..{16} D..{16} }
\hline\hline
  & \multicolumn{1}{c}{$Z^{(ud)s}_{++}$} & \multicolumn{1}{c|}{$Z^{(ud)s}_{--}$} 
  & \multicolumn{1}{c}{$Z^{(ud)d}_{++}$} & \multicolumn{1}{c}{$Z^{(ud)d}_{--}$} \\
\hline
24ID  & 0.6671(7)(60)(87)   & 0.6674(7)(51)(87)   & 0.6671(7)(60)(87)   & 0.6672(7)(49)(87)     \\
32ID  & 0.5895(11)(32)(77)  & 0.5896(9)(29)(77)   & 0.5893(11)(33)(77)  & 0.5897(9)(36)(77) \\
\hline\hline
\end{tabular}
\end{table}

\section{Results
  \label{sec:result}}
\subsection{Hadron spectrum
  \label{sec:result_spectrum}}
The first step of the analysis is to extract energies of proton and meson states and their overlaps
with the lattice operators from their two-point correlators.
We perform multi-state fits in order to control systematic effects arising from hadron excited
states.
Statistical precision of our data and coarse step in the time direction are sufficient to constrain
effectively only one excited state in each case.
Energy gaps between the ground and the excited state have the most impact on correct removal of
excited-state contamination from matrix elements determined from three-point correlation functions.

To find approximate values of the ground state parameters, we first perform 1-state fits with $\tmin$
sufficiently large to yield good $p$-values for all momenta $p^2=(0\ldots4)(2\pi/L)^2$.
We then perform series of two-state fits~(\ref{eqn:twopt_Pi_exp},\ref{eqn:twopt_N_exp}) with varying
time ranges $[\tmin, \tmax]$.
We use values of the ground-state overlaps $C_0$ and the energies $E_0$ from the 1-state
fits with $\tmin=6a$ for 24ID and $\tmin=7a$ for 32ID to impose Gaussian prior
constraints in order to stabilize the two-state fits.
To ensure that these priors are non-informative, we set their normal widths equal to
$(3\ldots5)\times$ their statistical uncertainties in the 1-state fits.
In addition, we impose priors on the energy gap $\Delta E_1$ with a wide log-normal prior
distribution
\begin{equation}
p_\text{log-N}(\Delta E_1) = \exp\Big[-\frac12 P_\text{log-N}(\Delta E_1) \Big]
  = \exp\Big[-\frac1{2\lambda_{\Delta E_1}} \Big(\log\frac{ \Delta E_1}{
          \widetilde{\Delta E_1}}\Big)^2\Big]
\end{equation}
with the mode $\widetilde{\Delta E_1}=0.5\,\mathrm{GeV}$ and the log-width $\lambda_{\Delta E_1}=3$.
We also impose constraints on the excited-state overlaps $C_1>0$, since the source and the sink
operators are the same.

To perform the fits, we use the ``augmented'' $\tilde\chi^2$ function 
\begin{equation}
\begin{aligned}
\tilde\chi^2 &= \chi^2 + \sum_{k} P_k(p_k)\,,
\\
\chi^2 &= \sum_{t,t^\prime} \big(y_t - \tilde C_{2pt}(t)\big) 
  S_{t,t^\prime}^{-1} 
  \big(y_{t^\prime} - \tilde C_{2pt}(t^\prime)\big)\,,
\end{aligned}
\end{equation}
where $P_k(p_k)$ are log-likelihood weights of the prior constraints imposed on parameters $p_k$,
and the regular $\chi^2$ is computed with the sample covariance matrix $S$.
The fits are performed by minimizing this ``augmented least-squares'' with trusted-region
Levenberg-Marquardt algorithm.

\begin{table}
\caption{\raggedright
  Ground-state energies of the pion, the kaon, and the nucleon extracted from the 2-state
  fits of the two-point functions with $tmin=2$ and used in three-point function fits.
  The columns correspond hadron momenta $p^2=(2\pi/L)^2 n^2$.
  \label{tab:energies}}
\begin{tabular}{cc| D..{10} D..{10} D..{10} D..{10} D..{10} }
\hline\hline
& & \multicolumn{1}{c}{$n^2=0$} & \multicolumn{1}{c}{$n^2=1$} 
& \multicolumn{1}{c}{$n^2=2$} & \multicolumn{1}{c}{$n^2=3$} & \multicolumn{1}{c}{$n^2=4$} \\
\hline
24ID
& $\pi$ & 0.13983(72) & 0.29951(95) & 0.3997(18) &  0.4799(39) &  0.531(22)  \\
& $K$   & 0.5079(25)  & 0.5719(28)  & 0.6285(33) &  0.6788(41) &  0.7232(66) \\
& $n$   & 0.953(19)   & 0.971(19)   & 0.998(22)  &  1.041(19)  &  1.073(25)  \\ 
\hline
32ID
& $\pi$ & 0.13889(68) & 0.3049(11)  & 0.4085(19)  & 0.4913(41) & 0.566(37)  \\
& $K$   & 0.48817(76) & 0.55725(78) & 0.61868(97) & 0.6746(15) & 0.7260(30) \\
& $n$   & 0.936(20)   & 0.983(15)   & 1.032(11)   & 1.067(13)  & 1.115(11)  \\
\hline\hline
\end{tabular}
\end{table}

\begin{figure}[ht!]
\centering
\begin{subfigure}[b]{.32\textwidth}
  \includegraphics[width=\textwidth]{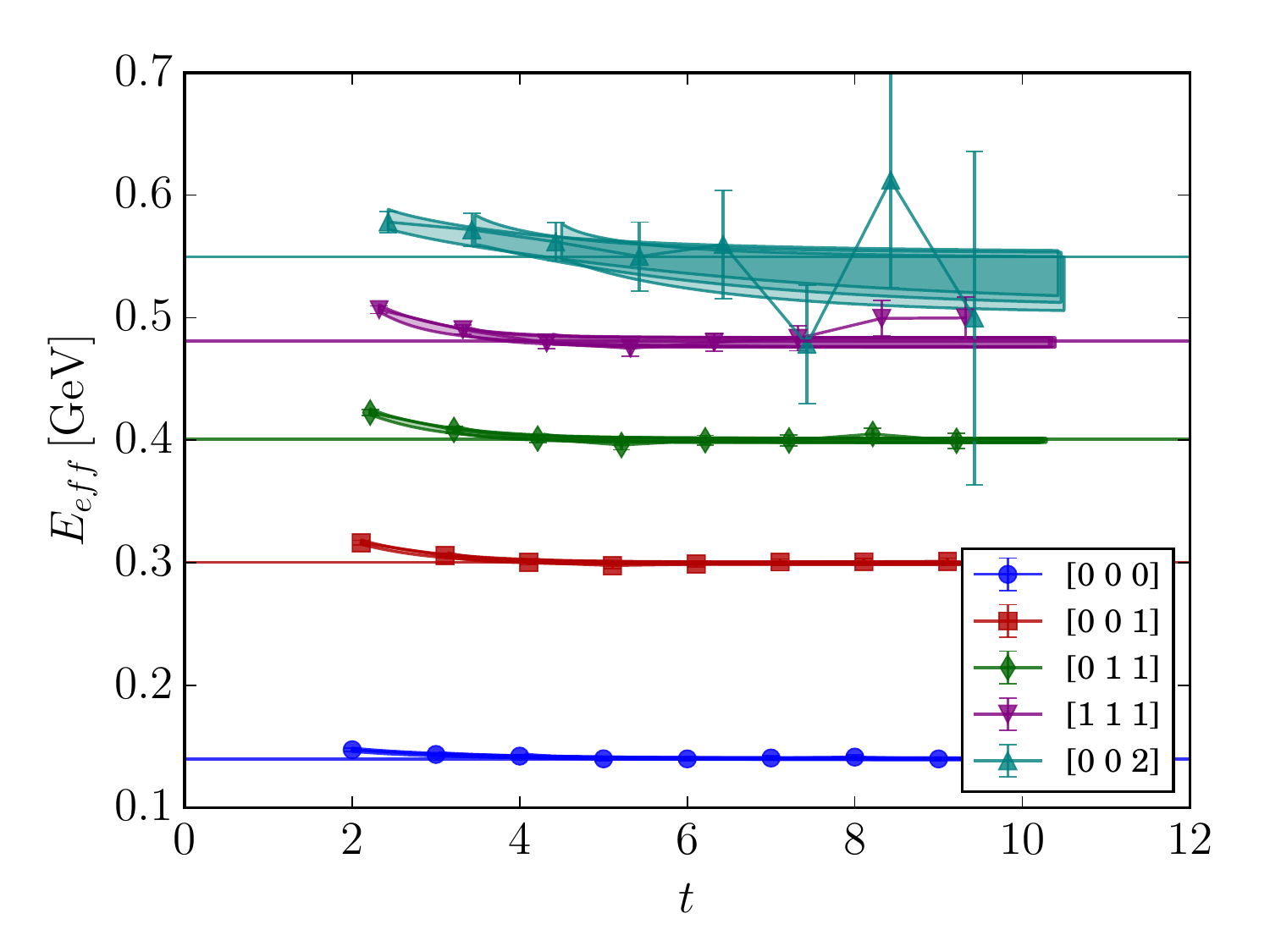}
  \includegraphics[width=\textwidth]{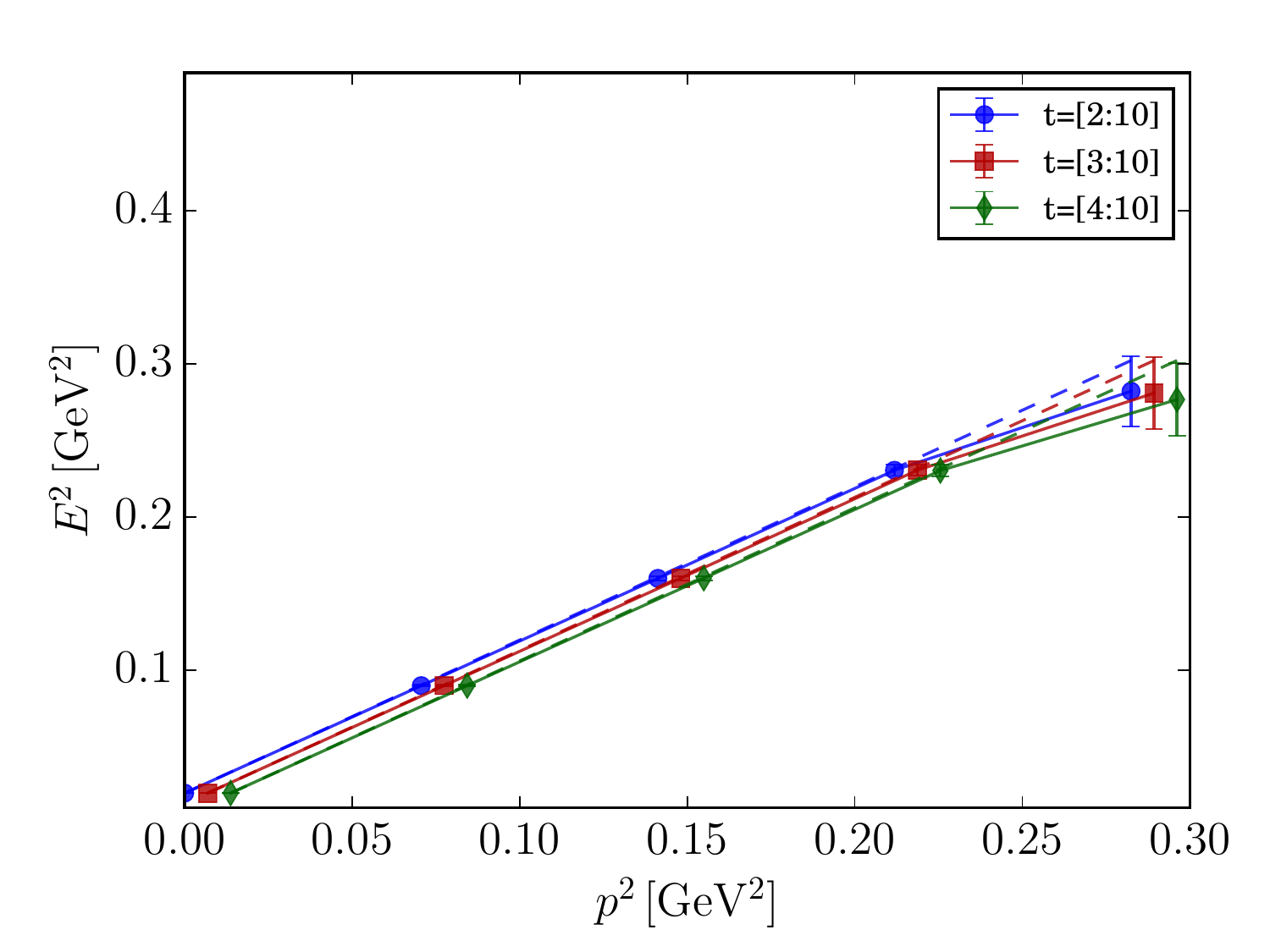}
  \caption{ Pion }
\end{subfigure}%
\begin{subfigure}[b]{.32\textwidth}
  \includegraphics[width=\textwidth]{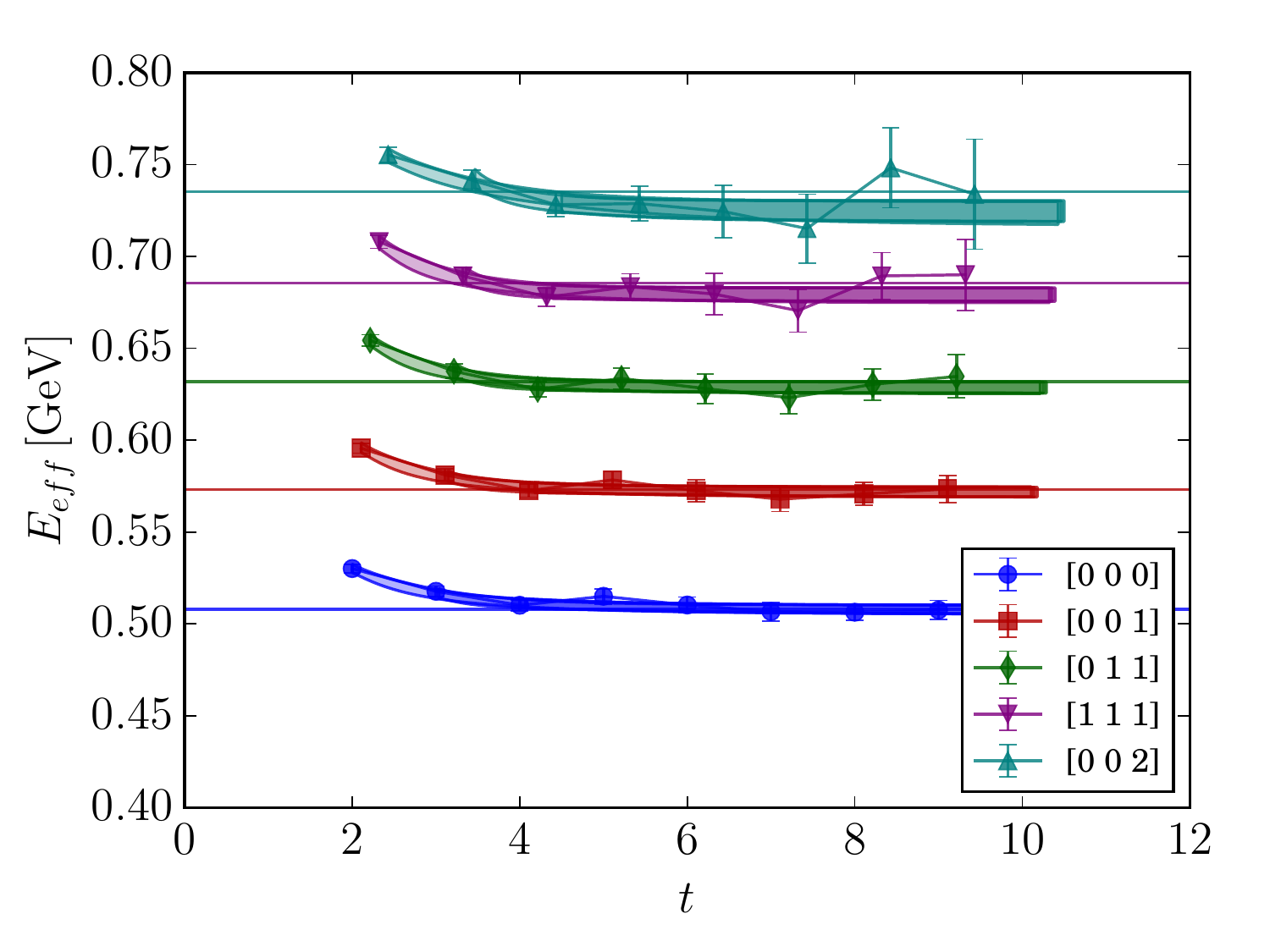}
  \includegraphics[width=\textwidth]{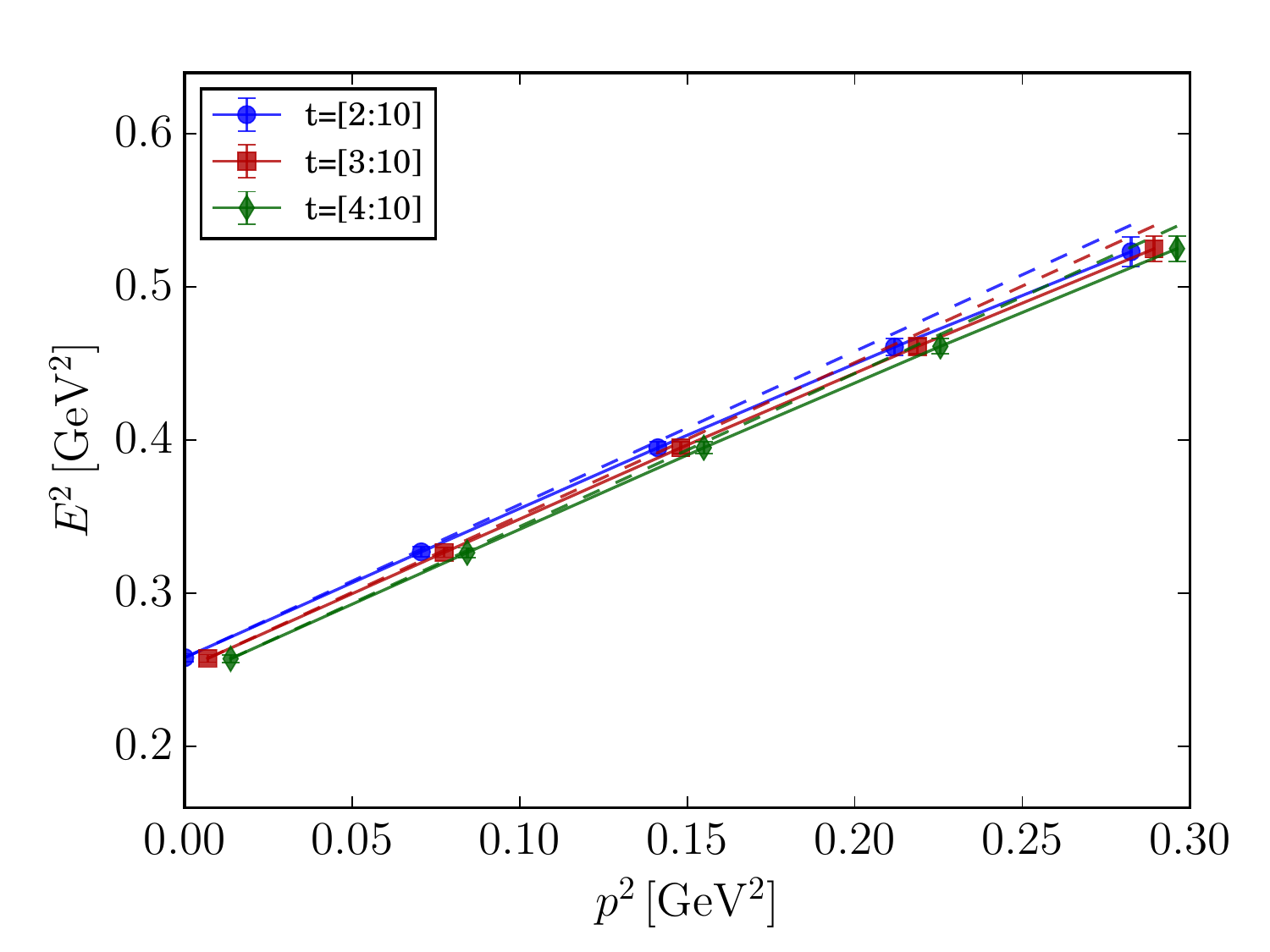}
  \caption{ Kaon }
\end{subfigure}%
\begin{subfigure}[b]{.32\textwidth}
  \includegraphics[width=\textwidth]{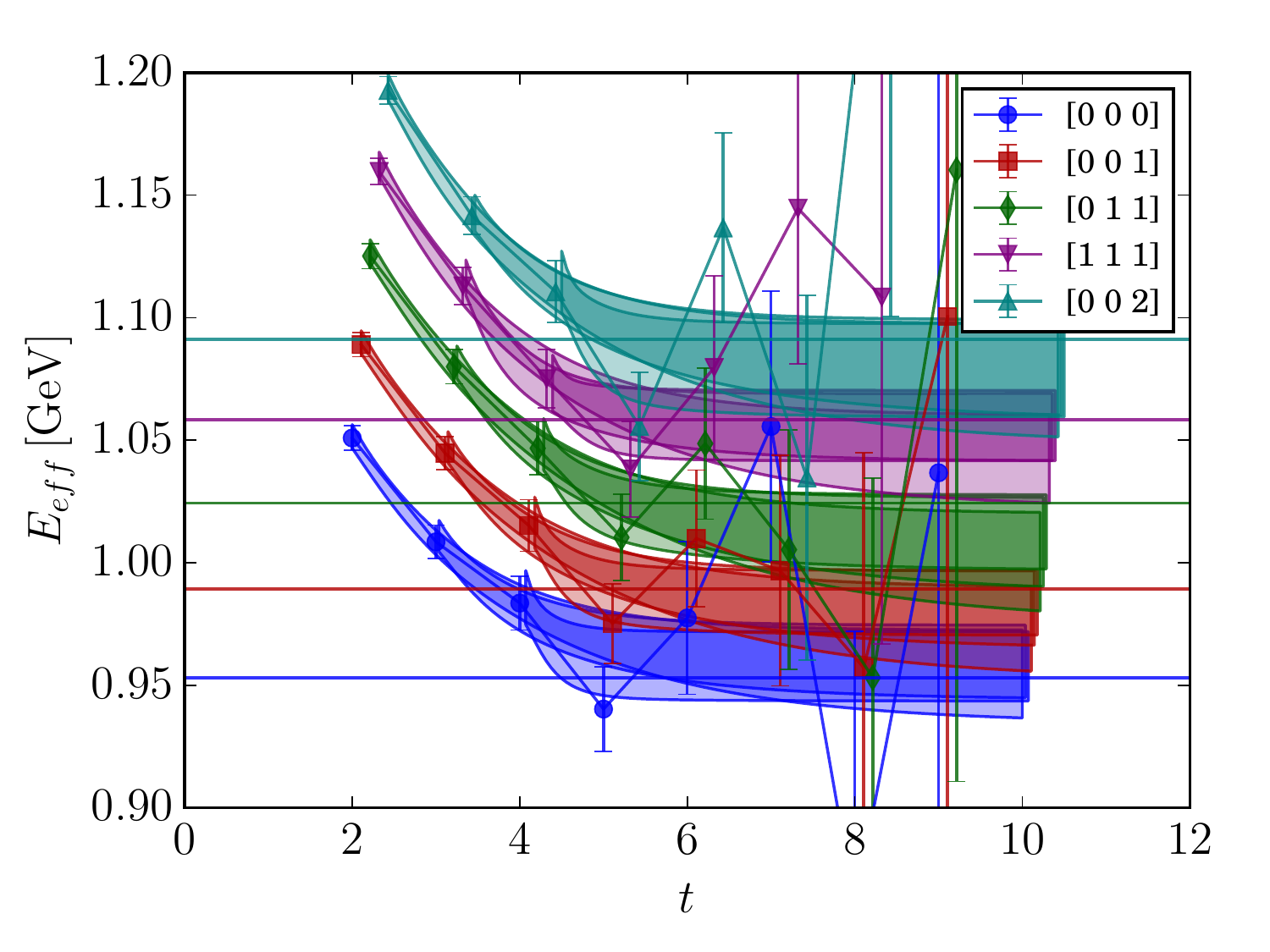}
  \includegraphics[width=\textwidth]{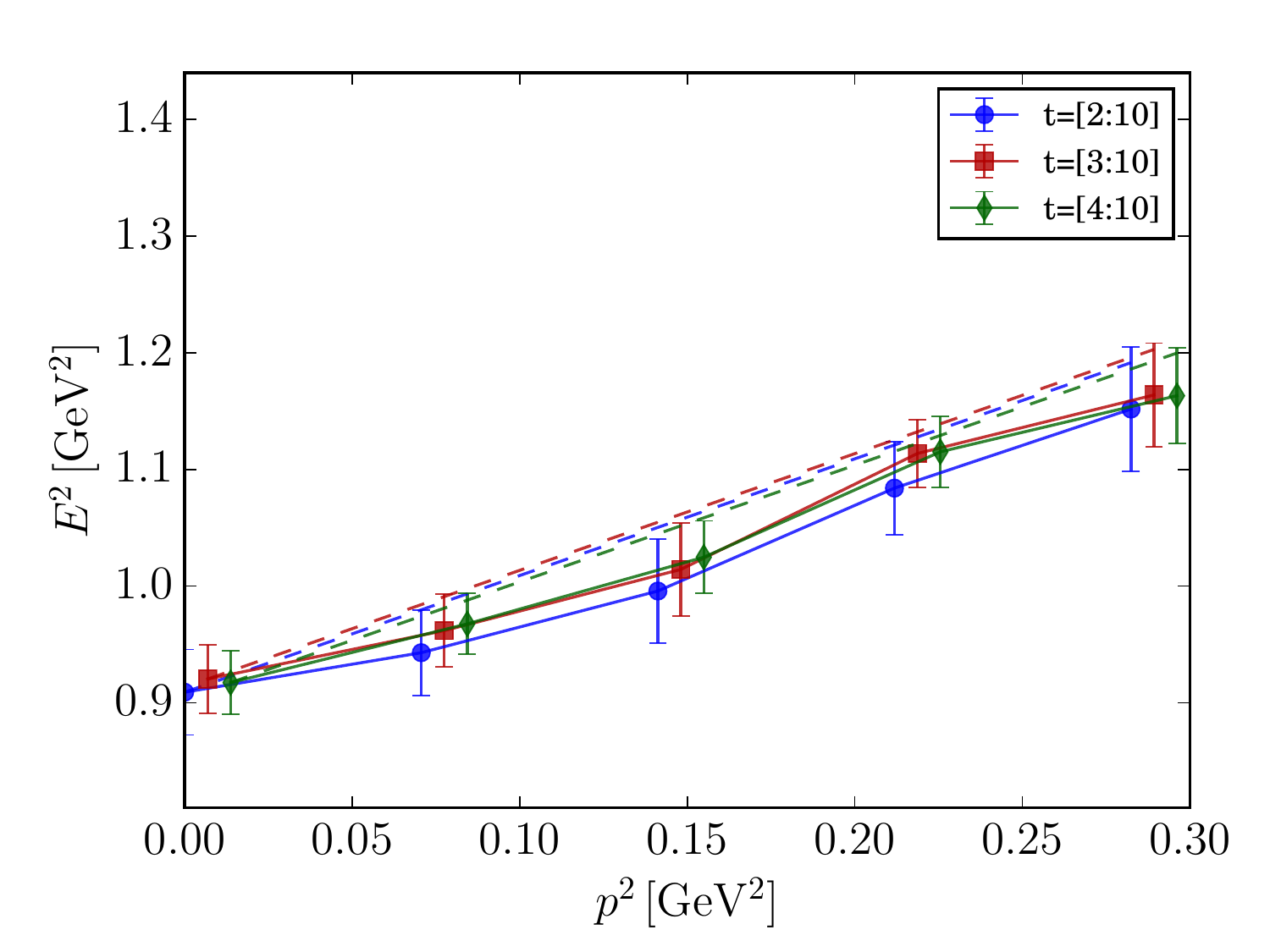}
  \caption{ Nucleon }
\end{subfigure}%
\caption{\raggedright
  Two-state fits to pion, kaon, and nucleon two-point correlation functions shown as effective
  energies (top)   and ground-state dispersion relations (bottom) on the 24ID ensemble. 
  The continuum dispersion relations $E^2(p)=E^2(0) + p^2$ for ground-state energies
  are shown with horizontal lines in the top panels and  with dashed lines in the bottom panels.
  \label{fig:fits_dispersion_ID24}}
\end{figure}

\begin{figure}[ht!]
\centering
\begin{subfigure}[b]{.32\textwidth}
  \includegraphics[width=\textwidth]{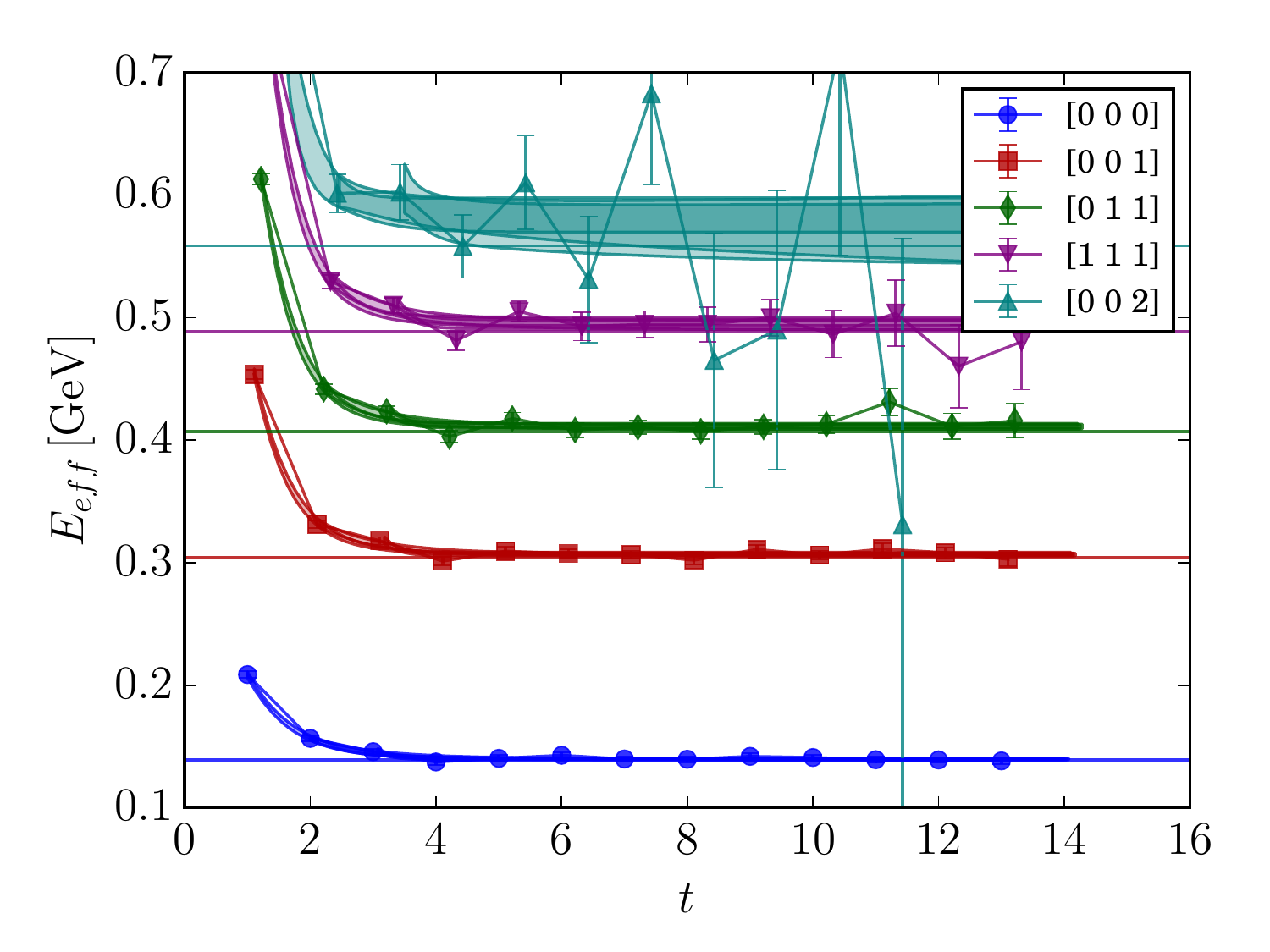}
  \includegraphics[width=\textwidth]{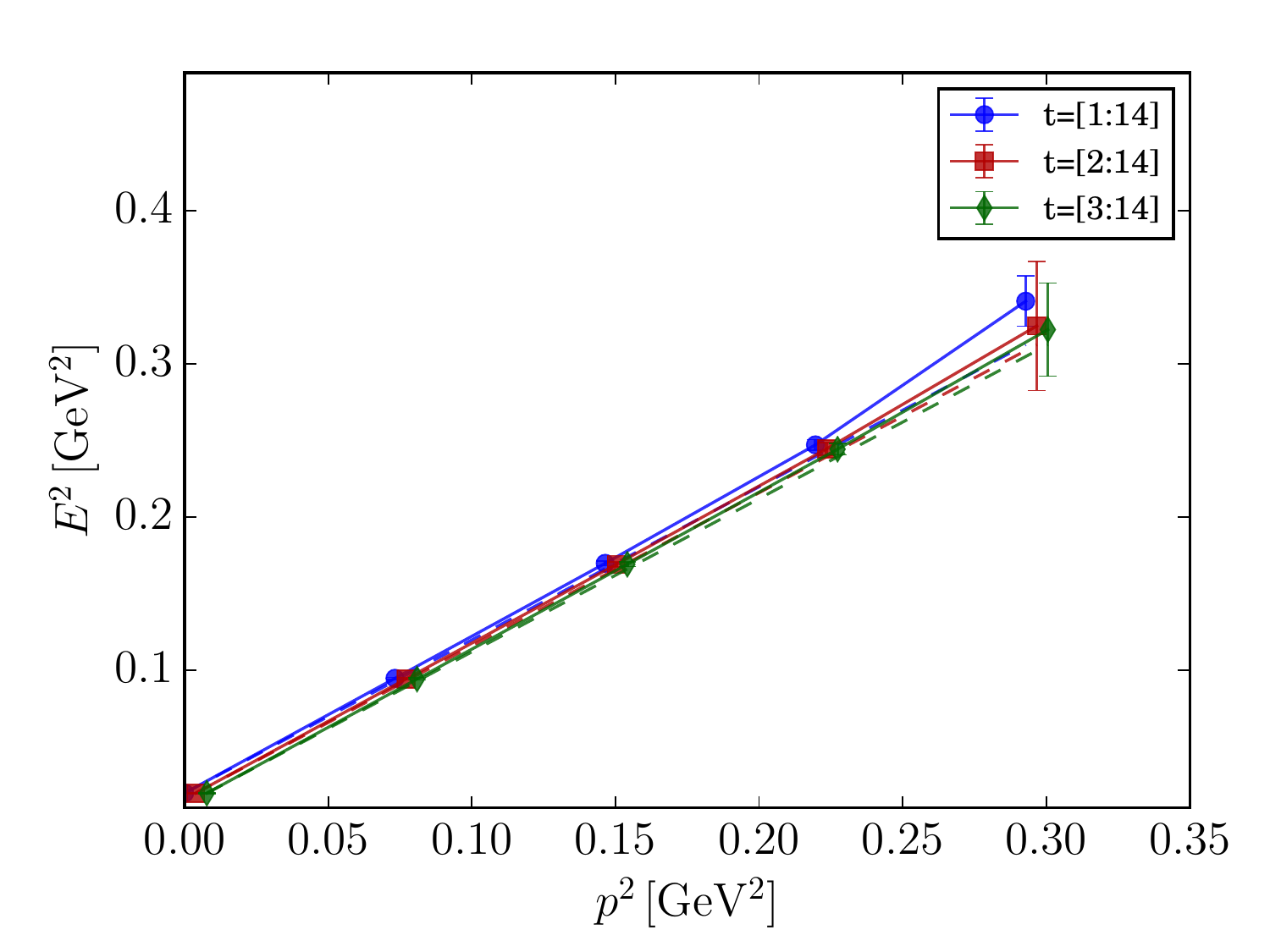}
  \caption{ Pion }
\end{subfigure}%
\begin{subfigure}[b]{.32\textwidth}
  \includegraphics[width=\textwidth]{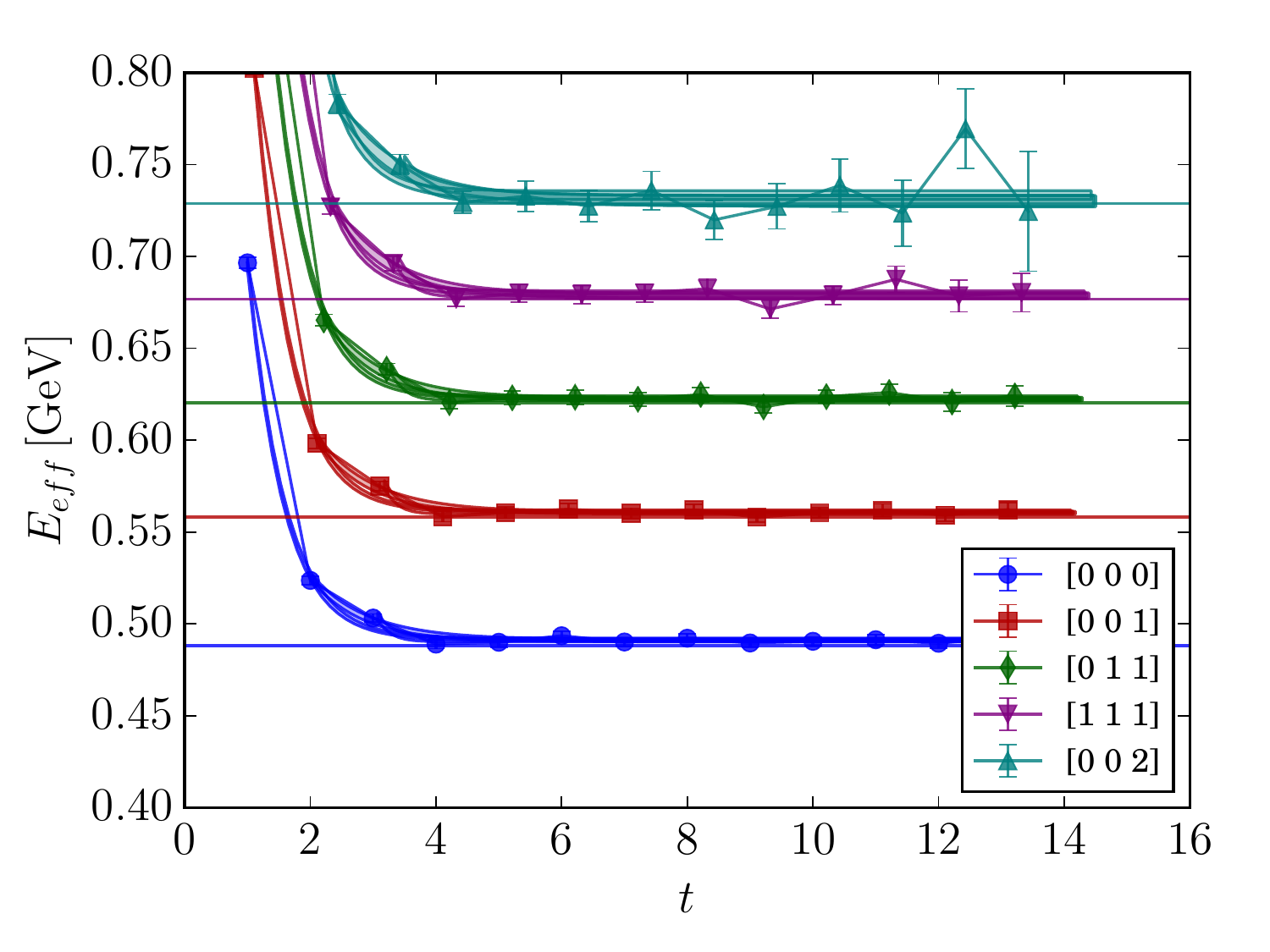}
  \includegraphics[width=\textwidth]{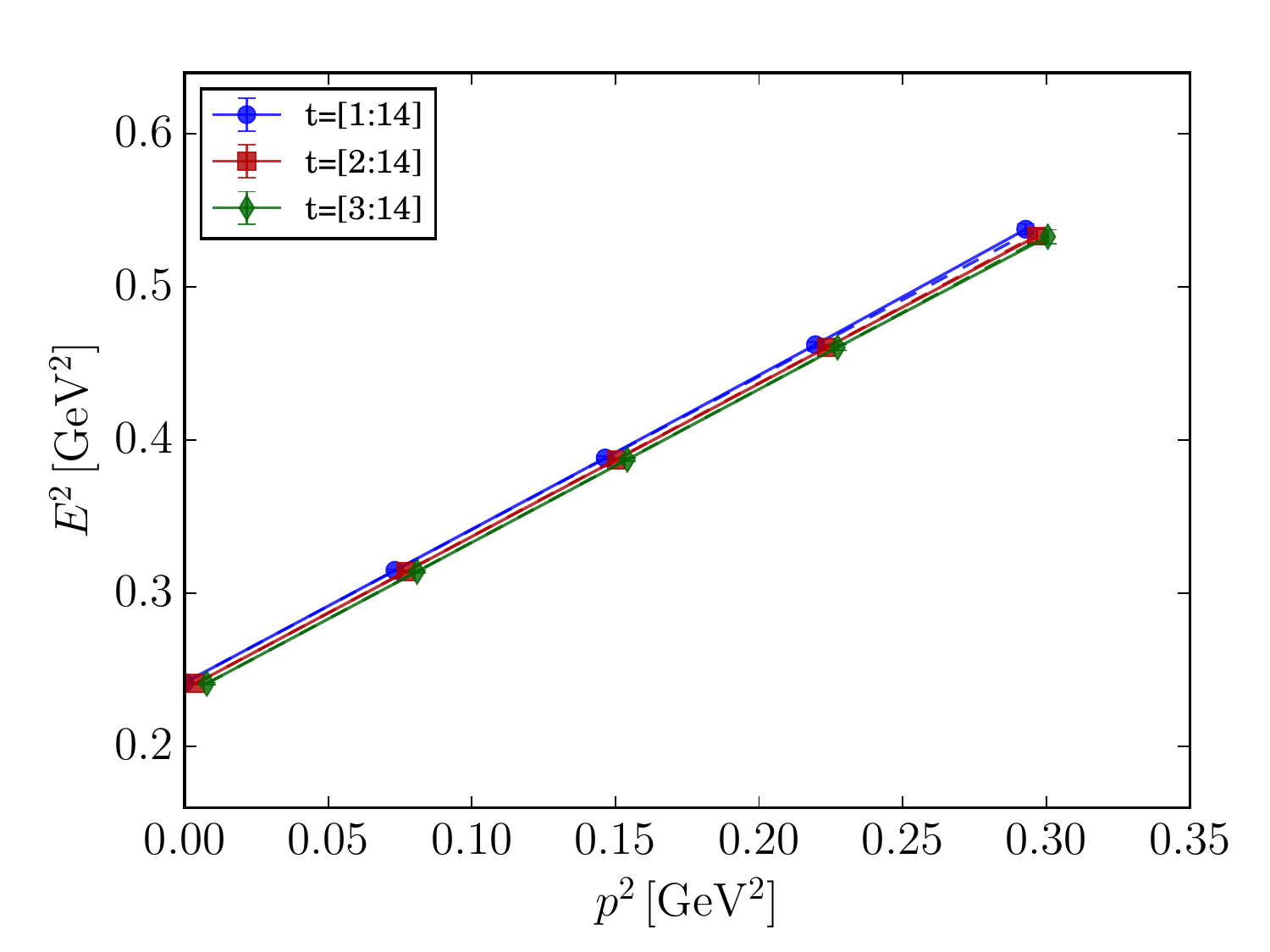}
  \caption{ Kaon }
\end{subfigure}%
\begin{subfigure}[b]{.32\textwidth}
  \includegraphics[width=\textwidth]{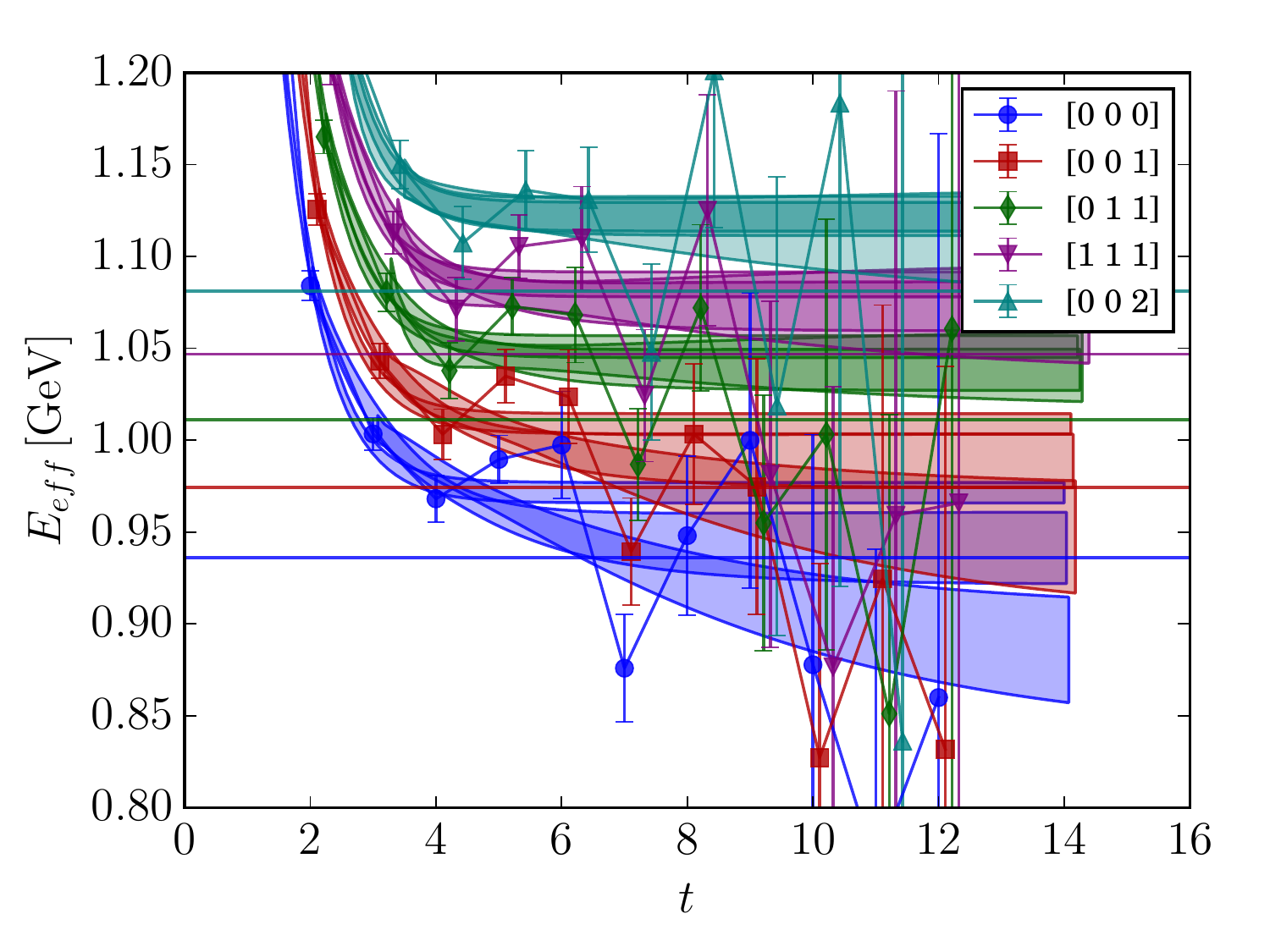}
  \includegraphics[width=\textwidth]{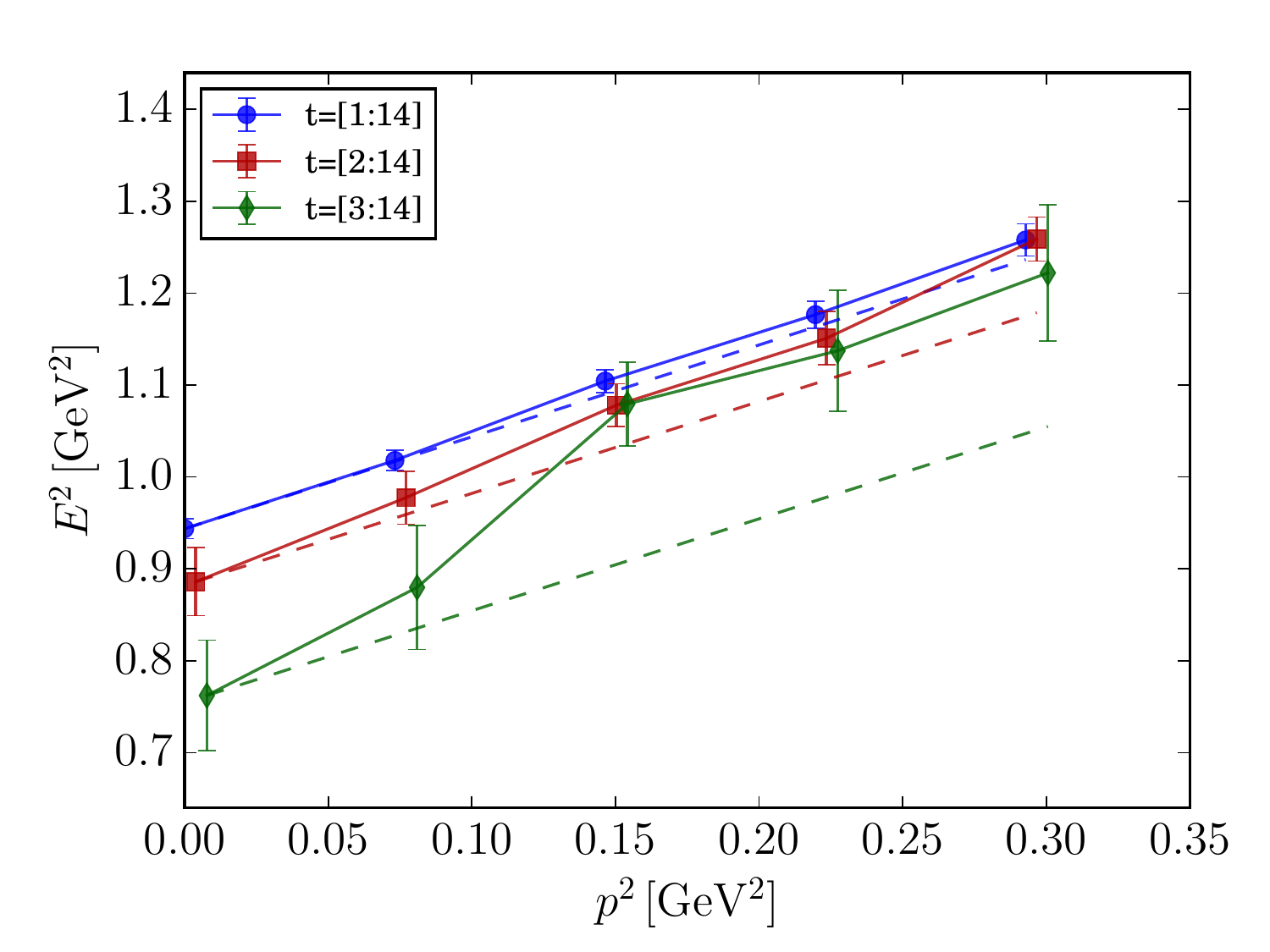}
  \caption{ Nucleon }
\end{subfigure}%
\caption{\raggedright
  Two-state fits to pion, kaon, and nucleon two-point correlation functions shown as effective
  energies (top)   and ground-state dispersion relations (bottom) on the 32ID ensemble. 
  See caption to Fig.~\ref{fig:fits_dispersion_ID24} for description.
  \label{fig:fits_dispersion_ID32}}
\end{figure}

The fits for pion, kaon, and nucleon at all relevant momenta are summarized in 
Figs.~(\ref{fig:fits_dispersion_ID24},\ref{fig:fits_dispersion_ID32}).
In the bottom panels, we compare the resulting energies $E^2(p^2)$ to the continuum dispersion
relation $E^2=m^2+p^2$ using masses $m=E(0)$ obtained on the lattice.
Uncertainties in all fits are estimated using bootstrap resampling with $N_\text{boot}=256$ samples.

Kaon correlation functions are the most statistically precise and their 2-state fits exhibit
remarkable consistency between $\tmin=(2\ldots4)a$ on 24ID and $\tmin=(2\ldots3)a$ on 32ID ensembles;
the corresponding ground-state energies align perfectly with the continuum dispersion relation.
Pion correlation functions are less statistically precise, perhaps due to larger fluctuations of
light-quark propagators. 
They display especially large fluctuation at the largest momentum $(0,0,2)$, which may indicate that
our approximation of light-quark propagators with a truncated CG and low-eigenmode deflation does
not perform as well for higher-momentum hadron states.
For all lower momenta, however, the data are precise, the fits are stable, and the agreement with
the continuum dispersion relation is remarkable. 
The largest-momentum pion data are used only in one of the three kinematic points, and its low
precision has very limited adverse effect on the final results.
The nucleon data are the least precise and the fits show some dependency on the fitting range.
On the 24ID ensemble, all fits with $\tmin=(2\ldots4)$ produce consistent results that agree with
the continuum dispersion relations.
On the 32ID ensemble, however, the results at some of the momenta depend on the fit range, 
albeit within statistical fluctuations.
We attribute this difference to over-smearing of the light-quark propagator sources on the 32ID
ensemble, where the larger statistical fluctuations make it difficult to constrain the
smaller excited-state contributions and specifically their energy gaps.
Although suppressing excited-state contributions is generally advantageous, poorly constrained
energy gaps may lead to larger fluctuations in the ground state matrix elements to be determined in
the next step.
For subsequent fitting of the 3-point functions on both 24ID and 32ID ensembles, we select parameters 
from 2-state fits with $\tmin=2a$, which all have satisfactory $p$-values.
These energies are collected in Tab.~\ref{tab:energies}.

Since we use the sample covariance matrix that may be poorly determined, assessing the fit quality
with the usual $\chi^2$ distribution may be misleading.
Instead, we judge the quality of the fits in two ways: 
(a) as $p$-values computed from the Hoteling distributions of the optimal ``$\chi^2$'' values, and 
(b) using empirical cumulative distribution of ``$\chi^2$'' computed with bootstrap variation of
the data around the optimal fit curve as detailed in Ref.~\cite{Kelly:2019wfj}.
Since the prior constraints are used only to stabilize the search of the optimal point, they are not
included in computing the ``$\chi^2$'' or the number of degrees of freedom.

\subsection{Proton decay form factors}
\begin{figure}[h!]
\centering
\includegraphics[width=.85\textwidth]{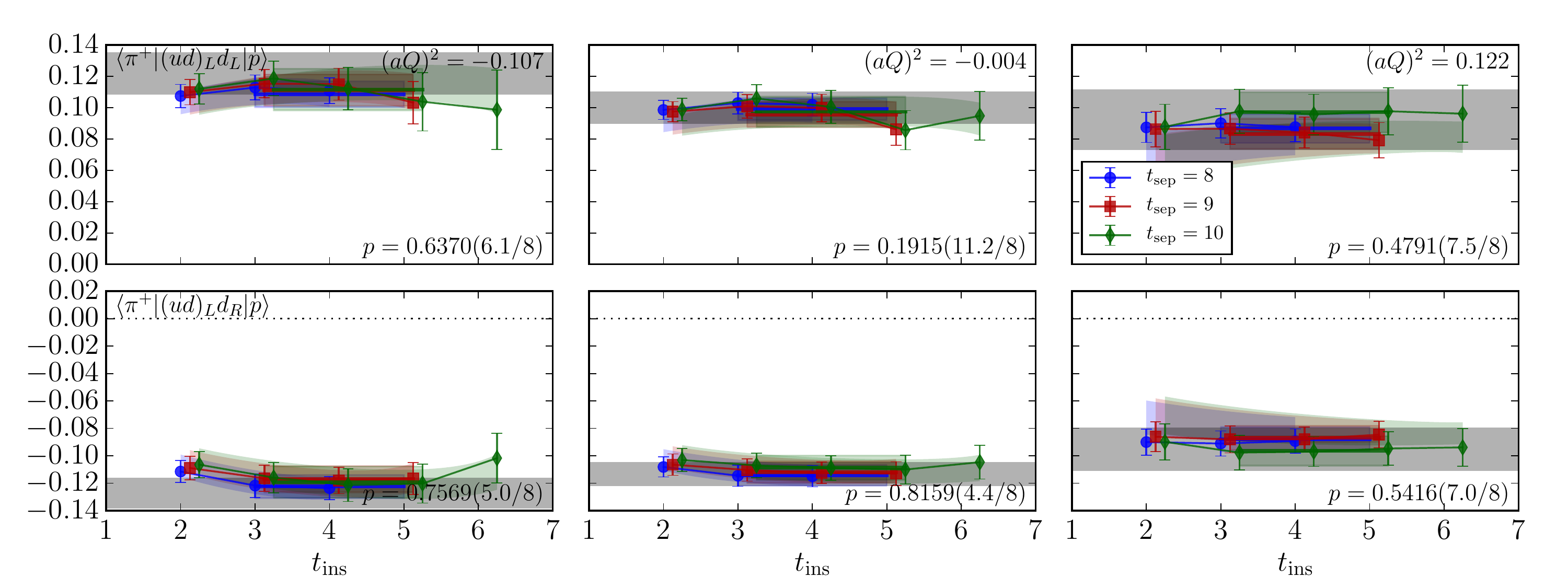}\\
\includegraphics[width=.85\textwidth]{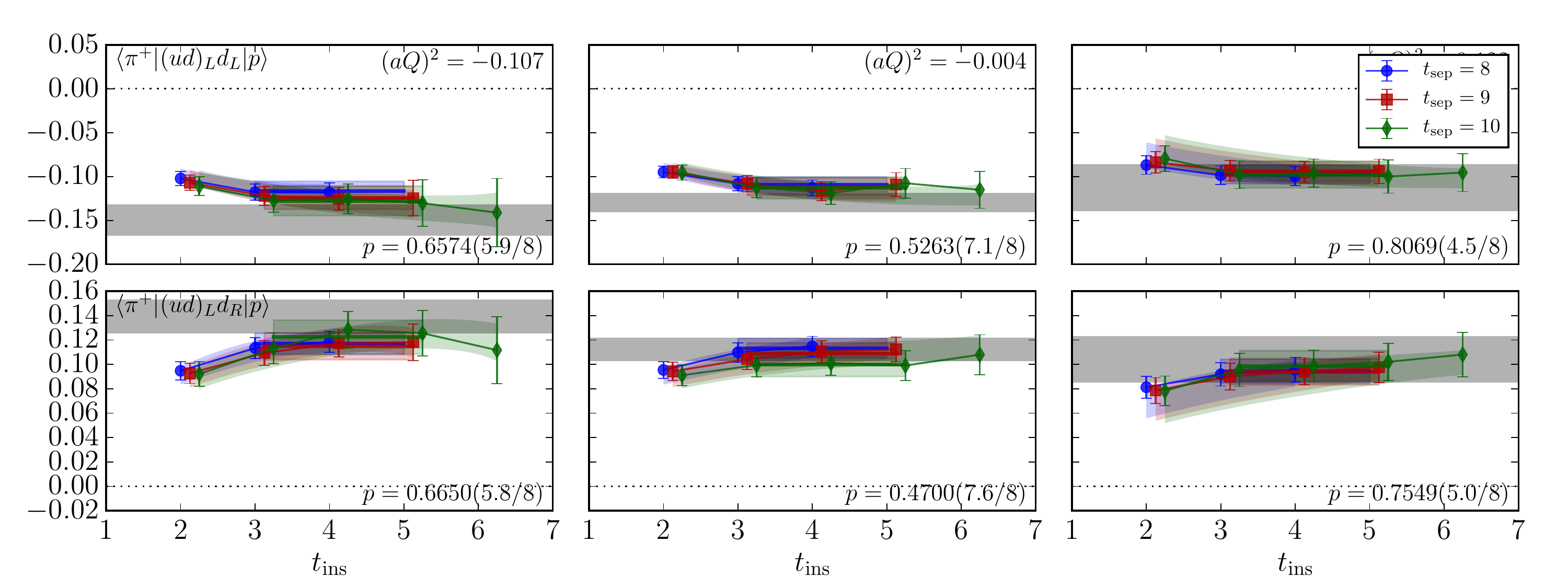}
\caption{\raggedright
  Two-state fits of renormalized ${N\to\pi}$ correlation functions for form factor $W_0$
  (top) and $W_1$ (bottom) at the three kinematic points on the 24ID lattices with
  $a\approx0.20\,\mathrm{fm}$.
  The data points show the ratio~(\ref{eqn:ratio}) and the bands show the corresponding fit
  functions as described in the text.
  The horizontal grey bands represent ground-state form factor values. 
  All error bars are  statistical and evaluated using bootstrap. 
  Fit quality ($p$-value) is estimated using the Hoteling distribution.
  \label{fig:c3fit_W01_N2pi_ID24}}
\end{figure}
\begin{figure}[h!]
\centering
\includegraphics[width=.85\textwidth]{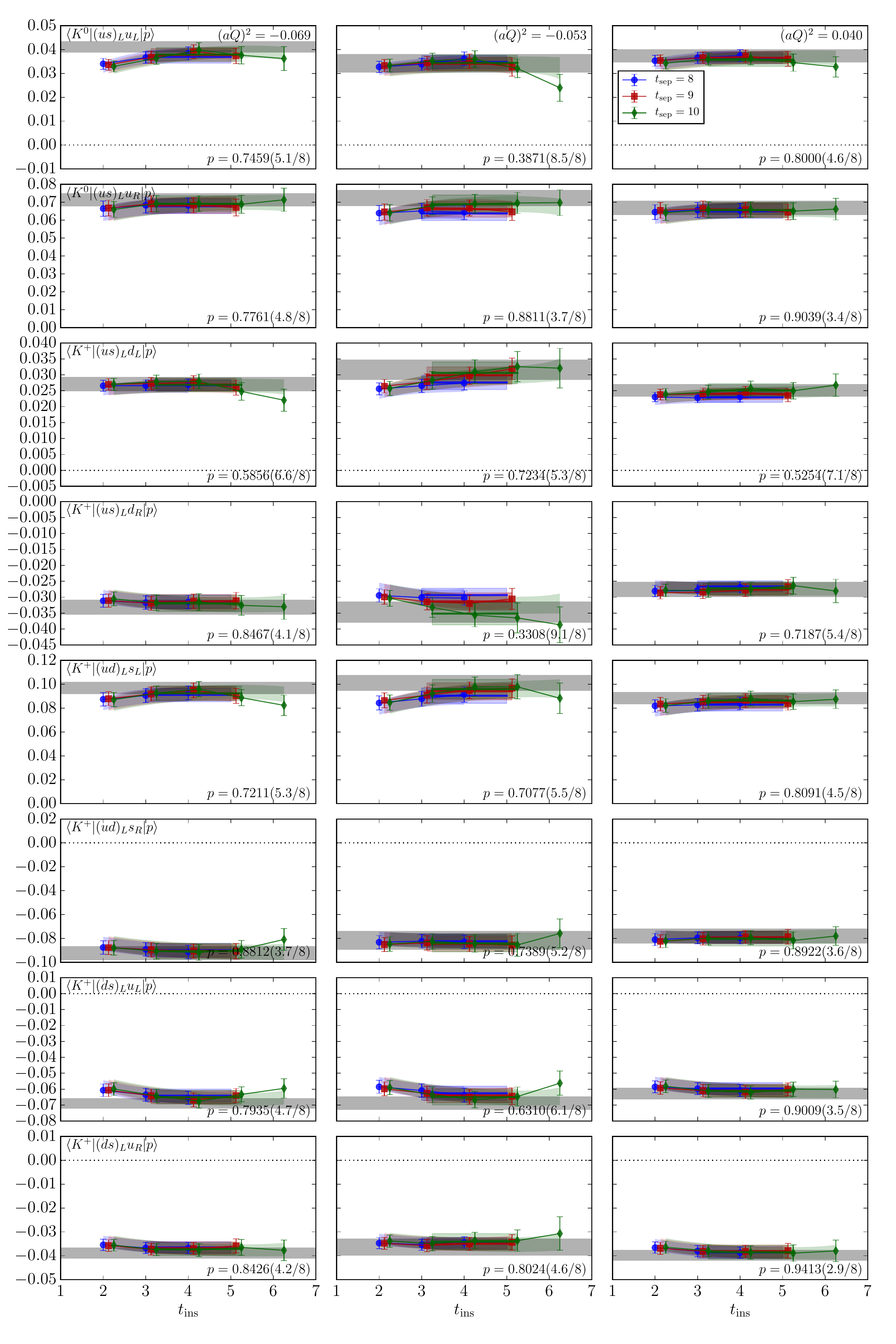}
\caption{\raggedright
  Two-state fits of the ${N\to K}$  form factor $W_0$ on 24ID.
  For explanation, see caption to Fig.~\ref{fig:c3fit_W01_N2pi_ID24}.
  \label{fig:c3fit_W0_N2K_ID24}}
\end{figure}
\begin{figure}[h!]
\centering
\includegraphics[width=.85\textwidth]{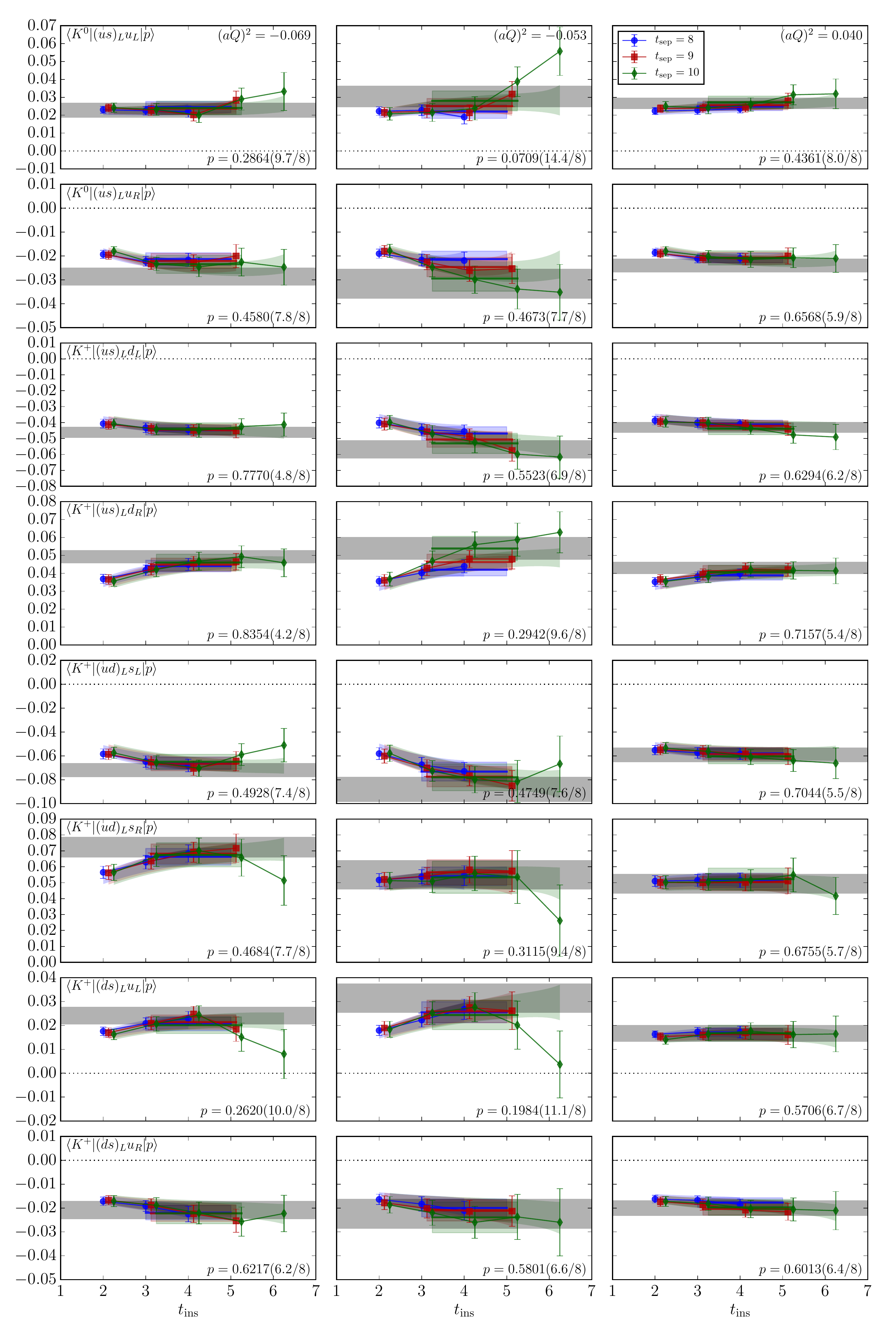}
\caption{\raggedright
  Two-state fits of the ${N\to K}$  form factor $W_1$ on 24ID.
  For explanation, see caption to Fig.~\ref{fig:c3fit_W01_N2pi_ID24}.
  \label{fig:c3fit_W1_N2K_ID24}}
\end{figure}
\begin{figure}[h!]
\centering
\includegraphics[width=.85\textwidth]{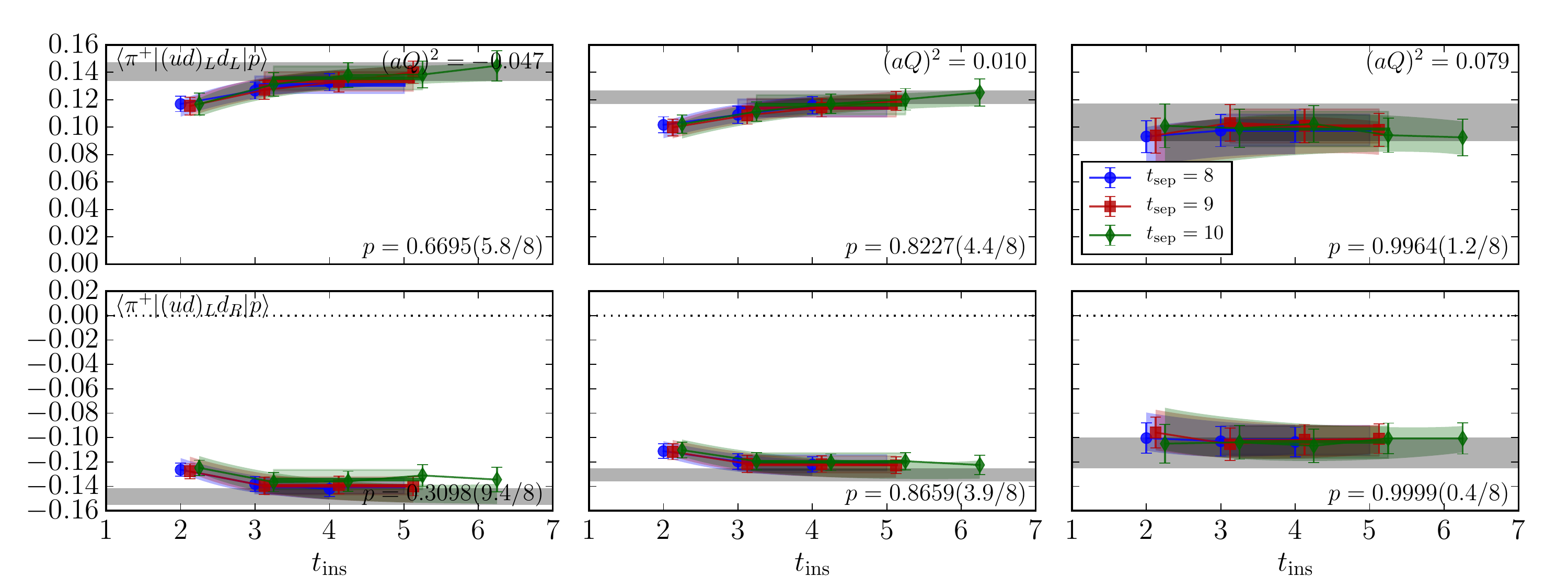}\\
\includegraphics[width=.85\textwidth]{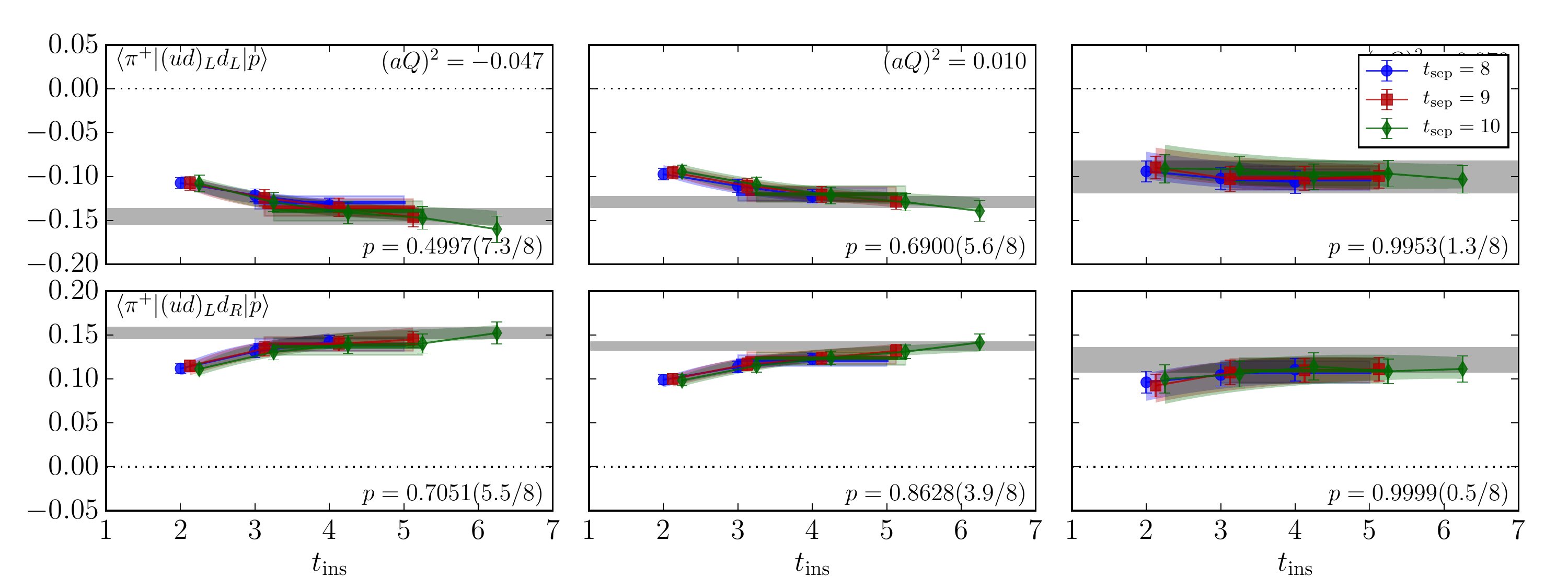}
\caption{\raggedright
  Two-state fits of the ${N\to\pi}$ correlation functions for form factor $W_0$ (top) and
  $W_1$ (bottom) at the three kinematic points on the 32ID lattices with $a\approx0.14\,\mathrm{fm}$.
  For explanation, see caption to Fig.~\ref{fig:c3fit_W01_N2pi_ID24}.
  \label{fig:c3fit_W01_N2pi_ID32}}
\end{figure}
\begin{figure}[h!]
\centering
\includegraphics[width=.85\textwidth]{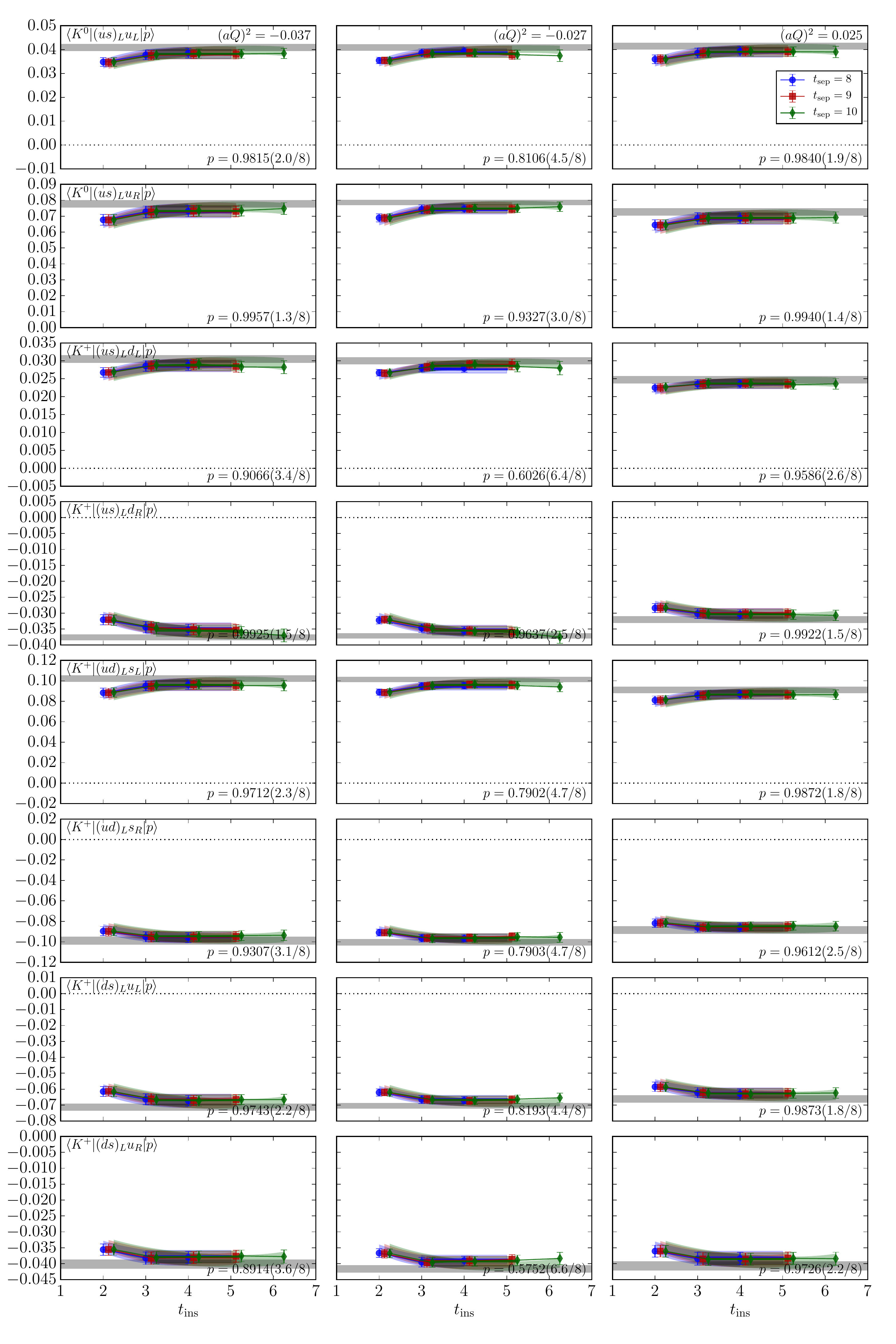}
\caption{\raggedright
  Two-state fits of the ${N\to K}$  form factor $W_0$ on 32ID.
  For explanation, see caption to Fig.~\ref{fig:c3fit_W01_N2pi_ID24}.
  \label{fig:c3fit_W0_N2K_ID32}}
\end{figure}
\begin{figure}[h!]
\centering
\includegraphics[width=.85\textwidth]{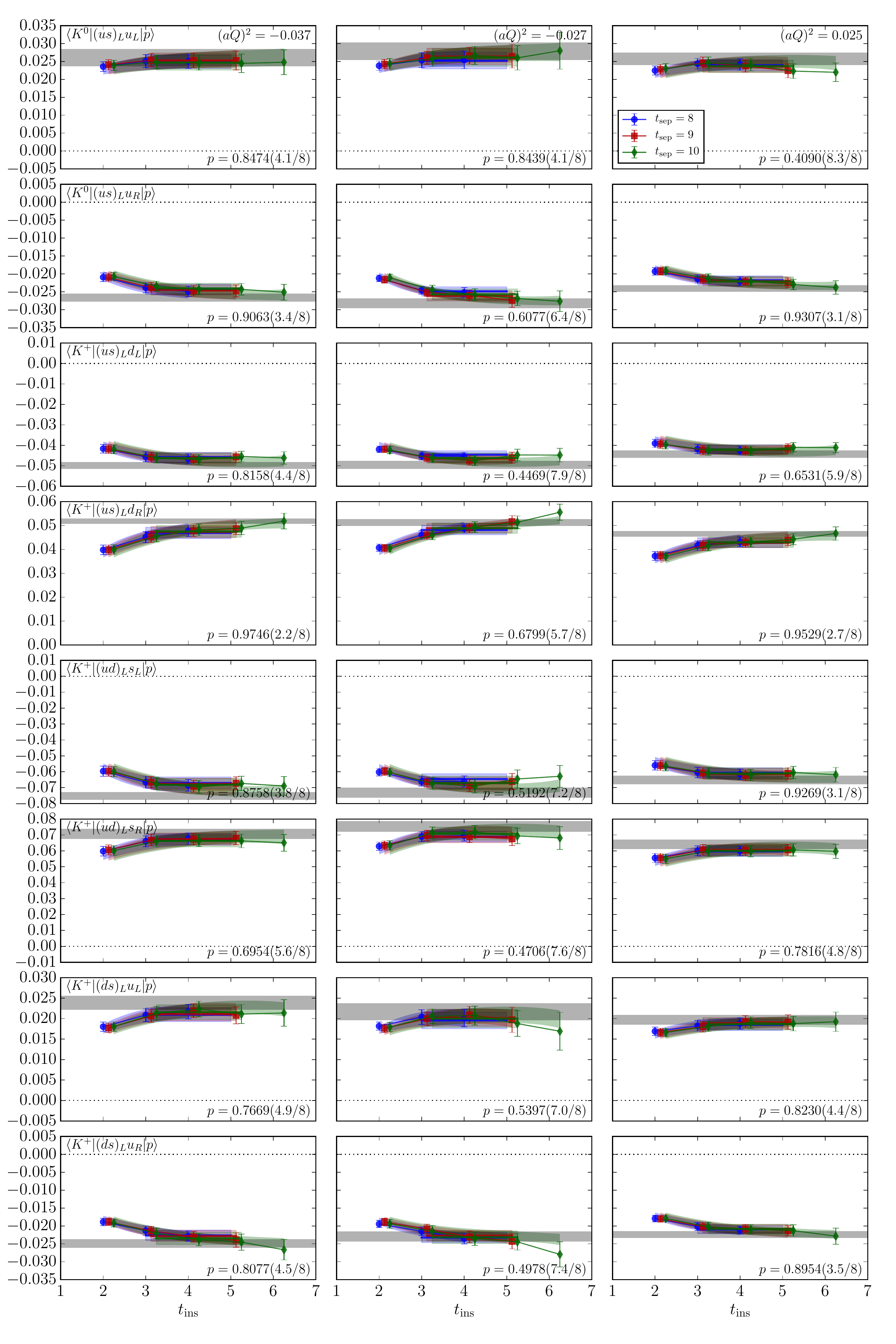}
\caption{\raggedright
  Two-state fits of the ${N\to K}$  form factor $W_1$ on 32ID.
  For explanation, see caption to Fig.~\ref{fig:c3fit_W01_N2pi_ID24}.
  \label{fig:c3fit_W1_N2K_ID32}}
\end{figure}

\begin{figure}[h!]
\centering
\includegraphics[width=.8\textwidth]{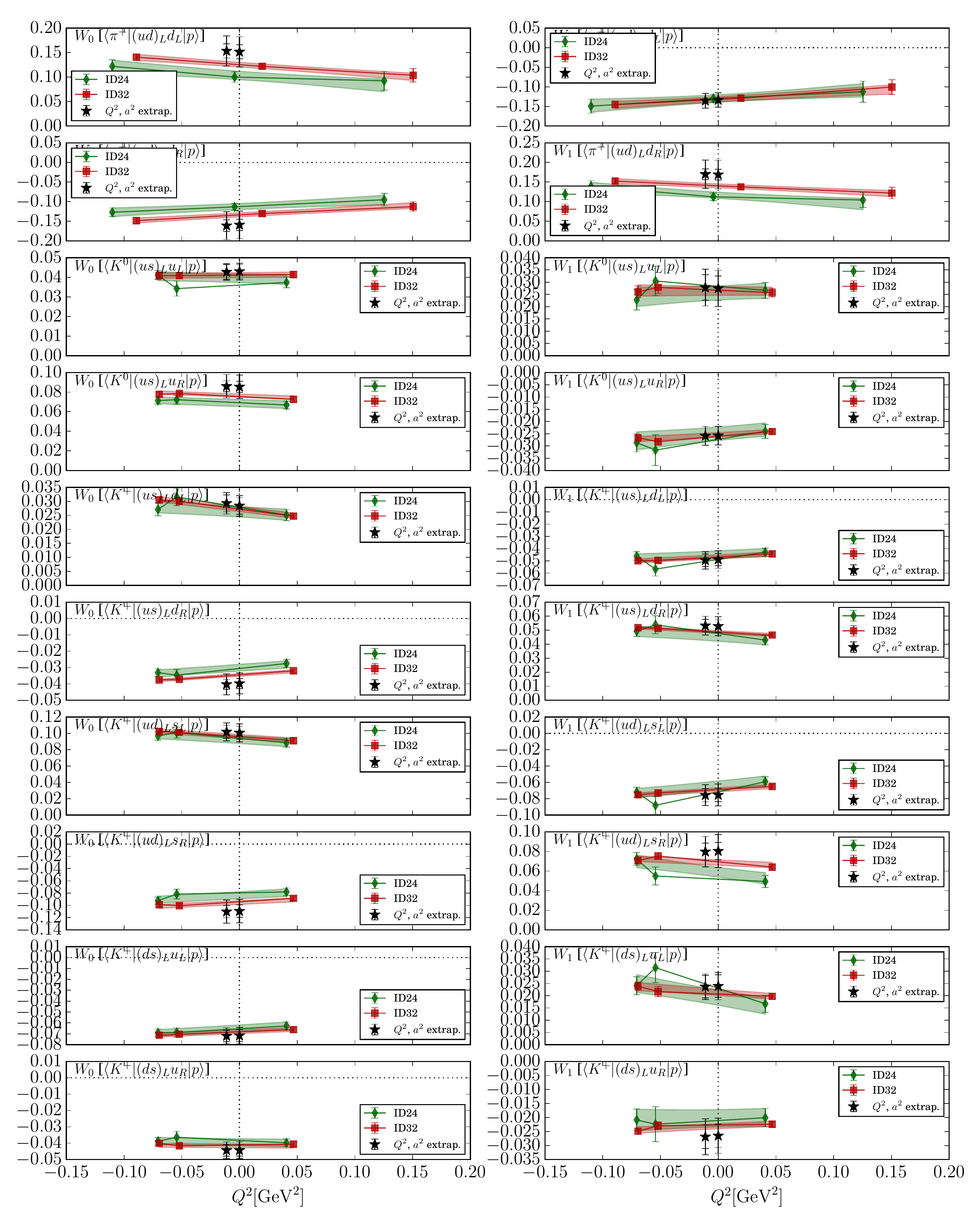}
\caption{\raggedright
  Linear interpolation of form factor data to the decay kinematic points $Q^2=-m_e^2\approx0$ 
  and $Q^2=-m_\mu^2$ (bands) for $W_0$ in the left column and $W_1$ in the right column
  followed by continuum extrapolation in $a^2$ (black stars).
  The 24ID and 32ID lattice data are shown with statistical uncertainties only, and the final
  extrapolated results are shown with statistical (smaller error bars) and the total (larger error
  bars) uncertainties including those from excited states and the continuum limit extrapolation
  as described in the text.
  \label{fig:Q2fit_a2fit_cmp}}
\end{figure}

\begin{table}[h!]
\caption{\raggedright
  Results for the form factors $W_{0,1}$ on the two ensembles and in the continuum limit at
  the two kinematic points $Q^2=0$ (first line) and $Q=-m_\mu^2$ (second line) renormalized to
  $\MSbar(2\,\mathrm{GeV})$. 
  The first uncertainty is statistical, the second is systematic due to excited states, and the
  third is the uncertainty of the continuum extrapolation.
  \label{tab:ff_results}}
\begin{tabular}{l| D..{16} D..{16} D..{16}}
\hline\hline
& \multicolumn{3}{c}{$W_0\,[\mathrm{GeV}^2]$} \\
& \multicolumn{1}{c}{24ID} & \multicolumn{1}{c}{32ID} & \multicolumn{1}{c}{cont.} \\
\hline
$\langle \pi^+|(ud)_L d_L|p\rangle$ &	0.1032(86)(26)	&	0.1252(48)(50)	&	0.151(14)(8)(26)	\\&  0.1050(87)(36)	&	0.1271(49)(50)	&	0.153(14)(7)(26)	\\
$\langle \pi^+|(ud)_L d_R|p\rangle$ &	-0.1125(78)(41)	&	-0.134(5)(11)	&	-0.159(15)(20)(25)	\\&  -0.1139(78)(45)	&	-0.136(5)(12)	&	-0.161(15)(20)(26)	\\
$\langle K^0|(us)_L u_L|p\rangle$   &	0.0395(22)(36)	&	0.0411(13)(25)	&	0.0430(38)(12)(19)	\\&  0.0397(22)(36)	&	0.0411(13)(25)	&	0.0427(37)(12)(16)	\\
$\langle K^0|(us)_L u_R|p\rangle$   &	0.0688(37)(19)	&	0.0764(17)(36)	&	0.0854(57)(55)(90)	\\&  0.0693(36)(20)	&	0.0769(17)(36)	&	0.0860(56)(55)(91)	\\
$\langle K^+|(us)_L d_L|p\rangle$   &	0.0263(19)(6)	&	0.0273(9)(11)	&	0.0284(30)(17)(12)	\\&  0.0266(19)(6)	&	0.0278(9)(11)	&	0.0293(30)(18)(15)	\\
$\langle K^+|(us)_L d_R|p\rangle$   &	-0.0301(21)(10)	&	-0.0345(9)(14)	&	-0.0398(31)(20)(52)	\\&  -0.0307(21)(10)	&	-0.0351(8)(15)	&	-0.0403(31)(20)(52)	\\
$\langle K^+|(ud)_L s_L|p\rangle$   &	0.0923(48)(35)	&	0.0961(26)(46)	&	0.1006(80)(60)(46)	\\&  0.0932(47)(37)	&	0.0972(26)(48)	&	0.1019(79)(60)(47)	\\
$\langle K^+|(ud)_L s_R|p\rangle$   &	-0.0835(58)(3)	&	-0.0954(32)(39)	&	-0.109(10)(8)(14)	\\&  -0.0846(58)(6)	&	-0.0964(32)(40)	&	-0.110(10)(8)(14)	\\
$\langle K^+|(ds)_L u_L|p\rangle$   &	-0.0651(33)(26)	&	-0.0681(18)(33)	&	-0.0717(54)(41)(35)	\\&  -0.0658(32)(28)	&	-0.0686(18)(34)	&	-0.0720(53)(40)(34)	\\
$\langle K^+|(ds)_L u_R|p\rangle$   &	-0.0394(22)(20)	&	-0.0417(11)(23)	&	-0.0443(35)(26)(27)	\\&  -0.0393(21)(21)	&	-0.0416(11)(23)	&	-0.0444(35)(26)(27)     \\
\hline\hline
& \multicolumn{3}{c}{$W_1\,[\mathrm{GeV}^2]$} \\
& \multicolumn{1}{c}{24ID} & \multicolumn{1}{c}{32ID} & \multicolumn{1}{c}{cont.} \\
\hline
$\langle \pi^+|(ud)_L d_L|p\rangle$ &	-0.130(10)(17)	&	-0.1316(67)(82)	&	-0.134(18)(2)(2)	\\&  -0.132(10)(17)	&	-0.1335(67)(81)	&	-0.136(19)(3)(2)	\\
$\langle \pi^+|(ud)_L d_R|p\rangle$ &	0.116(8)(11)	&	0.140(5)(14)	&	0.169(14)(18)(29)	\\&  0.118(8)(12)	&	0.142(5)(15)	&	0.170(14)(18)(28)	\\
$\langle K^0|(us)_L u_L|p\rangle$   &	0.0256(29)(4)	&	0.0264(18)(22)	&	0.0275(50)(53)(10)	\\&  0.0254(29)(4)	&	0.0265(19)(22)	&	0.0278(52)(52)(13)	\\
$\langle K^0|(us)_L u_R|p\rangle$   &	-0.0250(27)(30)	&	-0.0253(9)(18)	&	-0.0258(38)(3)(4)	\\&  -0.0254(28)(31)	&	-0.0256(9)(19)	&	-0.0259(38)(4)(2)	\\
$\langle K^+|(us)_L d_L|p\rangle$   &	-0.0448(30)(13)	&	-0.0467(17)(27)	&	-0.0489(51)(44)(22)	\\&  -0.0453(30)(16)	&	-0.0472(16)(28)	&	-0.0496(51)(43)(23)	\\
$\langle K^+|(us)_L d_R|p\rangle$   &	0.0452(31)(23)	&	0.0487(10)(25)	&	0.0529(45)(28)(42)	\\&  0.0458(31)(23)	&	0.0492(10)(26)	&	0.0532(45)(29)(40)	\\
$\langle K^+|(ud)_L s_L|p\rangle$   &	-0.0638(54)(24)	&	-0.0691(23)(52)	&	-0.0754(82)(86)(63)	\\&  -0.0653(54)(32)	&	-0.0701(23)(55)	&	-0.0757(80)(82)(57)	\\
$\langle K^+|(ud)_L s_R|p\rangle$   &	0.0588(50)(11)	&	0.0687(28)(43)	&	0.080(9)(8)(12)	        \\&  0.0605(50)(15)	&	0.0693(28)(43)	&	0.080(9)(8)(10)	        \\
$\langle K^+|(ds)_L u_L|p\rangle$   &	0.0192(31)(15)	&	0.0213(13)(16)	&	0.0239(46)(18)(26)	\\&  0.0201(31)(19)	&	0.0217(13)(17)	&	0.0237(46)(15)(19)	\\
$\langle K^+|(ds)_L u_R|p\rangle$   &	-0.0203(31)(5)	&	-0.0231(9)(12)	&	-0.0265(42)(32)(34)	\\&  -0.0204(31)(7)	&	-0.0233(9)(12)	&	-0.0269(42)(33)(35)     \\
\hline\hline
\end{tabular}
\end{table}


Values of the proton decay form factors $W_0$ and $W_1$ are extracted from the
three-point correlation functions~(\ref{eqn:pdecay_threept}) as follows:
\begin{enumerate}
\item
The projected lattice three-point functions~(\ref{eqn:pdecay_threept_proj}) are fitted to the
two-state Ansatz~(\ref{eqn:threept_exp2}) with proton and meson ground and excited state energies
fixed at values determined in the two-point function fits described above.
This linear fit yields nucleon-meson decay matrix elements up to the hadron operator normalization
factors $Z_{N,\Pi}$, which are also determined from the two-point function fits.
\item 
Form factors $W_{0,1}(Q^2)$ are computed From the ground-state matrix elements, 
at three kinematic points.
\item
On each ensemble, form factor data are interpolated to the points $Q_e^2=-m_e^2\approx0$ and
$Q_\mu^2=-m_\mu^2$ that correspond to the $N\to\Pi e$ and $N\to\Pi\mu$ decays, respectively (see
Fig.~\ref{fig:Q2fit_a2fit_cmp}).
\item
At each physical-decay kinematic point $Q_{\mu,e}^2$, linear extrapolations in $a^2$ are performed
to obtain the continuum-limit value (also shown in Fig.~\ref{fig:Q2fit_a2fit_cmp}).
\end{enumerate}

Due to the coarse lattice spacings, the fit ranges resulting in stable fits of the
excited state energy are very limited.
We find that nucleon and meson excitation energies $\Delta E_1$ obtained from fits with $\tmin=2$
lead to the most robust fits of the three-point functions on both ensembles.
In order to minimize excited state effects in the three-point functions, we omit
$\tskip^N=2$ points at the proton source and $\tskip^\Pi=4$ points at the meson sink, although
results are stable with respect to varying these numbers by $\pm1$.
To avoid unrealistic large fluctuations in $\chi^2$ values and fit parameters, the covariance matrix
for the fit is ``shrunk'' to its diagonal part
\begin{equation}
\tilde S(\lambda) = (1-\lambda)S + \lambda \mathrm{diag}(S).
\end{equation}
with ``shrinkage'' parameter $\lambda=0.1$. 
This is necessary due to strong correlations of data with different $\tsep$ that lead to
poorly-conditioned correlation matrices with eigenvalues as small as $10^{-5}$.

In each channel and at each momenta, we study two
projections~(\ref{eqn:ff_coeff},\ref{eqn:ff_coeff3}), from which the two form factors $W_{0,1}$
factors are computed directly.
Using parity, we take the average the left- and right-handed matrix elements, i.e., $LL$ with $RR$,
and $LR$ with $RL$\footnote{
  On the 24ID ensemble, the precision of the AMA approximation is different for the left-handed and
  right-handed components due to the asymmetric zMobius-action coefficients in the fifth dimension.
  To accommodate that, we compute the average of the left- and right-handed matrix elements weighted
  with $\propto\sigma^{-2}$, where $\sigma$ is the statistical fluctuation.
}.
Separate fits are performed independently for all channels and kinematic points.
In Figures~\ref{fig:c3fit_W01_N2pi_ID24}--\ref{fig:c3fit_W1_N2K_ID32}, we show results of these
fits in terms of the form factor $W_{0,1}$ values in the $\MSbar(2\,\mathrm{GeV})$
scheme and using physical units $\mathrm{GeV}^2$.
The time-dependent ``ratio'' data points are computed using Eq.~(\ref{eqn:ratio}), and the plateau
averages are computed over $\tins=(3\ldots5)a$.
To examine the agreement between the data and the fits, we also show similar ratios reconstructed from
the fit functions~(\ref{eqn:c2fit_Pi},\ref{eqn:c2fit_N},\ref{eqn:threept_exp2}).
In each panel, we also show Hoteling $p$-values along with the respective values of
$\chi^2/\mathrm{d.o.f.}$ we use to assess the fit quality. 

We observe close agreement between the plateau and the ground-state fit values indicating that
excited-state contributions are negligible.
The statistical uncertainties of the ground-state fit values are close to
those of the plateau values at the largest source-sink separation and are thus conservative.
In channels with the final state $\pi(0,0,2)$ on the 24ID ensemble, fluctuations are larger due to the
larger uncertainty in the corresponding two-point functions.
Systematic uncertainties due to excited states are conservatively estimated from the differences
between values obtained from the fits and the plateau averages at the largest source-sink separation
$\tsep=10a$.
These systematic errors are propagated forward to the final $Q^2$- and continuum-extrapolated
results.

Using data at the three kinematic points, we perform linear interpolation in $Q^2$ to
obtain values at $Q^2=-m_e^2\approx0$ and $Q^2=-m_\mu^2$.
The decay-kinematic data points are then extrapolated to the continuum limit as 
$W(a) \sim W^\text{cont} + W^\prime a^2$.
Such expected scaling of discretization errors is justified by the automatic $O(a^2)$ improvement
due to chiral symmetry of the fermion action.
Having only two values of the lattice spacing, it is impossible to estimate systematic uncertainty of such
extrapolation in a robust way; therefore, we resort to a conservative estimate from the discrepancy
between the continuum-extrapolated results and the results from the finer 32ID ensemble.
The momentum interpolations are shown in Fig.~\ref{fig:Q2fit_a2fit_cmp} for both ensembles,
together with the final continuum-extrapolated values.
The individual lattice data points and their $Q^2$ fit bands are shown only with statistical
uncertainties, while the extrapolated values are shown with statistical and total uncertainties.
The latter include systematic uncertainties due to the excited states and the continuum
extrapolation.

Our final results for both form factors $W_0$ and $W_1$ are collected in Tab.~\ref{tab:ff_results}.
For completeness, we include values on both ensembles as well as their continuum-extrapolated values.
We quote separate statistical, excited-state and continuum-extrapolation systematic uncertainties
where appropriate.

\subsection{Proton decay amplitudes}
In this section, we present our determination of proton decay constants~(\ref{eqn:pdecay_lec}).
Combined with phenomenological constants $D$ and $F$ from spin physics, these parameters yield
leading-order ChPT estimates of proton-meson decay amplitudes (see Appendix~\ref{sec:app_indirect}).

\begin{figure}[h!]
\centering
\includegraphics[width=.49\textwidth]{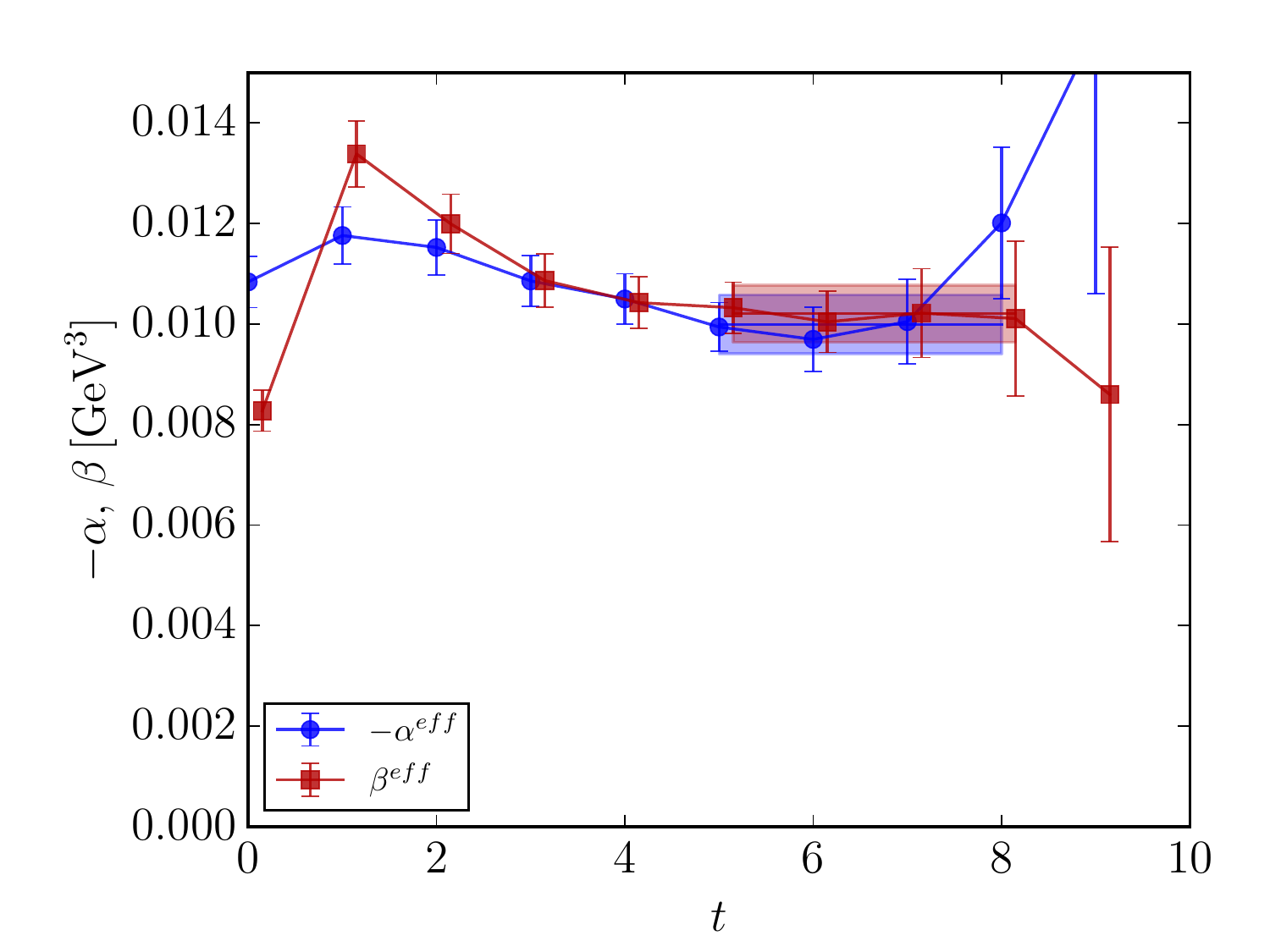}~
\includegraphics[width=.49\textwidth]{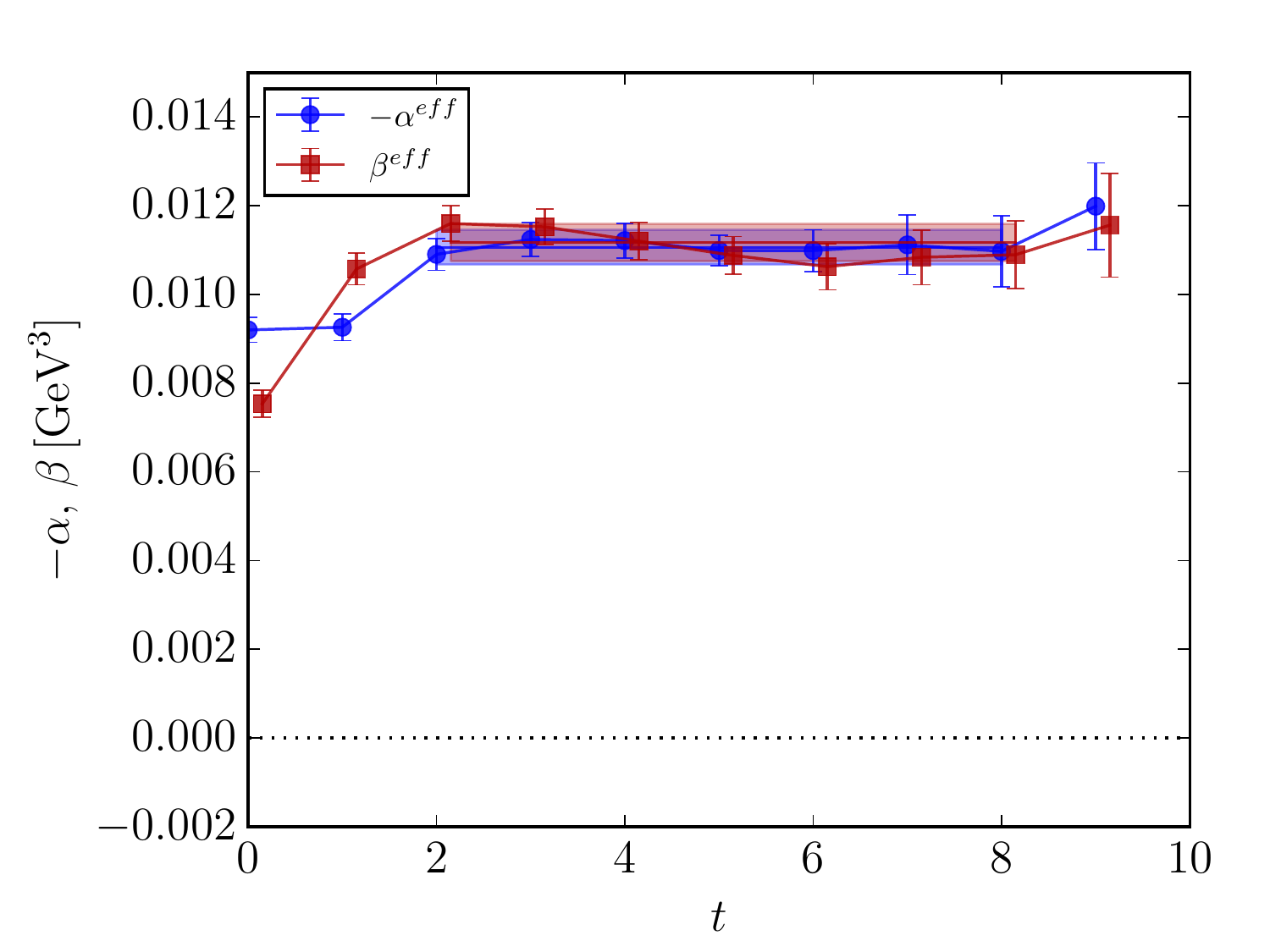}\\
\caption{\raggedright
  Ratios~(\ref{eqn:pdecay_c2lec_ratio}) determining the proton decay constants $(-\alpha)$ 
  and $\beta$ on 24ID (left) and 32ID (right) ensembles.
  \label{fig:pdecay_lec}}
\end{figure}
\begin{figure}[h!]
\centering
\includegraphics[width=.5\textwidth]{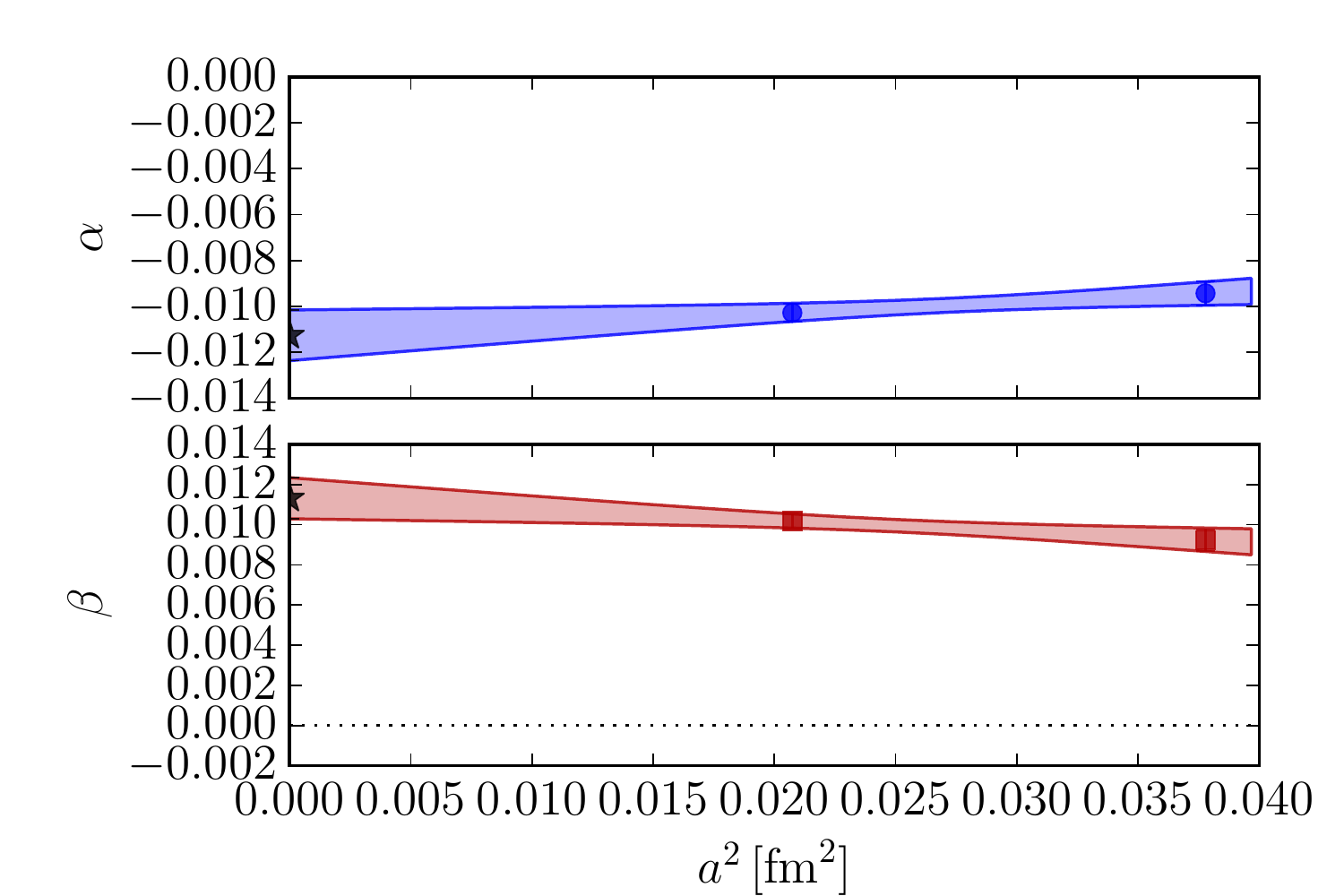}
\caption{\raggedright
  Continuum $O(a^2)$ extrapolations of the proton decay constants $(-\alpha)$ and $\beta$.
  \label{fig:pdecay_lec_cont}}
\end{figure}

We determine the proton decay constants from the two-point correlation functions of the proton
creation and proton decay operators:
\begin{equation}
\label{eqn:pdecay_c2lec_corr}
C^{\mcO\bar{N}}_+(\vec k, t) 
= \sum_{\vec x} \, e^{-i\vec k\vec x} \, \frac{(1+\gamma_4)_{\beta\alpha}}2
  \langle\mcO_\alpha(\vec x, t) \, \bar{N}(0)\rangle
\overset{t\to\infty} = \frac{f_N}{\sqrt{Z_N(\vec k)}} C_+^{N\bar{N}}(\vec k,t)\,,
\end{equation}
where $f_N=\{\alpha,\beta\}$ for $\mcO=\mcO^{(ud)u}_{RL,LL}$, respectively.
We extract these constants using the ratio
\begin{equation}
\label{eqn:pdecay_c2lec_ratio}
R^{\mcO \bar{N}}(\vec k, t) 
  = \sqrt{Z_N(\vec k)}\,\frac{C_+^{\mcO\bar{N}}(\vec k, t)}{C_+^{N\bar{N}(\vec k, t)}}
\overset{t\to\infty} = f_N\,,
\end{equation}
where $Z_N(\vec k)$ is obtained from two-state fits~(\ref{eqn:Zdef}).
These ratios are shown in Fig.~\ref{fig:pdecay_lec} for both ensembles for $\vec k=0$.
Although the proton decay constants can be extracted from correlators with any momentum $\vec k$, 
we study only zero-momentum ($\vec k=0$) data that has the highest statistical precision.
We observe much less excited-state effects in the case of 32ID ensemble, which we 
attribute to over-smearing of quark sources on this ensemble.
This over-smearing leads to stronger suppression of the excited states in the ``smeared-point''
correlator~(\ref{eqn:pdecay_c2lec_ratio}), while its statistical fluctuations are mostly cancelled
in the combination with the ``smeared-smeared'' two-point function~(\ref{eqn:twopt_N}) and its
parameter $Z_N$.
We estimate the decay constant values from plateaus in the time range $5\le t/a\le8$ for the 24ID
ensemble and $2\le t/a\le8$ for the 32ID ensemble.

\begin{table}
\caption{\raggedright
  Results for the proton decay constants $\alpha$, $\beta$ on the two ensembles and in the continuum
  limit.
  The first uncertainty is statistical, the second is systematic due to  the continuum extrapolation.
  \label{tab:pdecay_lec_results}}
\begin{tabular}{l| D..{12} D..{12} D..{12}}
\hline\hline
& \multicolumn{1}{c}{24ID} & \multicolumn{1}{c}{32ID} & \multicolumn{1}{c}{cont.} \\
\hline
$\alpha$ &
-0.0999(59)	&	-0.01106(39)	&	-0.01257(111)	\\
$\beta$ & 
0.01020(57)	&	0.01117(42)	&	 0.01269(107) \\
\hline\hline
\end{tabular}
\end{table}

The proton decay constant results are collected in Tab.~\ref{tab:pdecay_lec_results}.
Continuum extrapolations $\sim(f_N + f_N^\prime a^2)$ of the proton decay constants $\alpha,\beta$
are shown in Fig.~\ref{fig:pdecay_lec_cont}.
Similarly to the proton decay amplitudes, uncertainties from the continuum extrapolation are
estimated as the difference between the extrapolated results and the values on the finer 32ID ensemble.
For both constants, the statistical as well as systematic uncertainties from continuum extrapolation
are roughly $10\%$, so that the total uncertainties are comparable to those in the direct determination 
of the proton decay form factors $W_0$.

\section{Discussion
  \label{sec:concl}}
\begin{figure}
\centering
\includegraphics[width=0.49\textwidth]{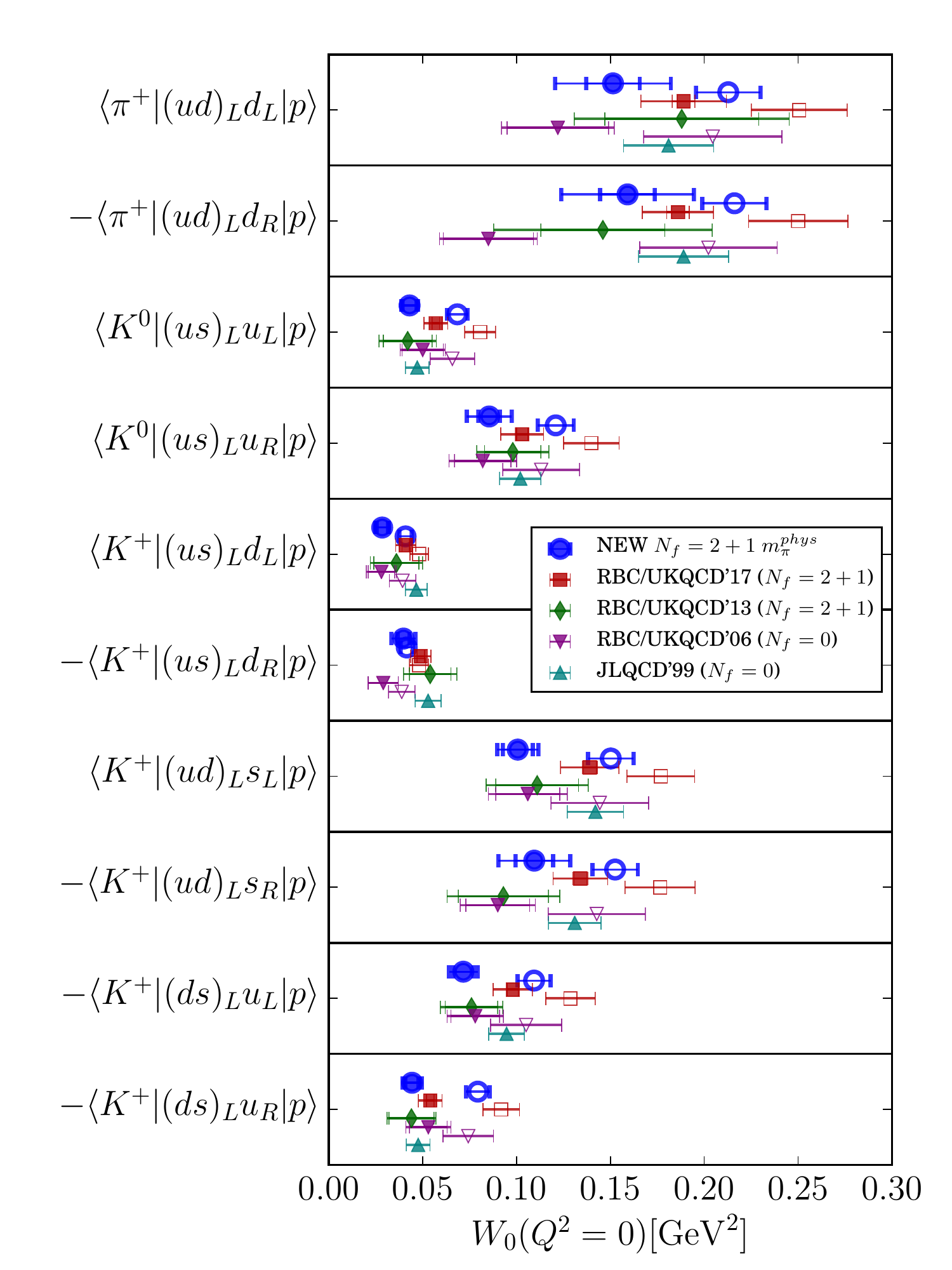}
\caption{\raggedright
  Comparison of our results (``NEW'') for the proton decay amplitudes $W_0(0)$ computed directly 
  (filled symbols) and indirectly (open symbols) to previous 
  determinations~\cite{Aoki:2017puj,Aoki:2006ib,Aoki:1999tw}.
  All results are renormalized to the $\MSbar(2\,\mathrm{GeV})$ scheme.
  \label{fig:cmp_prev}}
\end{figure}

The main finding of our paper is that proton decay amplitudes \emph{are not suppressed} as the
quark masses decrease and approach their physical values, and thus findings in previous lattice
calculations (e.g., Ref.~\cite{Aoki:2017puj}) are valid.
Using physical quark masses and absence of chiral extrapolation have resulted in a dramatic improvement 
of precision, yielding results that are perfectly consistent with those of Ref.~\cite{Aoki:2017puj}.
Therefore, dynamical suppression of proton decay amplitudes due to nonperturbative QCD
dynamics, as suggested in Ref.~\cite{Martin:2011nd}, is unlikely, at least at the physical
$u$-,$d$-quark masses, and \emph{the stringent constraints on the Grand-Unified Theories remain
unchanged}.

In this work, we have used the $N_f=(2+1)$-flavor chirally-symmetric Domain Wall fermion action with
physical quark masses on a lattice to compute transition matrix elements from proton to pion or kaon
(``direct method'').
We have omitted the $\eta$-channel decay amplitudes because they require evaluation of disconnected
contractions to the two- and three-point functions, without which the results would be totally
misleading at the physical point.
Lattice calculations in this work have been performed in the exact isospin limit and without QED
corrections, which is unlikely to introduce significant systematic bias compared to the current
level of precision.
Respective matrix elements for the neutron decays of which can also be potentially observed inside
nuclei are related to those of the proton by isospin symmetry. 
Additinally, we have also calculated the proton (neutron) decay constants that can be used for
computing rates of non-hadronic proton decays such as $p\to3\ell$.

We have obtained proton decay form factors at the relevant kinematic points $Q^2=-m_\ell^2$
by computing them at three small values of the lepton 4-momentum squared
 $|Q^2|\lesssim0.15\text{ GeV}^2$ and performing linear interpolations.
Form factor values are reported at the kinematic points with an electron and a muon in the final
state, although the differences are insignificant.
Our results are nonperturbatively renormalized using a variant of SMOM scheme suitable for our
coarse lattice spacings and converted to the $\MSbar$ scheme using $O(\alpha_S^3)$ perturbative
calculations, which is expected to have only negligible systematic uncertainties.
We find no signs of mixing between operators constructed from chiral fermion fields; absence of such
mixing indicates that chiral symmetry is preserved in our calculations.

We compare our results to earlier studies in Fig.~\ref{fig:cmp_prev}, where we show results from
direct and indirect calculations of the $p\to\pi\bar\ell$ and $p\to K\bar\ell$ proton decay
amplitudes (assuming $m_\ell\approx0$).
Our results are in very good agreement with earlier direct calculations that used dynamical Domain
Wall quark action at heavier pion masses~\cite{Aoki:2017puj}.
Also, our results are in reasonable agreement with quenched calculations that used Domain
Wall~\cite{Aoki:2006ib} and Wilson~\cite{Aoki:1999tw} fermions.
We have also found reasonable agreement of our indirect determination of the amplitudes with the 
analogous quenched results obtained earlier~\cite{Aoki:2006ib}.
In comparison with the direct determination, the indirect determination have been found to yield 
results systematically higher in magnitude; similar pattern was observed in Ref.~\cite{Aoki:2006ib}.

The precision of our results can be improved with additional statistics to reduce the stochastic
uncertainty, which would also help further constrain excited-stated effects and systematic errors
associated with them.
Although we generally observe nearly-perfect agreement between fits and ``plateaus'', we opt to
estimate excited-state effects in a very conservative fashion.
For this reason, these effects dominate the total uncertainty in some instances.
Further, since we used relatively long Euclidean time source-sink separations
$(t_\Pi-t_N)=1\ldots2\,\mathrm{fm}$, it is extremely unlikely that the true excited-state systematic
effects exceed our estimates.

Finite-volume effects may contribute to systematic uncertainty because both our ensembles have
similar lattice volume $\approx(4.6\,\mathrm{fm})^3$ that correspond to $m_\pi
L\approx3.3$.
A naive estimate suggests that these effects are of the order of $e^{-m_\pi L}\approx=0.04$ 
which is substantially below the combined quoted uncertainties in Tab.~\ref{tab:ff_results}.
Until a study with a different physical volume(s) is performed, it is impossible to estimate finite
volume effects with better certainty.

The largest potential sources of systematic uncertainty are discretization effects.
We use two ensembles with different, albeit coarse, values of the lattice spacing.
Due to our improved gauge action and chirally symmetric fermion action, discretization effects must
vanish as even powers of the lattice spacing $c_2 a^2 + c_4 a^4 +\ldots$.
With two lattice spacings, only $O(a^2)$ effects can be evaluated and removed.
Our results may be subject to the higher $O(a^4)$ discretization effects, which are impossible to
control without additinoal calculations with different lattice spacings.
Although we observe very good scaling of our results indicating that discretization errors are
generally small, we estimate our discretization uncertainties in a conservative fashion, which is
robust unless there is significant cancellation between $O(a^2)$ and $O(a^4)$ or higher effects.
However, such scaling violations are extremely unlikely since other observables computed on these
lattices are consistent with calculations on finer lattices~\cite{Blum:2014tka,Tu:2020vpn}, and the
hadron dispersion relations are accurately reproduced on both ensembles
(see Figs.~\ref{fig:fits_dispersion_ID24},\ref{fig:fits_dispersion_ID32}).

Despite conservative and likely overestimated systematic uncertainties, we have been able to
determine the nucleon decay constants $\alpha,\beta$ and form factors $W_{0,1}$ with $10-20\%$
precision, including the stochastic uncertainty.
This finding  definitively excludes suppression of nucleon decay matrix elements at light quark
masses, and thus removes the remaining systematic uncertainty in constraining some
Grand-Unified theories and completely excluding others such as (SUSY) $SU(5)$.

\begin{acknowledgments}
The authors would like to thank Christoph Lenher, Tom Blum, Eigo Shintani, Hooman Davoudiasl, and
Robert Shrock for many useful discussions.
During this work, S.S. and J.Y. were supported by the National Science Foundation under CAREER Award
PHY-1847893.
S.S. has also been supported by the RHIC Physics Fellow Program of the RIKEN BNL Research Center.
T.I. was supported through Brookhaven National Laboratory, the Laboratory Directed Research and
Development (LDRD) program No.~21-043, and by Program Development Fund No.~NPP~PD~19-025.
P.B. acknowledges Wolfson Fellowship WM160035, an Alan Turing Fellowship, and STFC grants ST/P000630/1,
ST/M006530/1, ST/L000458/1, ST/K005790/1, ST/K005804/1, ST/L000458/1. 
P.B., T.I., and A.S.  have also been supported in part by the U.S. Department of Energy, Office of
Science, Office of Nuclear Physics under the Contract No.~DE-SC-0012704 (BNL).
Y.A. acknowledges JSPS KAKENHI Grant No. 16K05320.
The computations were performed using the Qlua software suite~\cite{qlua-software} with (z)M\"obius
solvers from the Grid library~\cite{grid-software}.
The gauge configurations and (z)M\"obius eigenvectors have been generously provided by the RBC/UKQCD
collaboration.
Computations for this work were carried out on facilities of the USQCD Collaboration, which
are funded by the Office of Science of the U.S. Department of Energy.

\end{acknowledgments}

\appendix
\section{Conventions}\label{sec:app_convent}
In this Appendix section, we summarize the conventions that clarify definitions of operators and 
matrix elements throughout the paper.
The Euclidean $\gamma$-matrices we use, $\gamma_\mu=\gamma_\mu^\dag$, satisfy the same relations
as in, e.g., Ref.~\cite{Abramczyk:2017oxr}.
Positive-parity spinors are governed by the continuum limit of the lattice Dirac equation,
\begin{equation}
(i\slashed{p} + m) u(\vec p)= 0,\quad 
\bar{u}(\vec p) (i\slashed{p} + m) = 0\,,
\end{equation}
where $i\slashed{p} = -E(\vec p) \gamma_4 + i \vec\gamma \cdot \vec p$ and $E(p)=\sqrt{\vec p^2+m^2}$ 
is the on-shell energy, and the momentum states are defined in accordance with 
Eqs.~(\ref{eqn:pdecay_threept},\ref{eqn:twopt_Pi},\ref{eqn:twopt_N}).

The charge-conjugated spinors
\begin{equation}
v^C = C^{-1} (\bar{v})^T,\quad \bar{v}^C = - v^T C
\end{equation}
satisfy equations
\begin{equation}
(i\slashed{p} - m) v^C(\vec p)= 0,\quad 
\bar{v}^C(\vec p) (i\slashed{p} - m) = 0\,.
\end{equation}
where the Euclidean charge-conjugation matrix $C=\gamma_2\gamma_4$ satisfies
\begin{equation}
C \gamma_\mu C^{-1} = -\gamma_\mu^T\,, 
\quad C \sigma_{\mu\nu} C^{-1} = -\sigma_{\mu\nu}^T\,.
\end{equation}

Throughout the paper, we use the relativistic normalization of the particle states and matrix elements,
which is compatible with Eq.~\ref{eqn:decayrate} and is is typical for these quantities (see, e.g.,
Ref.~\cite{Aoki:2017puj}) and 
\begin{align}
\langle N(\vec k^\prime, s^\prime) |  N(\vec k, s) \rangle 
  &= \sqrt{2 E_{\vec k}} \, \delta^{(3)}(\vec k^\prime - \vec k) \, \delta^{s^\prime s}\,,\\
\langle \ell(\vec q^\prime, s^\prime) |  \ell(\vec q, s) \rangle 
  &= \sqrt{2 E_{\vec q}} \, \delta^{(3)}(\vec q^\prime - \vec q) \, \delta^{s^\prime s}\,,\\
\langle \Pi(\vec p^\prime) |  \Pi(\vec p) \rangle 
  &= \sqrt{2 E_{\vec p}} \, \delta^{(3)}(\vec p^\prime - \vec p)\,,
\end{align}
With this convention, the form factors $W_{0,1}$~(\ref{eqn:pdecay_ff_def}) have mass dimension 2
and the low-energy constants $\alpha,\beta$~(\ref{eqn:pdecay_lec}) have dimension 3.

\section{Proton decay amplitudes in ChPT}\label{sec:app_indirect}
According to the chiral Lagrangian method~\cite{Claudson:1981gh,Aoki:1999tw},
each decay matrix element can be calculated using the proton decay constants
$\alpha,\beta$ as follows:
\begin{align}
\langle\pi^+| (ud)_L u_L |p\rangle &= \phantom{-}\frac{\beta}{f}  (1+D+F),\\
\langle\pi^+| (ud)_L u_R |p\rangle &= \phantom{-}\frac{\alpha}{f} (1+D+F),\\
\langle K^0 | (us)_L u_L |p\rangle &= \phantom{-}\frac{\beta}{f}  \left(1-(D-F)\frac{m_N}{m_B}\right),\\
\langle K^0 | (us)_L u_R |p\rangle &=           -\frac{\alpha}{f} \left(1+(D-F)\frac{m_N}{m_B}\right),\\
\langle K^+ | (us)_L d_L |p\rangle &= \phantom{-}\frac{\beta}{f}  \left(\frac{2D}{3}\frac{m_N}{m_B}\right) \\
\langle K^+ | (us)_L d_R |p\rangle &= \phantom{-}\frac{\alpha}{f} \left(\frac{2D}{3}\frac{m_N}{m_B}\right) \\
\langle K^+ | (ud)_L s_L |p\rangle &= \phantom{-}\frac{\beta}{f}  \left(1+\left(\frac{D}{3}+F\right)\frac{m_N}{m_B}\right),\\
\langle K^+ | (ud)_L s_R |p\rangle &= \phantom{-}\frac{\alpha}{f} \left(1+\left(\frac{D}{3}+F\right)\frac{m_N}{m_B}\right),\\
\langle K^+ | (ds)_L u_L |p\rangle &=           -\frac{\beta}{f}  \left(1-\left(\frac{D}{3}-F\right)\frac{m_N}{m_B}\right),\\
\langle K^+ | (ds)_L u_R |p\rangle &= \phantom{-}\frac{\alpha}{f} \left(1+\left(\frac{D}{3}-F\right)\frac{m_N}{m_B}\right),
\end{align}
where D=0.8, F=0.47, $ m_N = 0.94 $GeV, $ m_B = 1.15 $GeV, and $ af = 0.13055 $. 

\section{Perturbative renormalization}\label{sec:app_renorm}
Throughout the paper, the uniform convention for renormalization factors of quark fields and
operators is to convert from bare to renormalized quantities,
\begin{equation}
\mcO^R(\mu) = Z_\mcO^{R[\text{reg.}]}(\mu)\mcO^{[\text{reg.}]}\,,
\quad
q^R(\mu) = \sqrt{Z_q^{R[\text{reg.}]}(\mu)} q^{[\text{reg.}]}\,,
\end{equation}
where $\mu$ is the scale associated with the renormalization scheme $R$ 
and ``reg.'' is the regulator $\epsilon$ (dim.reg.) or $a$ (lattice).
The anomalous dimensions are defined as 
\begin{equation}
\gamma_X = \frac{d\log Z_X}{d\log\mu}
\end{equation}
for $X=\mcO$  or $q$.
These conventions differ from some of the references.

To convert operators normalization from the $\SMOMg/\SYMpd$ to the $\MSbar$ scheme, we use the
$\MSbar$-renormalized amputated Green's function of the three-quark operator with external quark
fields with the $\SYMpd$ momenta $p^2=k^2=r^2=\mu^2$~\cite{Gracey:2012gx}
\begin{equation}
\label{eqn:pert_lambda3q}
\lp[\Lambda_\pm^{\MSbar}\rp]_{\SYMpd} 
  =  1 + 0.989426\lp(\frac{\alpha_S}{4\pi}\rp) 
    + (41.53105\mp1.69085 -3.91418 N_f) \lp(\frac{\alpha_S}{4\pi}\rp)^2\,.
\end{equation}
The multiplicatively renormalized (diagonal) Green's functions~(\ref{eqn:pert_lambda3q}) are obtained
for operators $\mcO_\pm = \mcO^{3q}_{SS} \pm \mcO^{3q}_{PP}$ with spin-color projectors
$\Pi_\pm=\frac12(\Pi^{3q}_{SS} \pm \Pi^{3q}_{PP})$, respectively (see Eq.~(\ref{eqn:proj_new})). 
Since the quark field is also $\MSbar$-renormalized in Eq.~(\ref{eqn:pert_lambda3q}), the difference
from the lattice scheme for $Z_q$ must be taken into account to get perturbative conversion factors 
for the three-quark operators,
\begin{equation}
C_\pm^{\MSbar\leftarrow \SMOMg/\SYMpd}
  = \lp(\frac{ Z^{\MSbar}_\pm(|p|)}{ Z^{\SMOMg/\SYMpd}_\pm(|p|) }\rp) 
  = \lp[\Lambda_\pm^{\MSbar}\rp]_{\SYMpd} \cdot C_q^{\MSbar\leftarrow\SMOMg}\,,
\end{equation}
where the field conversion factor has been computed in Ref.~\cite{Almeida:2010ns}
\begin{equation}
C_q^{\MSbar\leftarrow \SMOMg/\SYMpd}
  = \lp(\frac{ Z_q^{\MSbar} }{ Z_q^{\SMOMg} }\rp) 
  = 1 + \frac43 \lp(\frac{\alpha_S}{4\pi}\rp) + (9.59901 + 0.185185 N_f) \lp(\frac{\alpha_S}{4\pi}\rp)^2\,.
\end{equation}

Finally, the anomalous dimensions for operators $\mcO^{3q}_\pm$ are also provided in
Ref.~\cite{Gracey:2012gx} to the $O(\alpha_S^3)$ order
,
\begin{align}
\gamma^{\MSbar}_+ 
&=-4 \lp(\frac{\alpha_S}{4\pi}\rp)
  +\frac29(-2N_f -21) \lp(\frac{\alpha_S}{4\pi}\rp)^2
  +\frac1{81}\big(260 N_f^2 +(4320\zeta_3 -4740)N_f +2592\zeta_3 +22563 \big) \lp(\frac{\alpha_S}{4\pi}\rp)^3
\\
\gamma^{\MSbar}_- 
&=-4 \lp(\frac{\alpha_S}{4\pi}\rp) 
  +\frac29(-2N_f -81) \lp(\frac{\alpha_S}{4\pi}\rp)^2
  +\frac1{81}\big(260 N_f^2 +(4320\zeta_3 -4572)N_f + 24399 \big) \lp(\frac{\alpha_S}{4\pi}\rp)^3
\end{align}

\bibliography{main}

\end{document}